\documentclass[trackchanges,resetfootnote]{aastex7}

 \usepackage[T1]{fontenc}
 \usepackage{comment}
 \usepackage{array}
 \usepackage{amsmath}
 \usepackage{txfonts}

\begin{document}

\title{Milliarcsecond-resolution Radio Imaging Survey of Blazar Candidates at 4 < z < 5.4}

\author[0000-0002-8813-4884]{M\'at\'e Krezinger}
\affiliation{Konkoly Observatory, HUN-REN Research Centre for Astronomy and Earth Sciences, Konkoly Thege Mikl\'os \'ut 15-17, 1121 Budapest, Hungary}
\affiliation{CSFK, MTA Centre of Excellence, Konkoly Thege Mikl\'os \'ut 15-17, 1121 Budapest, Hungary}
\email{krezinger.mate@csfk.org}

\author[0000-0002-2339-8264]{Alessandro Caccianiga}
\affiliation{INAF -- Osservatorio Astronomico di Brera, via Brera 28, 20121 Milan, Italy}
\email{alessandro.caccianiga@inaf.it}

\author[0000-0003-1246-6492]{Daniele Dallacasa}
\affiliation{Dipartimento di Astronomia, Universit\`a di Bologna, via Ranzani 1, 40127 Bologna, Italy}
\affiliation{INAF -- Istituto di Radioastronomia, Via Gobetti 101, 40129 Bologna, Italy}
\email{ddallaca@ira.inaf.it}

\author[0000-0003-1516-9450]{Luca Ighina}
\affiliation{Center for Astrophysics -- Harvard \& Smithsonian, 60 Garden Street, Cambridge, MA 02138, USA}
\affiliation{INAF -- Osservatorio Astronomico di Brera, via Brera 28, 20121 Milan, Italy}
\email{luca.ighina@cfa.harvard.edu}

\author[0000-0003-3079-1889]{S\'andor Frey}
\affiliation{Konkoly Observatory, HUN-REN Research Centre for Astronomy and Earth Sciences, Konkoly Thege Mikl\'os \'ut 15-17, 1121 Budapest, Hungary}
\affiliation{CSFK, MTA Centre of Excellence, Konkoly Thege Mikl\'os \'ut 15-17, 1121 Budapest, Hungary}
\affiliation{Institute of Physics and Astronomy, ELTE E\"otv\"os Lor\'and University, P\'azm\'any P\'eter s\'et\'any 1/A,
1117 Budapest, Hungary}
\email{frey.sandor@csfk.org}

\author[0000-0002-9770-0315]{Alberto Moretti}
\affiliation{INAF -- Osservatorio Astronomico di Brera, via Brera 28, 20121 Milan, Italy}
\email{alberto.moretti@inaf.it}

\author[0000-0002-0658-644X]{Sonia Ant\'on}
\affiliation{Centro de F\'isica da UC, Departamento de Física, Universidade de Coimbra, 3004-516 Coimbra, Portugal}
\email{santon@uc.pt}

\author[0000-0003-4747-4484]{Silvia Belladitta}
\affiliation{Max-Planck-Institut f\"ur Astronomie, Königstuhl 17, 69117 Heidelberg, Germany}
\affiliation{INAF -- Osservatorio di Astrofisica e Scienza dello Spazio, via Gobetti 93/3, 40129 Bologna, Italy}
\email{belladitta@mpia.de}

\author[0000-0003-0522-6941]{Claudia Cicone}
\affiliation{Institute of Theoretical Astrophysics, University of Oslo, P.O. Box 1029, Blindern, 0315 Oslo, Norway}
\email{claudia.cicone@astro.uio.no}

\author[0000-0003-1020-1597]{Krisztina \'E. Gab\'anyi}
\affiliation{Konkoly Observatory, HUN-REN Research Centre for Astronomy and Earth Sciences, Konkoly Thege Mikl\'os \'ut 15-17, 1121 Budapest, Hungary}
\affiliation{CSFK, MTA Centre of Excellence, Konkoly Thege Mikl\'os \'ut 15-17, 1121 Budapest, Hungary}
\affiliation{Department of Astronomy, Institute of Physics and Astronomy, ELTE E\"otv\"os Lor\'and University, P\'azm\'any P\'eter s\'et\'any 1/A, 1117 Budapest, Hungary}
\affiliation{HUN-REN--ELTE Extragalactic Astrophysics Research Group, E\"otv\"os Lor\'and University, P\'azm\'any P\'eter s\'et\'any 1/A, 1117 Budapest, Hungary}
\affiliation{Institute of Astronomy, Faculty of Physics, Astronomy and Informatics, Nicolaus Copernicus University, Grudzi\c adzka 5, 87-100 Toru\'n, Poland}
\email{gabanyi.krisztina@ttk.elte.hu}

\author[0000-0002-5251-5538]{M. J. M. March\~a}
\affiliation{Deptartment of Physics \& Astronomy and Deptartment of Computer Science, University College London, Gower Street, London, WC1E 6BT, UK}
\email{m.marcha@ucl.ac.uk}

\author[0000-0002-6044-6069]{Krisztina Perger}
\affiliation{Konkoly Observatory, HUN-REN Research Centre for Astronomy and Earth Sciences, Konkoly Thege Mikl\'os \'ut 15-17, 1121 Budapest, Hungary}
\affiliation{CSFK, MTA Centre of Excellence, Konkoly Thege Mikl\'os \'ut 15-17, 1121 Budapest, Hungary}
\email{perger.krisztina@csfk.org}

\correspondingauthor{M\'at\'e Krezinger}
\email{krezinger.mate@csfk.org}

\begin{abstract}

We present a milliarcsecond-resolution radio survey of 17 high-redshift ($4 \leq z \leq 5.4$) blazar candidates observed with the European Very Long Baseline Interferometry (VLBI) Network at 5~GHz. The primary objective of this study was to investigate the nature of these distant active galactic nuclei (AGN) and to confirm their blazar nature. Utilizing the technique of VLBI, we obtained high-resolution radio images of compact core and core--jet structures. To confirm the classification of these sources, we collected multi-band archival data, including total radio flux densities from single-dish and low-resolution interferometric surveys, optical astrometric positions from \textit{Gaia}, and X-ray data. These diagnostics collectively help distinguish between blazars and misaligned jetted AGN. We were able to measure the core brightness temperatures and found that 11 objects show the Doppler-boosted emission expected from blazars. For five additional sources, we do not see evidence of Doppler boosting even if X-ray data suggest that the source is a blazar. These could be either borderline objects or variability may have affected the classifications, considering that VLBI and X-ray data are not simultaneous. Finally, for the two remaining objects the data suggest a non-blazar classification. Our findings confirm that a significant fraction of these high-redshift radio-loud quasars are blazars and mainly characterized by compact core structures. Overall, the VLBI classifcations are consistent with the X-ray classes. This study further increases the sample of VLBI-imaged radio quasars at $z \geq 4$ by $\sim 10\%$, offering valuable on the population of AGN in the early Universe.

\end{abstract}



\section{Introduction}
Understanding the formation and evolution of the earliest supermassive black holes (SMBHs) and their host galaxies remains one of the central challenges in observational cosmology \citep[e.g.][]{2010AARv..18..279V}. Quasars, powered by accreting SMBHs, at high redshifts ($z > 4$) provide a direct observational window into the first billion years of the Universe. The number of high-$z$ active galactic nuclei (AGN) discovered has been constantly increasing, thanks to recent surveys \citep{1999AJ....118....1F,2016ApJS..227...11B, 2016ApJ...833..222J,2019ApJ...884...30W,2022MNRAS.511..572O,2024AJ....168...58D}. 

About $10\%$ of AGN show a high radio-to-optical flux density ratio\footnote{Radio loudness $R$ is conventionally defined as $R_{\mathrm{4400\,\AA}} = L_{\mathrm{5\,GHz}}/L_{\mathrm{4400\,\AA}}$ \citep{1989AJ.....98.1195K, 2016ApJ...831..168K} or as $R_{\mathrm{2500\,\AA}} = L_{\mathrm{5\,GHz}}/L_{\mathrm{2500\,\AA}}$ \citep{1980ApJ...238..435S}}, i.e. $R>10$ \citep{1984RvMP...56..255B,2002AJ....124.2364I,2025ApJS..280...23A} and are called radio-loud quasars (RLQs). The ratio of RLQs does not strongly evolve with redshift \citep{2015ApJ...804..118B,2024MNRAS.528.5692K}. The radio emission of RLQs originates mainly from the powerful relativistic plasma jets \citep[e.g.][]{2015MNRAS.452.1263P,2019ARA&A..57..467B}. AGN where the jet orientation is observed close to the line of sight (i.e. within $1/\Gamma$, where $\Gamma$ is the bulk Lorentz factor of the jet; for $\Gamma=5$, the viewing angles of blazars are $<10^{\circ}$) are called blazars \citep{1995PASP..107..803U}. Relativistic beaming makes the blazar jet emission Doppler-boosted and the strong enchanced synchrotron emission easily detected in the radio bands, even from large cosmological distances. The characteristic properties of blazars are their X-ray and $\gamma$-ray emission \citep[e.g.][]{2013MNRAS.433.2182S,2015MNRAS.452.3457G}, flat radio spectrum, variability, and compact radio structure with brightness temperatures exceeding the $T_\mathrm{b,int}$ intrinsic value indicating relativistic beaming. $T_\mathrm{b,int}$ is often assumed as the equipartition limit, $T_\mathrm{b,eq} \approx 5 \times 10^{10}$\,K \citep{1994ApJ...426...51R}, or it can be somewhat lower \citep[e.g.][]{2014JKAS...47..303L,2021ApJ...923...67H}. \citet{2021ApJ...923...67H} estimated $T_\mathrm{b,int} \approx 4 \times 10^{10}$~K, and the value of $T_\mathrm{b,int}$ appears to depend on the observing frequency \citep{2014JKAS...47..303L, 2020ApJS..247...57C}. A recent study showed that the intrinsic values for flat-spectrum radio quasars (FSRQs) are often below the equipartition \citep{2023ApJS..268...23F}.
Conversely, RLQs are referred to as ``misaligned'' when observed at inclination angles $> 1/\Gamma$. As the jets propagate outwards from the central engine, they interact with the ambient medium, creating radio-emitting shock-wave regions, i.e., hotspots and lobes. These features are more visible at large inclination angles, and in contrast to blazars, the radio emission of misaligned AGN are dominated by the outer regions of the jet. In such misaligned objects, the radio emission spatially coincident with the optical position often remains undetectable owing to the lack of significant Doppler boosting \citep[e.g.,][]{2012ApJ...760...77A,2021A&ARv..29....3O}. The characteristic features of misaligned sources are the extended kpc-scale radio emission, symmetric radio structures, and steep radio spectrum. They are usually non-variable and their emission is not significantly Doppler-boosted. 
 
Relativistic jets likely played a crucial role in both the 
growth of the central SMBHs \citep[e.g.][]{2008MNRAS.386..989J} and the evolution of their host galaxies, via feedback mechanisms \citep[e.g.][]{2019ARA&A..57..467B}. For this reason, a reliable assessment of the cosmological properties (luminosity function, evolution) of jetted SMBHs, in particlar at high redshift, is an essential step for understanding galaxy and black hole co-evolution. Crucially, blazars can be used to obtain a reliable and unbiased census of jetted AGN since their number ($N_\mathrm{blazar}$) is strictly related to the total number of jetted AGN ($N$) by the relation $N=2\Gamma^2 \times N_\mathrm{blazar}$ \citep[e.g.][]{2015MNRAS.452.1263P}. The most critical step for a reliable application of this technique, however, is the correct classification of a source as blazar. At high redshifts, it can be challenging to differentiate between blazars and AGNs with misaligned relativistic jets. In general, the jet orientation can be inferred from X-ray and radio measurements, since these are the spectral intervals where the jet emission can dominate over the non-jet components (accretion, host-galaxy emission). A strong and flat X-ray emission is considered a good proxy for the jet orientation \citep{2019MNRAS.489.2732I}.
In the radio band, the finest, milliarcsecond (mas) scale angular resolution required to study the inner jet features can be achieved at cm wavelengths by very long baseline interferometry (VLBI) imaging. VLBI is suitable for distinguishing between compact, high-brightness-temperature radio emission of blazars and the more extended structures of misaligned jetted AGN. The comparison of VLBI observations with lower-resolution radio images and data obtained at different wavebands is also essential in the classification process \citep{2021Galax...9...23S}. With VLBI, it is also possible to measure proper motions of components in jets, and to determine physical and geometric properties of resolved radio jets at high redshift, by means of monitoring over a sufficiently long time interval \citep[e.g.][]{2015MNRAS.446.2921F,2018MNRAS.477.1065P,2020SciBu..65..525Z,2022ApJ...937...19Z,2024MNRAS.530.4614K,2025Univ...11...91G}.

The study of the statistical properties of jetted AGN in the early Universe, inferred from the class of blazars, was the main goal of the work discussed in  \citet{2019MNRAS.484..204C}, which presented a well-defined, flux-density-limited sample of $z>4$ blazars. This sample was carefully selected from the Cosmic Lens All Sky Survey \citep[CLASS,][]{2003MNRAS.341....1M,2003MNRAS.341...13B} in order to achieve a high level of completeness and it was then used to estimate the space  density of high-$z$ jetted quasars \citep{2019MNRAS.484..204C} and to evaluate the impact of absorption in the early Universe \citep{2024AA...684A..98C}. The sources have been classified as blazars on the basis of their multi-wavelength properties, with a particular focus on the X-ray emission whose intensity is a proxy of the presence of beaming \citep{2019MNRAS.489.2732I}. However, the X-ray emission alone can fail, in some cases, to correctly classify the sources, as recently suggested by some authors \citep[e.g.][]{2017MNRAS.467..950C,2018evn..confE..31G,2021AN....342.1092G,2022ApJS..260...49K}. Given the statistical relevance of even a single blazar that implies the existence of hundreds of misaligned jetted quasars, it is of fundamental importance to achieve the most reliable classification for the entire CLASS sample. For this reason, we have carried out a VLBI follow-up of the enitre sample in order to support the X-ray classifications with the radio ones that are independent and complementary. The combination of these data will allow us to obtain the most accurate classification of the entire sample.

In this paper, we present VLBI observations of 17 high-redshift blazar candidates, whose redshifts range from $4.00$ to $5.36$. 
There is an 18th source in the sample, J1517$+$3753, which later turned out to be a low-redshift quasar at $z = 0.56$. For completeness, we present the results obtained for this object as well. Observing at 5~GHz with the European VLBI Network (EVN) allowed us to detect compact core--jet features characteristic of blazars. With the newly-obtained high-resolution and high-sensitivity VLBI data, we investigate the nature of these sources to decide whether they are truly blazars or misaligned jetted AGN. Radio and X-ray spectral data \citep{2019MNRAS.489.2732I} collected from the literature, as well as the accurate \textit{Gaia} \citep{2016A&A...595A...1G} Data Release 3 \citep[DR3,][]{2023A&A...674A...1G} optical positions of the sources are used to support the classification. 
In Section~\ref{sample}, we introduce the list of 18 objects targeted with the EVN. In Section~\ref{obs-reduc}, we provide details on the observations and data reduction. The results are given in Section~\ref{results}. Finally, our findings are discussed in Section~\ref{discussion}, along with notes on individual sources. Section~\ref{conclusion} gives a summary. Throughout this paper, we assume a standard flat $\Lambda$ Cold Dark Matter cosmological model with $\Omega_{\mathrm m} = 0.3$, $\Omega_{\mathrm \Lambda} = 0.7$, and $H_0 = 70$~km\,s$^{-1}$\,Mpc$^{-1}$, and used these parameters as inputs in the cosmology calculator of \citet{2006PASP..118.1711W}. In this model, the linear scales are about $(6-7)$~pc\,mas$^{-1}$ and the luminosity distances of our high-redshift sources are about $(35-50)$~Gpc.

\section{Sample Selection} \label{sample}
We selected a sample of 18 high-redshift blazar candidates from the CLASS sample for VLBI imaging observations, either to confirm or rule out their blazar nature. The goal was to observe these sources with VLBI to reveal their mas-scale morphology, and to determine the core properties such as the flux density, angular size, compactness, and brightness temperatures. VLBI observations provide independent estimates of the jet parameters which can be evaluated together with the results of previous modeling of spectral energy distributions \citep[SED, e.g.][]{2017MNRAS.467..950C}. The CLASS sample contains sources with a flat radio spectrum between the Green Bank 6-cm survey \citep[GB6, $4.85$~GHz,][]{1996ApJS..103..427G} and the National Radio Astronomy Observatory (NRAO) Very Large Array (VLA) Sky Survey \citep[NVSS, $1.4$~GHz][]{1998AJ....115.1693C}. The selected targets are a subsample of the objects presented in \citet{2019MNRAS.484..204C}, and were also part of our previous X-ray study, focusing on the CLASS blazar candidates \citep{2019MNRAS.489.2732I}. The list of the sources is presented in Table \ref{tab:targets}. Spectroscopic redshifts range from $z = 4$ to $z = 5.36$, except for the source J1517$+$3753, which later turned out to be a low-redshift quasar with $z = 0.56$, based on its LBT spectrum. All sources have a total flux density above 30~mJy at 5~GHz, according to the GB6 survey \citep{1996ApJS..103..427G}, and most of them are also detected in the Faint Images of the Radio Sky at Twenty-centimeters \citep[FIRST,][]{1997ApJ...475..479W} survey and the VLA Sky Survey \citep[VLASS,][]{2020PASP..132c5001L,2020RNAAS...4..175G}. At the time of the analysis, 11 of the targets have additional archival VLBI observations, either with the NRAO Very Long Baseline Array (VLBA) or with the EVN. Eight of the sources were observed with the VLBA, covering a wide frequency range ($2.3-8.7$~GHz), with their data available from the Astrogeo database\footnote{\url{https://astrogeo.org/}, maintained by L. Petrov} \citep{2025ApJS..276...38P}. As these observations typically have short on-source integration times, they are not sufficient to reach the imaging sensitivity required for our goals, and occasionally have poorer angular resolution ($5-6$~mas), depending on the frequency. Four targets have previous EVN observations at $1.7$ and $5$~GHz \citep{2022ApJS..260...49K}, or at $2.3$ and $8.4$~GHz \citep{2010AA...520A.113B}\footnote{The VLBI images were not published in this paper, nor can be found in the EVN archive.}. Most of these VLBI measurements were published after our EVN observations (Section~\ref{obs}) took place. 

\begin{deluxetable*}{llccc}
\tablecaption{The $z > 4$ radio quasar sample presented in this study.}
\tablewidth{0pt}
\tablehead{
\colhead{CLASS Name} & \colhead{Source ID} & \colhead{$z$} & \colhead{Linear scale} & \colhead{$D_\mathrm{L}$}  \\
\nocolhead{} & \nocolhead{} & \nocolhead{} & \colhead{[pc\,mas$^{-1}$]} & \colhead{[Mpc]} 
}
\decimalcolnumbers
\startdata
GB6~J003126+150729 & J0031$+$1507 & 4.29 & 6.75 & 38\,805  \\
GB6~J012126+034646 & J0121$+$0346 & 4.13 & 6.86 & 37\,037  \\
GB6~J012202+030951 & J0122$+$0309 & 4.00 & 6.94 & 35\,604  \\
GB6~J025758+433837 & J0257$+$4338 & 4.07 & 6.89 & 36\,343  \\
GB6~J083548+182519 & J0835$+$1825 & 4.41 & 6.67 & 40\,154  \\
GB6~J091825+063722 & J0918$+$0637 & 4.16 & 6.84 & 37\,319  \\
GB6~J102107+220904 & J1021$+$2209 & 4.26 & 6.77 & 38\,494  \\
GB6~J132512+112338 & J1325$+$1123 & 4.42 & 6.66 & 40\,210  \\
GB6~J134811+193520 & J1348$+$1935 & 4.40 & 6.68 & 40\,060  \\
GB6~J141212+062408 & J1412$+$0624 & 4.42 & 6.67 & 40\,252  \\
GB6~J153533+025419 & J1535$+$0254 & 4.39 & 6.69 & 39\,874  \\
GB6~J161216+470311 & J1612$+$4703 & 4.36 & 6.69 & 39\,600  \\
GB6~J162956+095959 & J1629$+$0959 & 5.00 & 6.28 & 46\,632  \\
GB6~J164856+460341 & J1648$+$4603 & 5.36 & 6.06 & 50\,602  \\
GB6~J171103+383016 & J1711$+$3830 & 4.00 & 6.95 & 35\,604  \\
GB6~J231449+020146 & J2314$+$0201 & 4.11 & 6.87 & 36\,817  \\
GB6~J235758+140205 & J2357$+$1402 & 4.34 & 6.72 & 39\,328  \\
\hline
GB6~J151736+375332 & J1517$+$3753 & 0.56 & 6.47 & 3\,247  \\
\enddata
\tablecomments{Col.~1 -- radio source name in the CLASS catalogue; Col.~2 -- shortened source name used throughout this paper; Col.~3 -- redshift; Col.~4 -- linear scale at the source redshift; Col.~5 -- luminosity distance.}
\label{tab:targets}
\end{deluxetable*}

\section{Observations and Data Reduction} \label{obs-reduc}
\subsection{EVN Observations} \label{obs}
We observed the selected radio sources with the EVN at a single frequency band centered around $4.95$~GHz. The observations were split into four sessions, EC062A--B and EC066A--B (PI.: A. Caccianiga). The dates of the observation for each session and the respective target sources are included in Table~\ref{tab:obs}. The observations were performed in phase-reference mode \citep{1995ASPC...82..327B} to increase the coherent integration time on the sources, while also relating the target positions to known ICRF \citep[Internatonial Celestial Reference Frame,][]{2020A&A...644A.159C} astrometric positions of the respective calibrator sources. In this observing mode, the radio telescopes nodded between a nearby bright compact calibrator source and the given target, in repeating, few-minute cycles. Phase-reference calibrators and their angular separations from the targets are listed in Table~\ref{tab:obs}. In all sessions, the total bandwidth of $128$~MHz was divided into eight $16$-MHz wide intermediate frequency channels (IFs), each further divided into $32$ spectral channels. The recording was done in left and right circular polarizations with 1024~Mbps data rate. Then, the data were processed at the Joint Institute for VLBI European Research Infrastructure Consortium (JIVE, Dwingeloo, The Netherlands) with the SFXC software correlator \citep{2015ExA....39..259K} with $2$~s integration time. The $15$ participating radio telescopes were the following: Jodrell Bank Mk2 ($38\,\mathrm{m} \times 25\,\mathrm{m}$ diameter, United Kingdom), Westerbork ($25$~m, The Netherlands), Effelsberg ($100$~m, Germany), Medicina ($32$~m, Italy), Noto ($32$~m, Italy), Onsala ($25$~m, Sweden), Toru\'{n} ($32$~m, Poland), Hartebeesthoek ($26$~m, South Africa), Yebes ($40$~m, Spain), Irbene ($16$~m, Latvia), Svetloe ($32$~m, Russia), Zelenchukskaya ($32$~m, Russia), Badary ($32$-m, Russia), Nanshan ($25$~m, China), and Tianma ($65$~m, China).

\begin{deluxetable*}{lcrcrc}
\tablecaption{Observing log.}
\tablewidth{0pt}
\tablehead{
\colhead{Session ID} & \colhead{Date of Observation} & \colhead{Target} & \colhead{On-source Time} & \colhead{Phase} & \colhead{Separation} \\
 & & \colhead{Source ID} & \colhead{[min]} & \colhead{Calibrator} & \colhead{[$\degr$]}
}
\decimalcolnumbers
\startdata
EC062A & 2017 Oct 25 & J0031$+$1507 & 80 & J0019+2021 & 5.9 \\
&                    & J0121$+$0346 & 70 & J0121+0422 & 0.6 \\
&                    & J0122$+$0309 & 73 & J0121+0422 & 1.2 \\
&                    & J0257$+$4338 & 91 & J0251+4315 & 1.2 \\
&                    & J2314$+$0201 & 56 & J2320+0513 & 3.5 \\
\hline 
EC062B & 2017 Oct 27 & J0835$+$1825 & 64 & J0842+1835 & 1.5 \\
&                    & J0918$+$0637 & 60 & J0909+0121 & 5.8 \\
&                    & J1021$+$2209 & 72 & J1016+2037 & 1.8 \\
&                    & J1325$+$1123 & 84 & J1309+1154 & 3.9 \\
&                    & J1348$+$1935 & 56 & J1357+1919 & 2.1 \\
&                    & J1412$+$0624 & 84 & J1405+0415 & 2.8 \\
&                    & J1535$+$0254 & 95 & J1534+0131 & 1.4 \\
&                    & J1612$+$4703 & 86 & J1625+4134 & 6.0 \\
\hline 
EC066A & 2018 Oct 18 & J2357$+$1402 & 74 & J0006+1235 & 2.5\\
\hline 
EC066B & 2018 Oct 26 & J1517$+$3753 & 70 & J1506+3730 & 2.3 \\
&                    & J1629$+$0959 & 60 & J1623+0741 & 2.7 \\
&                    & J1648$+$4603 & 74 & J1658+4737 & 2.2 \\
&                    & J1711$+$3830 & 74 & J1701+3954 & 2.3 \\
\enddata
\tablecomments{Col.~1 -- project code of the EVN observing session; Col.~2 -- observing date; Col.~3 -- target source designation; Col.~4 -- total time spent on the target source; Col.~5 -- phase-reference calibrator source designation; Col.~6 -- angular separation between the target and the phase calibrator.}
\label{tab:obs}
\end{deluxetable*}

\subsection{Data Reduction} \label{reduc}
The data were calibrated using the NRAO Astronomical Image Processing System (\textsc{Aips}) software package \citep{2003ASSL..285..109G}, following a standard procedure \citep[e.g.][]{1995ASPC...82..227D}. As a first step, the interferometric visibility amplitudes were calibrated using the antenna gain curves and the system temperatures measured at each telescope. In a few cases, nominal values were used when system temperature measurements were missing. The data were then corrected for the dispersive ionospheric delay, based on total electron content maps obtained from global navigation satellite systems data. Phase changes resulting from the time variation of the source parallactic angle were also corrected for radio telescopes with an azimuth--elevation mount. An initial correction of instrumental phases and delays was performed using a 1-min scan on a strong fringe-finder source. Bandpass calibration was done by removing the first and last $4$ spectral channels in each IF. Then global fringe-fitting \citep{1983AJ.....88..688S} was performed on the fringe-finder sources and phase-reference calibrators (EC062A: 3C\,454.3 and J0237+2848; EC062B: OJ287, 4C\,39.25, and 3C\,345; EC066A: 3C\,454.3; EC066B: 3C\,279 and 3C\,345). At this point, the calibrated visibility data were exported to the \textsc{Difmap} software package \citep{1994BAAS...26..987S}. We carried out hybrid mapping, involving several iterations of the \textsc{clean} deconvolution algorithm \citep{1974A&AS...15..417H} and phase-only self-calibration \citep{1984ARA&A..22...97P}, followed by a few rounds of amplitude and phase self-calibration with decreasing solution intervals.

During sessions EC062A--B, firmware amplitude issues were reported by the JIVE correlator, affecting several IFs for most of the antennas, causing jumps in the amplitudes across IFs. To avoid the large amplitude uncertainties these jumps would have caused later in these sessions, without averaging the visibility data in frequency, we determined the mean gain corretion factors for each antenna, IF, and polarization, based on mapping all individual calibrator and fringe-finder visibility data in \textsc{Difmap}. The median values of these antenna-based gain correction factors obtained source by source were applied to the visibility amplitudes over all the antennas, IFs, and polarizations, using a \textsc{ParselTongue} \citep{2006ASPC..351..497K} script, by running the \textsc{Aips} task \textsc{clcor}. In this way, we achieved a much smoother amplitude curve across the band, thus increasing the reliability of the subsequent flux density modeling for our target sources. 

The \textsc{clean} model components of the calibrators produced in \textsc{Difmap} were transferred back to \textsc{Aips} as correction inputs for a repeated fringe-fit to improve the phase solutons. The final fringe-fit solutions obtained for the phase-reference calibrators were interpolated to the respective target source data. Finally, the calibrated visibility data of the target sources were exported from \textsc{Aips} to \textsc{Difmap}.

The final target source images were produced with \textsc{Difmap}. We applied natural weighting (the weights were calculated as the reciprocal of the amplitude errors, \textsc{uvweight $0,-1$}) to reduce the image noise. The dirty images were shifted to make the brightness peak coincide with the origin at $(0,0)$ relative right ascension and declination. We followed a standard hybrid mapping procedure, with only a few rounds of \textsc{clean} iterations and phase-only self-calibration performed. For the bright sources ($\sim 50$~mJy), self-calibration was applied to all antennas, while for the faint sources ($\sim 10$~mJy), phases were corrected only for the most sensitive antennas (Effelsberg, Yebes, and Tianma). After obtaining residual images consistent with $\sim 5\sigma$ noise level, we performed a final \textsc{clean} iteration involving $1000$ steps with a very small loop gain $(0.01)$, to smooth the noise features. The final images are presented in Figure~\ref{fig:imgs} and their parameters are listed in Table~\ref{tab:imgparams}. The phase-referenced VLBI positions of the sources were determined from the image brightness peaks using the \textsc{Aips} task \textsc{maxfit}. 

To obtain the source physical parameters, we used two-dimensional Gaussian brightness distribution model components \citep{1995ASPC...82..267P} fitted directly to the visibility data in \textsc{Difmap}. This allowed us to quantitatively characterize component sizes and flux densities listed in Table~\ref{tab:phyparams}. Between model-fit iterations, we applied phase-only self-calibration in the same way as during \textsc{clean}ing, until reaching the $\sim 5\sigma$ peak-to-noise ratio in the residual images. The fitted Gaussians models were circular, except for J0257$+$4338, where an elliptical component described the core brightness distribution much better. In five cases (J0257$+$4338, J1021$+$2209, J1348$+$1935, J1612$+$4703, and J2357$+$1402), an additional jet component was found. The uncertainties of the parameters of the fitted Gaussian model components were estimated following the method of \citet{1999ASPC..180..301F}. For the flux densities, an additional $5\%$ error was added in quadrature, to account for the VLBI absolute amplitude calibration uncertainty \citep[e.g.][]{2012ApJS..198....5A,2015MNRAS.446.2921F}.

\section{Results} \label{results}
The naturally weighted 5-GHz clean maps are shown in Figure~\ref{fig:imgs}, while the corresponding image parameters are listed in Table~\ref{tab:imgparams}. The only non-detected source is J1517$+$3753 -- the one that is not a high-redshift quasar. For the detected $17$ sources, all at $z\ge4$, the calculated pyhsical properties are listed in Table~\ref{tab:phyparams}.

Following \citet{1982ApJ...252..102C}, we calculated the 5-GHz redshift-corrected brightness temperatures: 
\begin{equation} \label{eq:tb}
    T_{\mathrm{b}} = 1.22 \times 10^{12} \, (1 + z) \frac{S_\nu}{\theta^2 \nu^2} \,\, [\mathrm{K}].
\end{equation}
Here, $S_\nu$ is the integrated flux density of the core in Jy, $\theta$ the fitted circular Gaussian component diameter (full width at half maximum, FWHM) in mas, and $\nu$ the observing frequency in GHz. In case of an elliptical Gaussian component, $\theta$ equals to the geometric mean of the major and minor axis FWHM values. Equation~\ref{eq:tb} gives the lower limit to the brightness temperature if the source is unresolved by the interferometer. The minimum resolvable size were calculated following \citet{2005AJ....130.2473K}. If the minimum resolvable size exceeds the modeled component size, the former was used instead of $\theta$ in Equation~\ref{eq:tb}. We found $3$ sources (J1021$+$2209, J1348$+$1935, and J1612$+$4703) out of the detected $17$ to be unresolved. 

We calculated the 5-GHz redshift-corrected monochromatic radio powers using the formula of \citet{1979AA....80...13B}:
\begin{equation} \label{eq:power}
    P_{\nu} = 1.20 \times 10^{20} \, D_{\mathrm{L}}^2 S_{\nu} (1+z)^{-1-\alpha} \,\, [\mathrm{W}\,\mathrm{Hz}^{-1}], 
\end{equation}
where $D_\mathrm{L}$ refers to the source luminosity distance in Mpc, $\alpha$ is the radio spectral index (defined as $S_\nu \propto \nu^{\alpha}$), and $S_{\nu}$ the $5$-GHz flux density in Jy. There are four sources (J0121$+$0347, J0918$+$0637, J1325$+$1123, and J1412$+$0624) with published spectral index characterizing the compact radio emission, which we could adopt. For the rest of the sources, a flat spectrum with $\alpha = 0$ was assumed. 

To characterize the source compactness, we compared the ratio of the $5$-GHz VLBI flux densities ($S_\mathrm{VLBI}$) to the total flux densities ($S_\mathrm{total}$). The total $\sim 10^{\prime\prime}$-scale values are taken from the GB6 survey. The frequency difference between $4.85$ and $4.99$~GHz, the latter having been used for our EVN measurements, is negligible. The $S_{\rm VLBI}/S_{\rm total}$ ratio gives information about the relation between the pc- and kpc-scale radio emissions, assuming that the source has not varied. When the two values are consistent within about $10\%$, the source can be considered as core-dominated. An $S_{\rm VLBI}/S_{\rm total}$ ratio above $1.1$ indicates a variable source. Potential variability can also be assessed by calculating the ratio of the FIRST and NVSS flux densities, $S_\mathrm{FIRST}/S_\mathrm{NVSS}$ \citep[e.g.][]{2016MNRAS.463.3260C,2022ApJS..260...49K}, and also the FIRST and RACS-mid \citep[Rapid Australian Square Kilometre Array Pathfinder Continuum Survey,][]{2024PASA...41....3D} ratio $S_\mathrm{FIRST}/S_\mathrm{RACS}$. While their observing frequencies are the same, the angular resolution of the FIRST survey is much better than that of the NVSS or RACS. We consider a source variable if the FIRST flux density exceeds the NVSS or RACS value by more than $10\%$. The $S_\mathrm{VLBI}/S_\mathrm{total}$, $S_\mathrm{FIRST}/S_\mathrm{NVSS}$ and $S_\mathrm{FIRST}/S_\mathrm{RACS}$ ratios are listed in Table~\ref{tab:phyparams}. We note that the presence of variability cannot be excluded even in cases when either flux density ratio is significantly below unity. However, this can also be caused by resolution effects if the extended radio emission gets resolved out on the longer interferometer baselines.  

\begin{deluxetable*}{rrrcccc}
\tablecaption{Image parameters and source positions.}
\tablewidth{0pt}
\tablehead{
\colhead{Source ID} & \colhead{RA$_{\rm VLBI,5GHz}$} & \colhead{Dec$_{\rm VLBI,5GHz}$} & \colhead{Peak Intensity} & \colhead{$1\sigma$ Nosie} &\multicolumn2c{Restoring Beam} \\
 & \colhead{[h~m~s]} & \colhead{[$^\circ$~$^\prime$~$^{\prime\prime}$]} & [mJy~beam$^{-1}]$ & [mJy~beam$^{-1}$]  & \colhead{[mas$\times$mas]} & \colhead{[$^\circ$]} 
}
\decimalcolnumbers
\startdata
J0031$+$1507 & 00 31 26.80137 (0.00012) & 15 07 39.5018 (0.0018) & 49.4 & 0.75 & 1.7$\times$1.4 & 20.3 \\
J0121$+$0346 & 01 21 26.14774 (0.00005) & 03 47 06.7495 (0.0008) & 12.8 & 0.29 & 1.6$\times$1.3 & 30.8 \\
J0122$+$0309 & 01 22 01.90936 (0.00006) & 03 10 02.4168 (0.0008) & 30.7 & 0.52 & 1.7$\times$1.4 & 35.4 \\
J0257$+$4338 & 02 57 59.07765 (0.00006) & 43 38 37.6759 (0.0008) & 203.0 & 0.08 & 1.6$\times$1.0 & $-1.2$ \\
J0835$+$1825 & 08 35 49.42472 (0.00006) & 18 25 20.0531 (0.0009) & 29.1 & 0.31 & 1.8$\times$1.2 & $-0.9$ \\
J0918$+$0637 & 09 18 24.37967 (0.00012) & 06 36 53.4143 (0.0018) & 16.6 & 0.39 & 1.8$\times$1.2 & 2.5 \\
J1021$+$2209 & 10 21 07.57752 (0.00006) & 22 09 21.5568 (0.0009) & 65.3 & 0.73 & 1.8$\times$1.2 & 5.3 \\
J1325$+$1123 & 13 25 12.49324 (0.00008) & 11 23 29.8386 (0.0012) & 18.7 & 0.31 & 1.7$\times$1.2 & 5.2 \\
J1348$+$1935 & 13 48 11.26252 (0.00006) & 19 35 23.5885 (0.0011) & 11.5 & 0.31 & 1.9$\times$1.6 & 41.8 \\
J1412$+$0624 & 14 12 09.96963 (0.00007) & 06 24 06.8651 (0.0011) & 5.2 & 0.12 & 1.8$\times$1.3 & 9.4 \\
J1535$+$0254 & 15 35 33.88683 (0.00006) & 02 54 23.4032 (0.0009) & 18.9 & 0.41 & 1.7$\times$1.3 & 11.3 \\
J1612$+$4703 & 16 12 16.75108 (0.00012) & 47 02 53.6101 (0.0019) & 3.6 & 0.17 & 2.5$\times$1.4 & $-11.1$\\
J1629$+$0959 & 16 29 57.27858 (0.00007) & 10 00 23.5014 (0.0012) & 14.8 & 0.26 & 1.9$\times$1.6 & $-9.8$ \\
J1648$+$4603 & 16 48 54.52910 (0.00006) & 46 03 27.3871 (0.0011) & 20.6 & 0.29 & 3.1$\times$1.6 & $-7.3$ \\
J1711$+$3830 & 17 11 05.53102 (0.00007) & 38 30 04.3433 (0.0011) & 36.1 & 0.19 & 2.0$\times$1.5 & $-7.2$\\
J2314$+$0201 & 23 14 48.71726 (0.00007) & 02 01 51.0664 (0.0011) & 26.7 & 0.30 & 2.3$\times$1.2 & 6.8 \\
J2357$+$1402 & 23 57 58.55006 (0.00009) & 14 02 01.8538 (0.0013) & 37.5 & 0.14 & 2.1$\times$1.6 & $-2.2$ \\
\enddata
\tablecomments{Col.~1 -- source designation; Col.~2 -- right ascension of the 5~GHz VLBI radio position (errors are in parentheses); Col.~3 -- declination of the 5~GHz VLBI radio position; Col.~4 -- peak intensity in the \textsc{clean} map; Col.~5 -- $1\sigma$ noise in the \textsc{clean} map; Col.~6 -- size of the elliptical Gaussian restoring beam (FWHM); Col.~7 -- position angle of the restoring beam major axis, measured from north to east.}
\label{tab:imgparams}
\end{deluxetable*}

\begin{deluxetable*}{lrrrrcccccccr}
\tablecaption{Basic parameters and the derived physical properties of the sample.}
\tablewidth{0pt}
\tablehead{
\colhead{Source ID} &  \colhead{$S_\mathrm{VLBI,5~GHz}$} & \colhead{$\theta_\mathrm{VLBI,5~GHz}$} &  \colhead{$T_\mathrm{b,5~GHz}$} & \colhead{$P_\mathrm{5~GHz}$} & \colhead{$S_{\rm FIRST}/S_{\rm NVSS}$} & \colhead{$S_{\rm FIRST}/S_{\rm RACS}$} & \colhead{$S_{\rm VLBI}/S_{\rm total}$} & \colhead{$\alpha$} & \colhead{$S_0$} & \colhead{$\nu_0$} & \colhead{$\tilde{\alpha}_{\rm ox}$} & \colhead{$\delta$}  \\
 & \colhead{[mJy]} & \colhead{[mas]} & \colhead{[$10^9$~K]}  & \colhead{[$10^{26}$~W~Hz$^{-1}$]} & \nocolhead{} & \nocolhead{} & \nocolhead{} & \nocolhead{} & \colhead{[mJy]} & \colhead{[GHz]}  & \nocolhead{} & \nocolhead{}
}
\decimalcolnumbers
\startdata
J0031$+$1507 & 51.2 (5.7) & 0.48 (0.02) & 56.7 (8.0) & 17.5 (2.7) & 1.04 (0.03) & 0.87 (0.05) & 0.55 (0.08) & -- & 93.9 (11.3) & 7.3 (5.0) & $1.40^{+0.03}_{-0.02}$  & 2.8 \\
J0121$+$0346 & 19.9 (2.3) & 1.05 (0.06) & 4.5 (0.7) & 37.4 (5.9) & 0.98 (0.03) & 0.97 (0.06) & 0.39 (0.07) & $-0.52$ (0.09) & -- & -- & $1.41^{+0.03}_{-0.04}$  & 0.2 \\ 
J0122$+$0309 & 32.9 (3.7) & 0.34 (0.02) & 69.9 (10.2) & 10.0 (1.5) & 1.11 (0.03) & 1.40 (0.09) & 0.34 (0.05) & $0.03$ (0.05) & -- & -- & $0.79^{+0.03}_{-0.01}$  & 3.5 \\
J0257$+$4338 & 229.8 (13.5) & 0.46$^{*}$ (0.01) & 265.7 (15.9) & 72.1 (8.7) & -- & -- & 1.45 (0.15) & -- & 231.1 (24.5) & 1.6 (0.3) & $1.09^{+0.01}_{-0.01}$ & 13.3 \\ 
J0835$+$1825 & 21.4 (2.4) & 0.35 (0.01) & 47.1 (6.2) & 7.7 (1.2) & 0.98 (0.03) & 1.02 (0.06) & 0.53 (0.09) & $-0.25$ (0.04) & -- & -- & $1.03^{+0.05}_{-0.01}$  & 2.4 \\ 
J0918$+$0637 & 13.8 (1.9) & 0.52 (0.03) & 13.0 (2.3) & 5.1 (0.9) & 0.86 (0.03) & 0.61 (0.04) & 0.38 (0.08) & -- & 37.0 (2.3) & 1.7 (0.4) & $1.26^{+0.04}_{-0.04}$  & 0.6 \\ 
J1021$+$2209 & 49.4 (5.5) & 0.13 (0.01) & $>$394.0 & 16.7 (2.6) & 0.91 (0.03) & 0.88 (0.05) & 0.46 (0.07) & $-0.48$ (0.03) & -- & -- & $1.01^{+0.16}_{-0.09}$  & >19.7 \\
J1325$+$1123 & 20.5 (2.3) & 0.76 (0.03) & 9.5 (1.4) & 20.6 (3.4) & 0.87 (0.03) & 0.94 (0.06) & 0.33 (0.06) & -- & 73.7 (4.8) & 1.3 (0.2) & $1.31^{+0.05}_{-0.04}$  & 0.5 \\ 
J1348$+$1935 & 10.5 (1.5) & 0.17 (0.01) & $>$31.7 & 3.7 (0.7) & 0.97 (0.03) & 0.86 (0.05) & 0.28 (0.05) & $-0.07$ (0.06) & -- & -- & $1.23^{+0.05}_{-0.03}$ & $>$1.6 \\
J1412$+$0624 & 4.7 (0.6) & 0.38 (0.02) & 8.7 (1.5) & 2.3 (0.4) & 0.92 (0.03) & 1.06 (0.07) & 0.14 (0.03) & $-0.41$ (0.05) & -- & -- & $1.29^{+0.09}_{-0.04}$  & 0.4 \\
J1535$+$0254 & 16.8 (2.2) & 0.37 (0.02) & 32.5 (5.5) & 5.9 (1.0) & 1.35 (0.04) & 1.71 (0.11) & 0.32 (0.06) &  $-0.26$ (0.06) & -- & -- & $0.93^{+0.08}_{-0.03}$  & 1.6 \\
J1612$+$4703 & 3.1 (0.6) & 0.46 (0.04) & $>$3.1 & 1.1 (0.2) & 1.03 (0.03) & 0.91 (0.05) & 0.03 (0.01) & $-0.38$ (0.07) & -- & -- & $1.39^{+0.09}_{-0.05}$ & $>$0.2 \\
J1629$+$0959 & 16.6 (1.9) & 0.44 (0.02) & 25.3 (3.8) & 7.2 (1.2) & 1.02 (0.03) & 1.16 (0.07) & 0.50 (0.10) & $-0.44$ (0.06)  & -- & -- & $1.09^{+0.04}_{-0.01}$  &  1.6 \\
J1648$+$4603 & 21.3 (2.3) & 0.41 (0.02) & 39.8 (5.5) & 10.3 (1.7) & 0.96 (0.03) & 0.95 (0.06) & 0.71 (0.12) & $-0.19$ (0.02)  & -- & -- & $1.34^{+0.14}_{-0.05}$  & 2.0\\
J1711$+$3830 & 44.8 (3.6) & 0.85 (0.02) & 15.1 (1.4) & 13.7 (1.8) & 1.10 (0.03) & 0.98 (0.06) & 1.24 (0.20) & $0.04$ (0.07) & -- & -- & $1.15^{+0.05}_{-0.04}$  & 0.8 \\
J2314$+$0201 & 28.5 (2.9) & 0.30 (0.01) & 81.9 (10.3) & 9.1 (1.3) & 0.97 (0.03) & 1.02 (0.06) & 0.34 (0.05) & $-0.15$ (0.03) & -- & -- & $1.25^{+0.06}_{-0.05}$  & 4.1 \\ 
J2357$+$1402 & 39.4 (3.1) & 0.43 (0.01) & 54.8 (4.9) & 13.7 (1.8) & -- & -- & 0.50 (0.07) & $-0.23$ (0.04) & -- & -- & $1.02^{+0.05}_{-0.01}$ & 2.7 \\
\hline
J1517$+$3753 & $<0.2^{\dagger}$ & -- & -- & -- & 0.93 (0.04) & -- &  & $-0.66$ (0.08) & -- & -- & --  & -- \\
\enddata
\tablecomments{$^{*}$ The fitted elliptical Gaussian component size (FWHM) is $0.66\,\mathrm{mas} \times 0.32\,\mathrm{mas}$ with $\mathrm{PA}= -17.9\degr$, and we used the geometric mean of the major and minor axes in the calculations.  $^{\dagger}$ $5\sigma$ upper limit of the peak brightness in mJy~beam$^{-1}$. Col.~1 -- source designation; Col.~2 -- fitted core flux density at 5~GHz (errors are in parentheses); Col.~3 -- fitted core component size (FWHM) at 5~GHz; Col.~4 -- core brightness temperature; Col.~5 -- core 5-GHz radio power (assuming $\alpha=0$, when compact spectral index is not available}); Col.~6 -- ratio of the FIRST and NVSS flux densities; Col.~7 -- ratio of the FIRST and RACS flux densities; Col.~8 -- ratio of the 5-GHz VLBI and total GB6 flux densities; Col.~9 -- the total radio continuum spectral index where power-law function was fitted; Col.~10--11 -- peak flux density and peak frequency, where log-parabolic function was fitted; Col.~12 -- the $\tilde{\alpha}_{\rm ox}$ values adopted from \citet{2019MNRAS.489.2732I}; the corrected uncertainties are calculated following \citet{2024AA...684A..98C}; Col.~13 -- Doppler factor.
\label{tab:phyparams}
\end{deluxetable*}

\begin{figure*}
\centering
\gridline{  \includegraphics[width=0.40\textwidth]{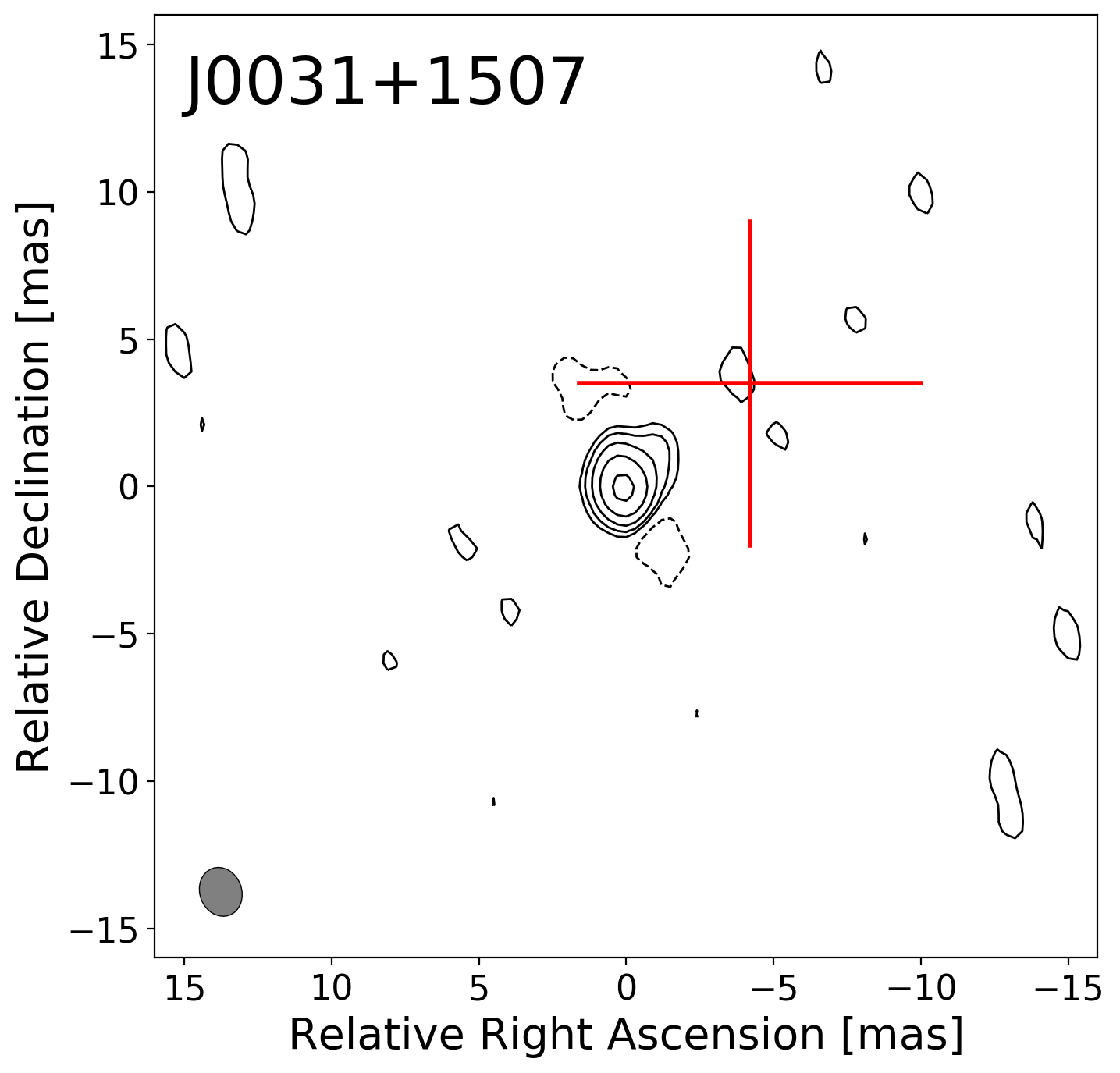}
            \includegraphics[width=0.40\textwidth]{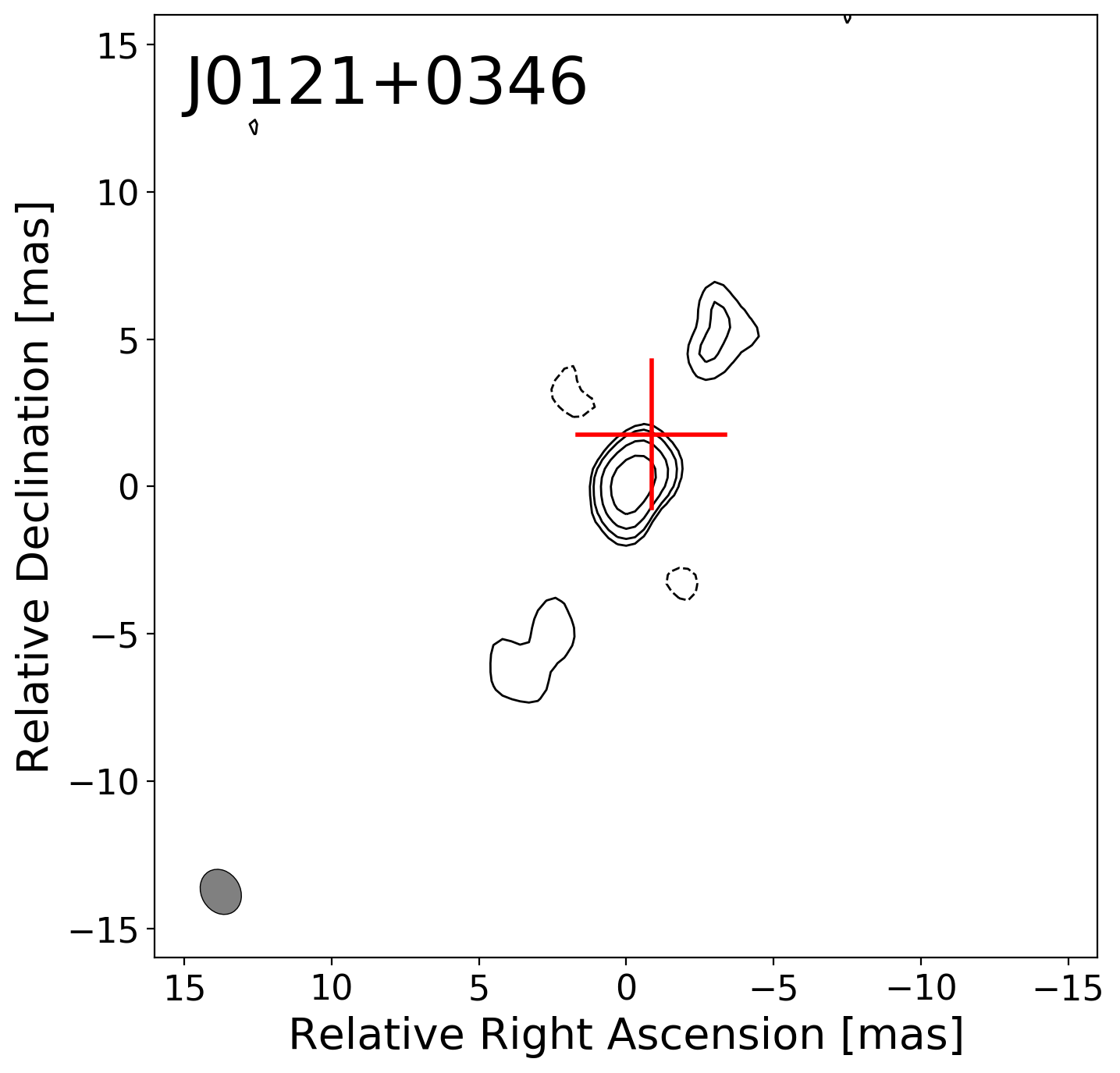}
            }
\gridline{  \includegraphics[width=0.40\textwidth]{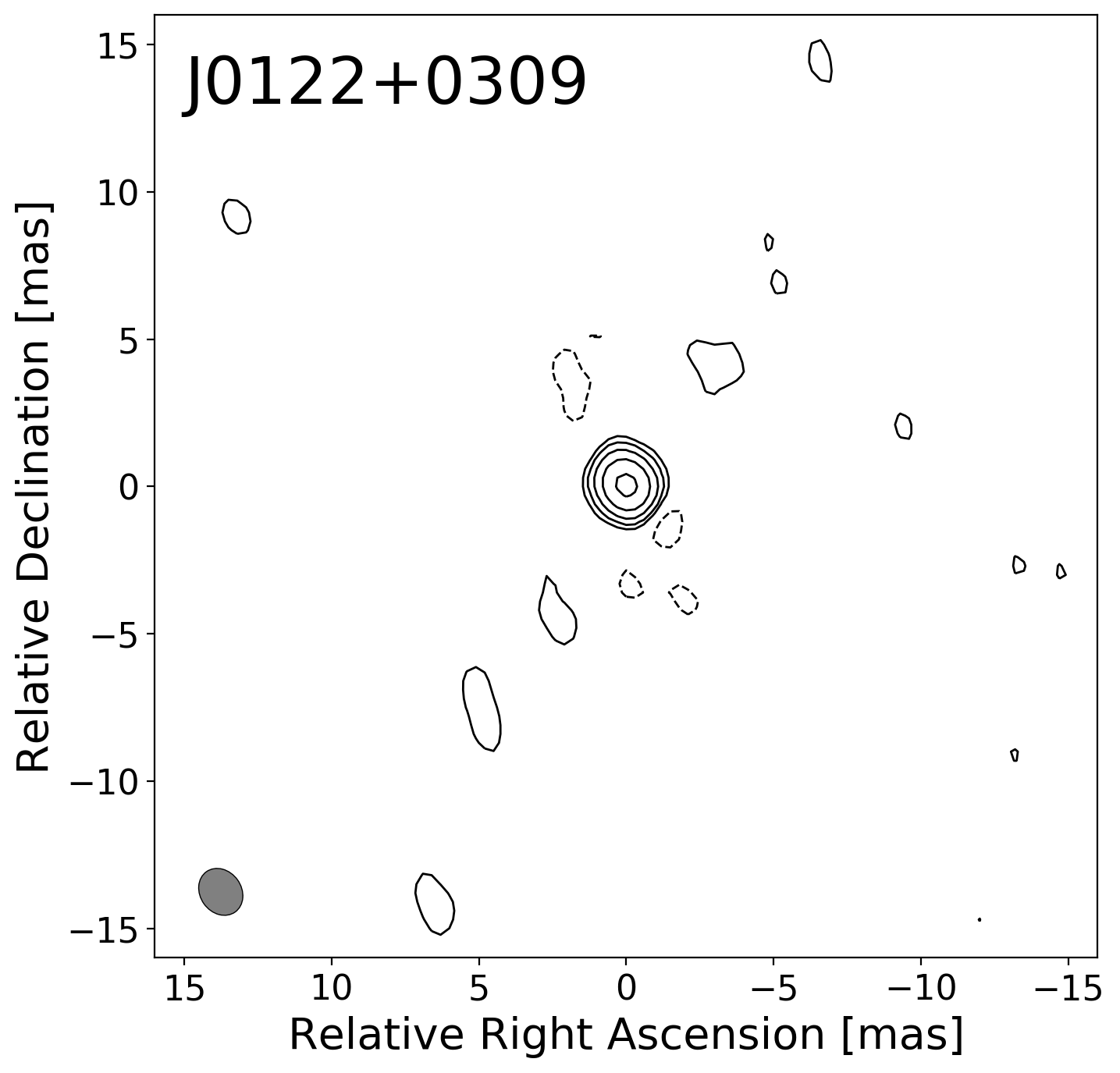}
            \includegraphics[width=0.40\textwidth]{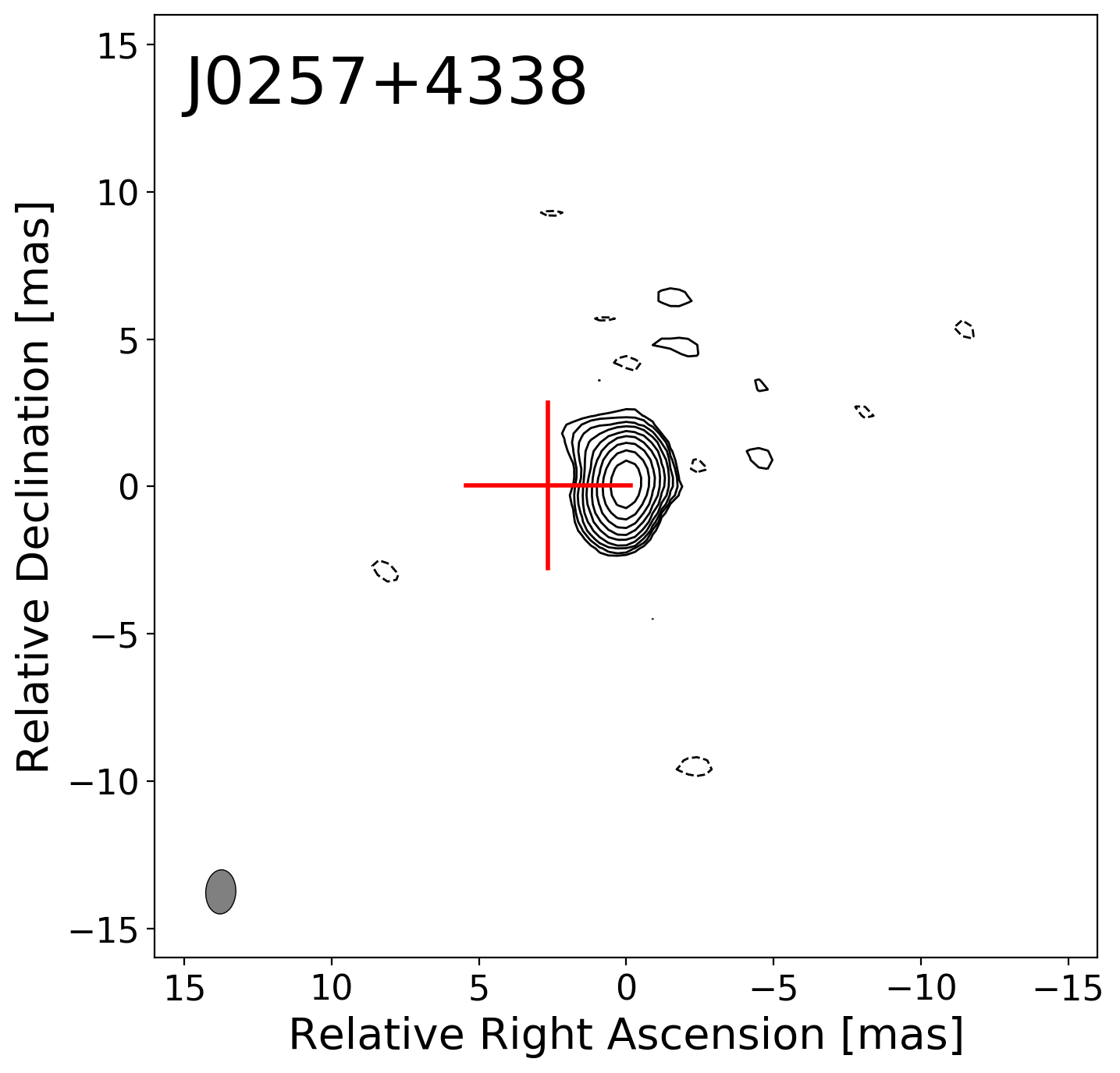}
            }
\gridline{  \includegraphics[width=0.40\textwidth]{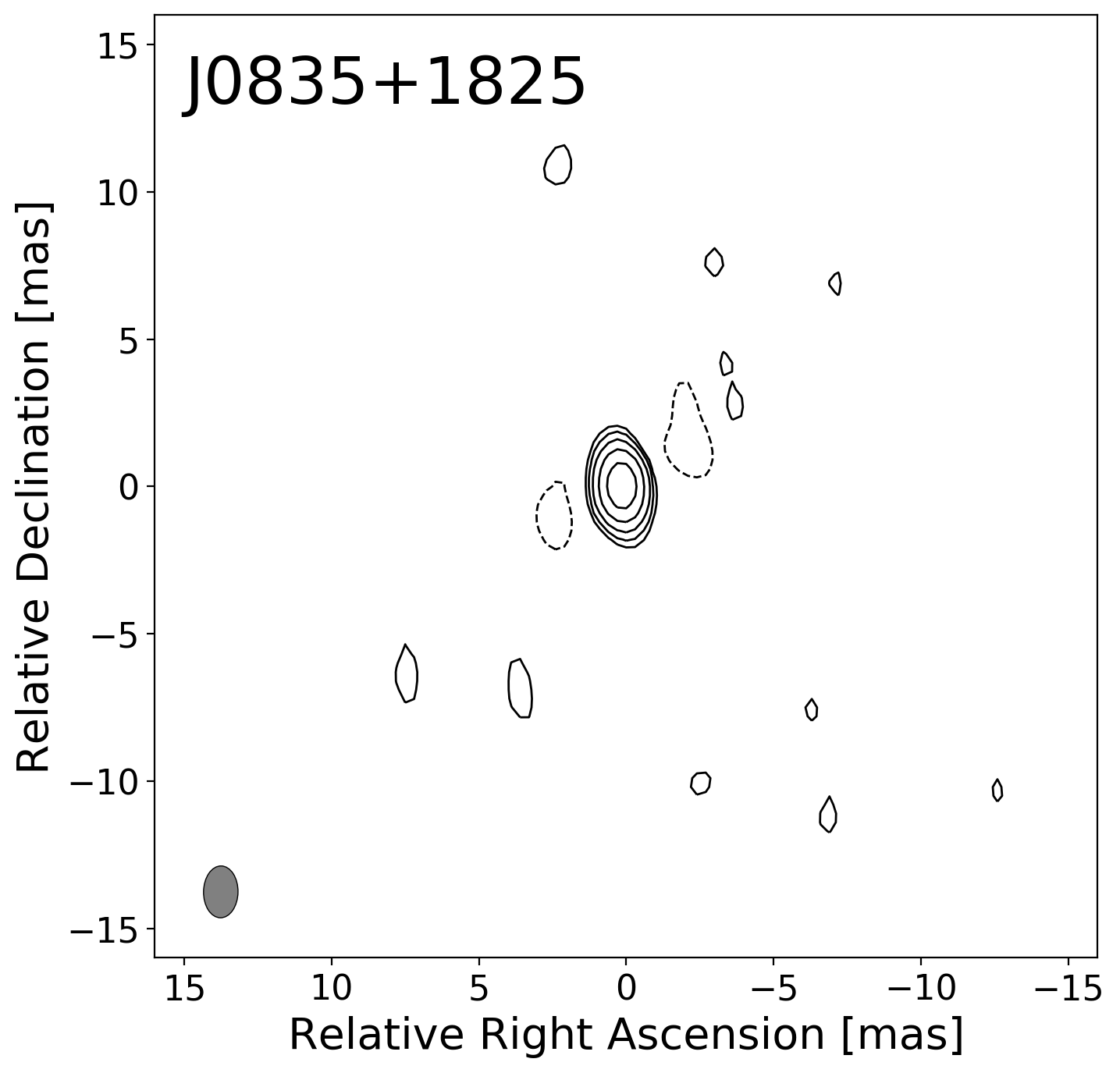}
            \includegraphics[width=0.40\textwidth]{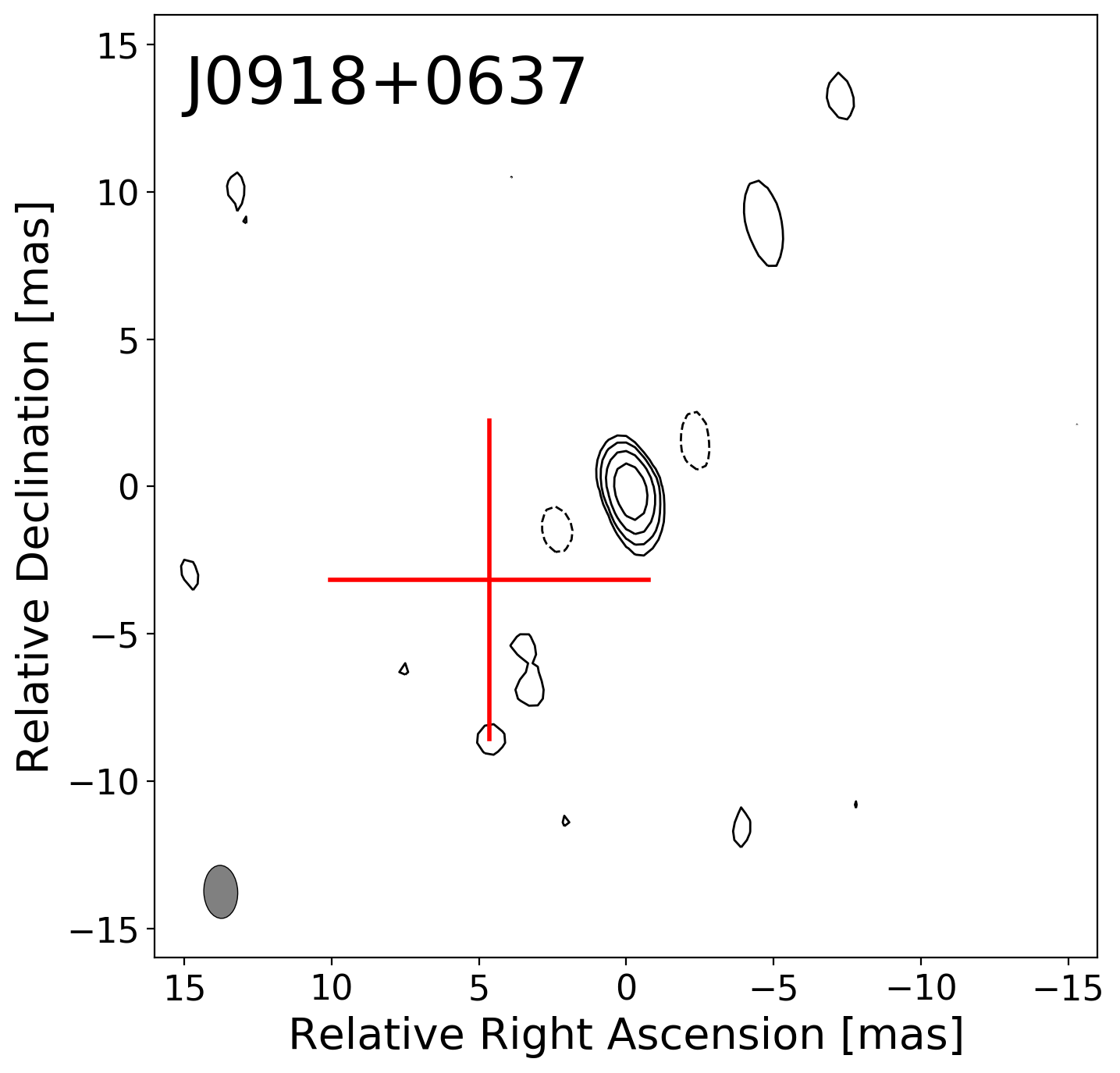}
            }
    \caption{Naturally weighted $5$-GHz EVN images of $17$ high-redshift quasars. The lowest contours are drawn at $\pm3$ times the image noise, the positive contours increase by a factor of 2. The restoring beam is shown in the bottom-left corner. Table~\ref{tab:imgparams} contains the image parameters and the coordinates corresponding to the image centers. Red crosses mark the \textit{Gaia} DR3 optical position in the cases where it is available. The size of the crosses indicates the $3\sigma_{\mathrm{pos}}$ uncertainty.} 
    \label{fig:imgs}
\end{figure*}
\begin{figure*}
    \setcounter{figure}{0}
\gridline{  \includegraphics[width=0.40\textwidth]{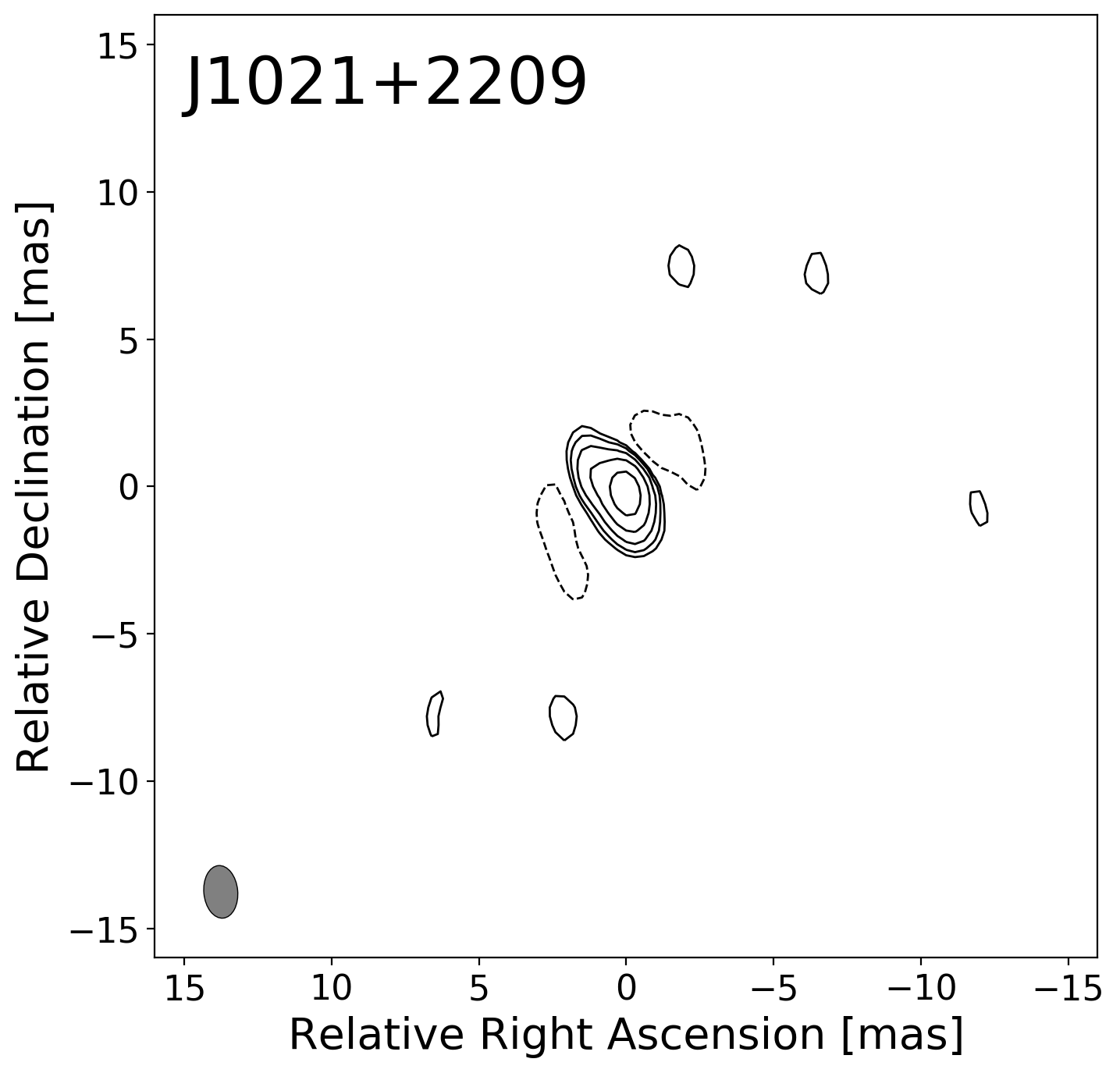}
            \includegraphics[width=0.40\textwidth]{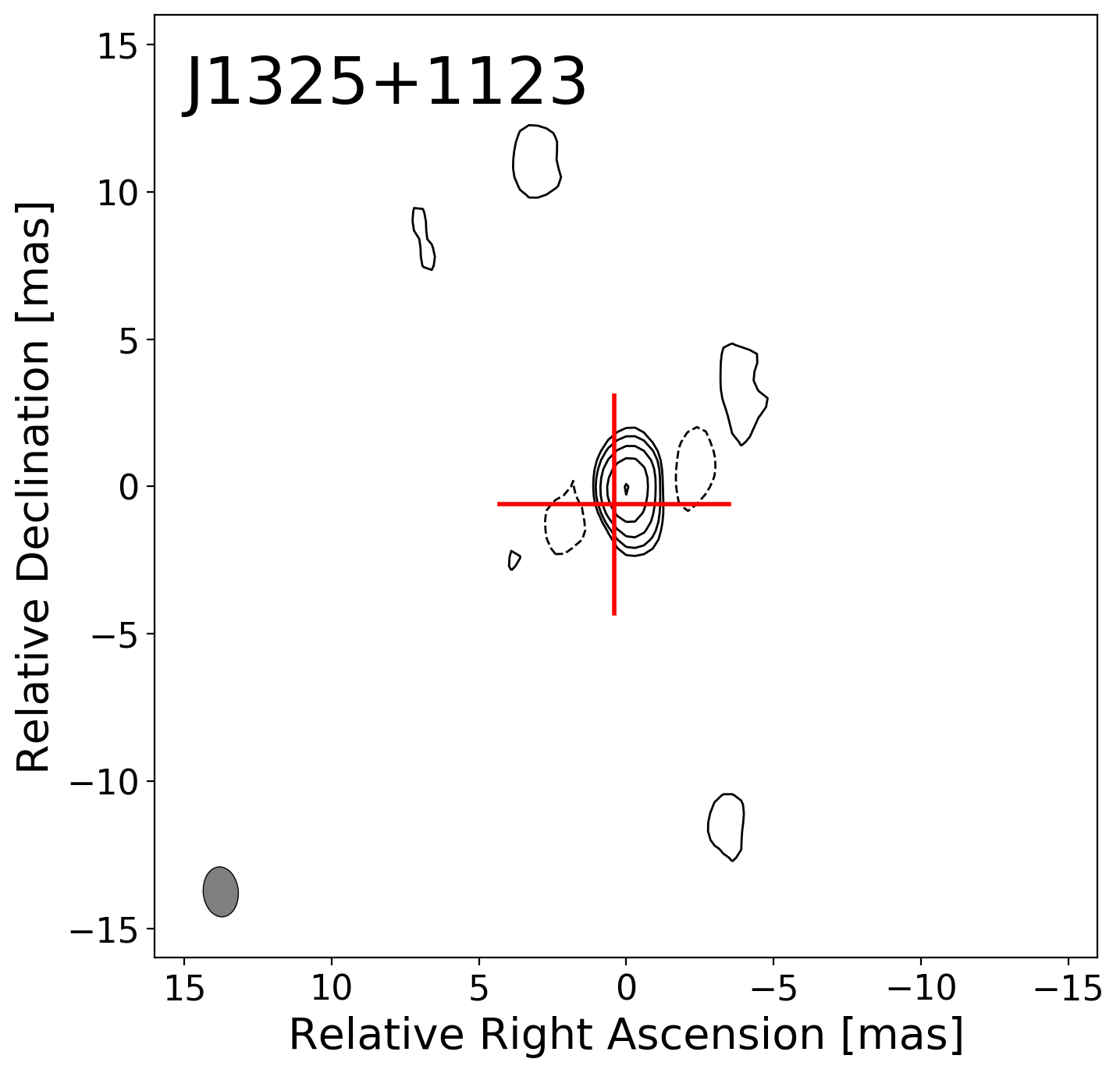}
            }
\gridline{  \includegraphics[width=0.40\textwidth]{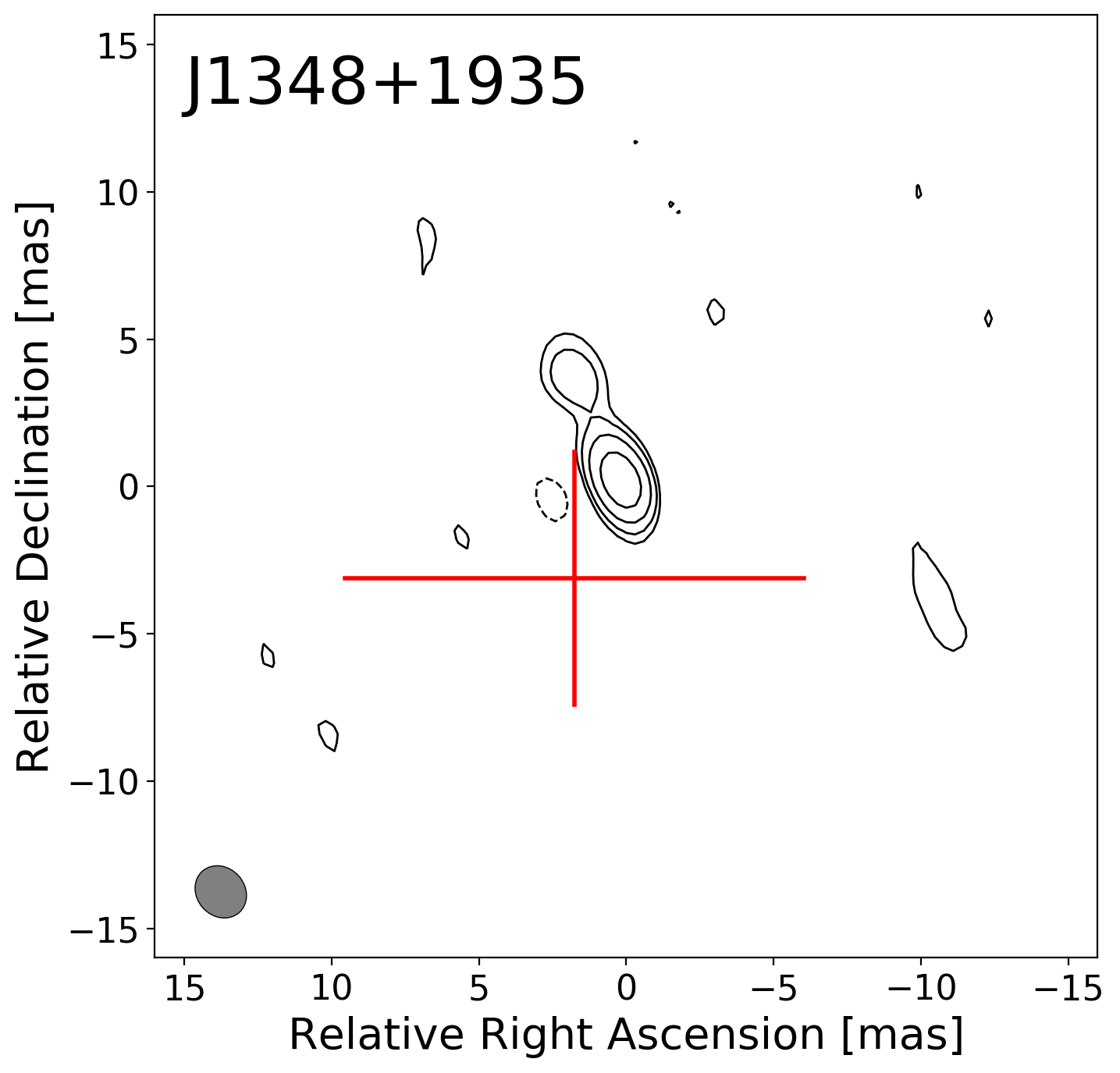}
            \includegraphics[width=0.40\textwidth]{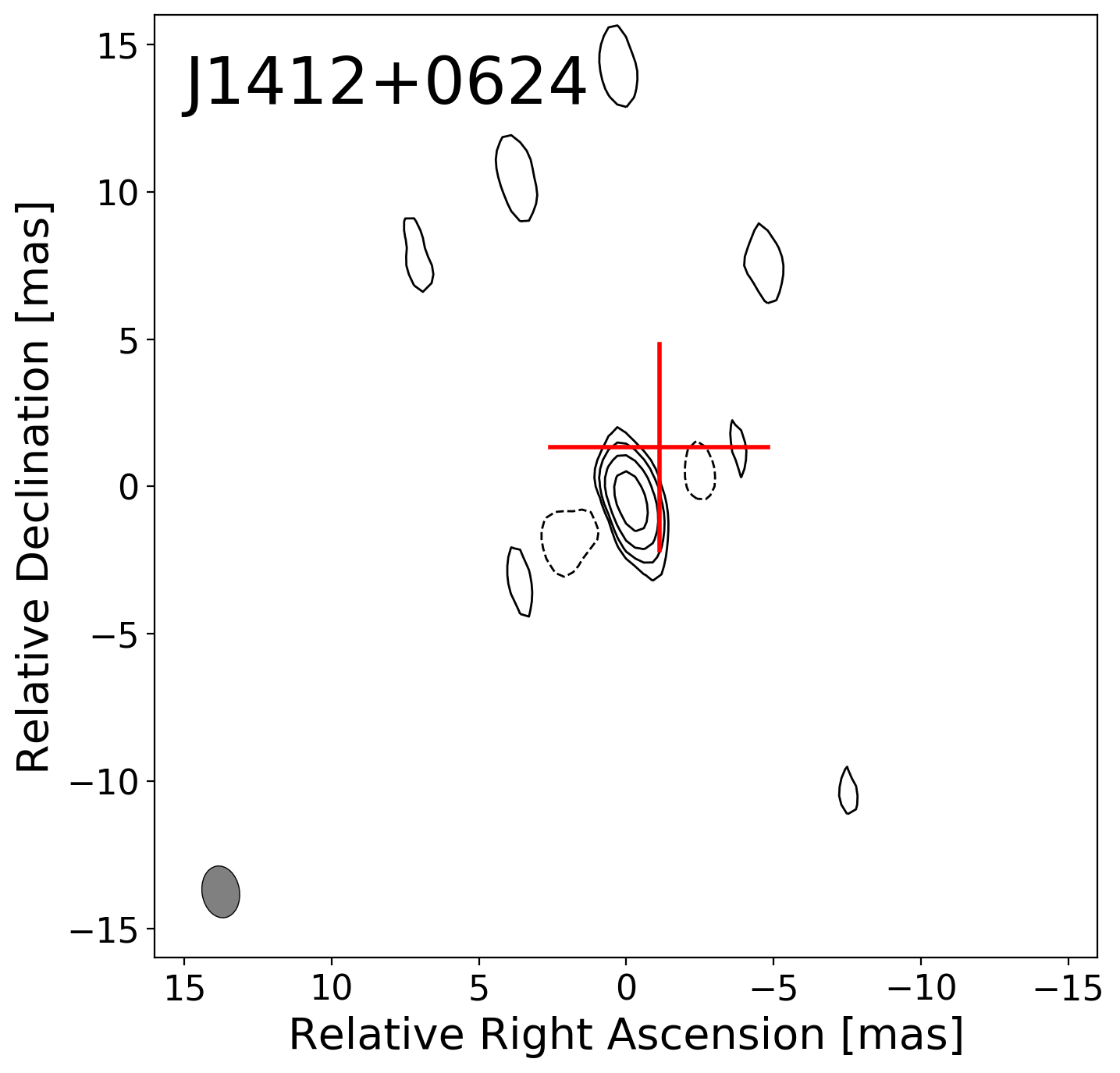}
            }
\gridline{  \includegraphics[width=0.40\textwidth]{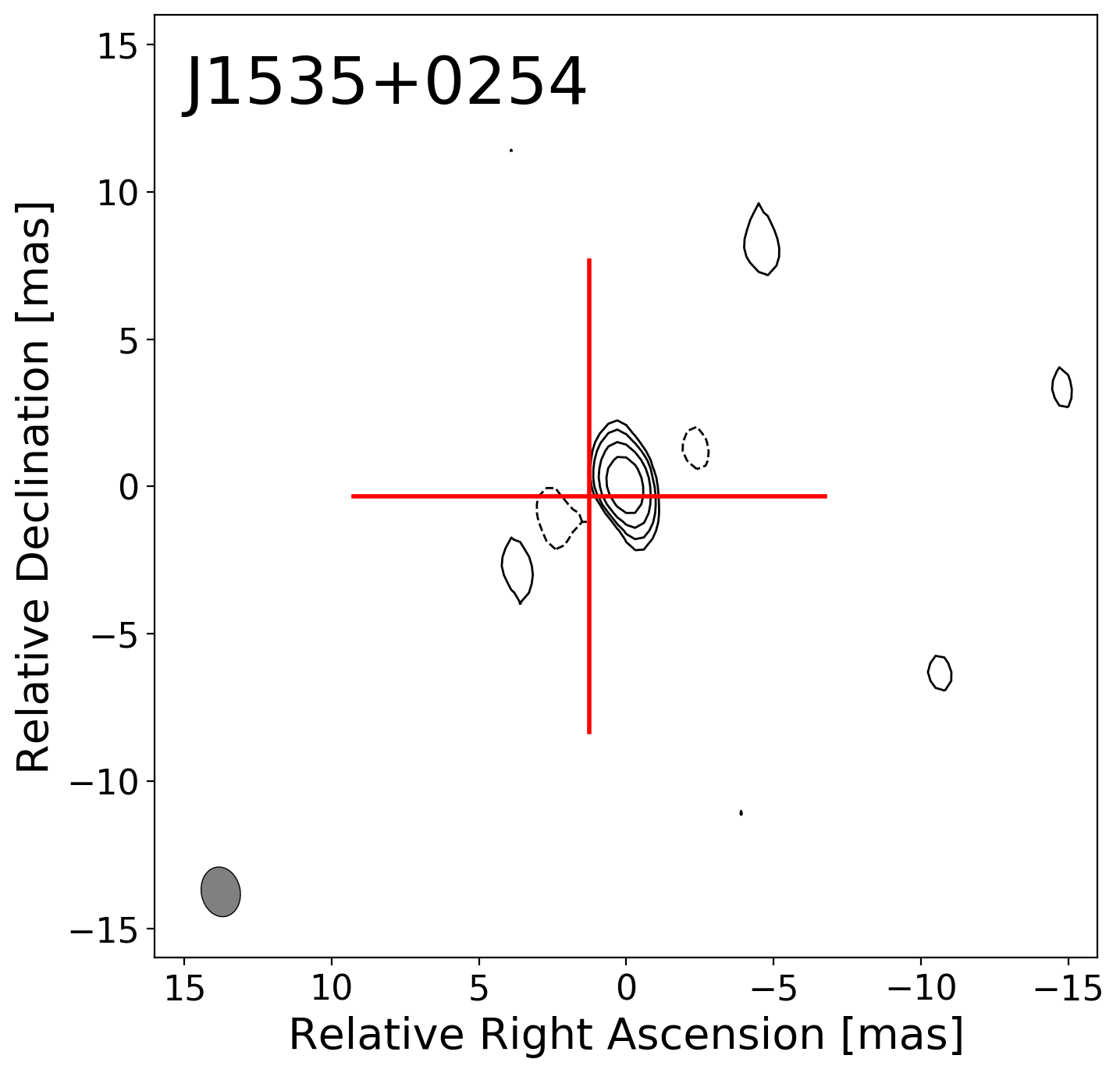}
            \includegraphics[width=0.40\textwidth]{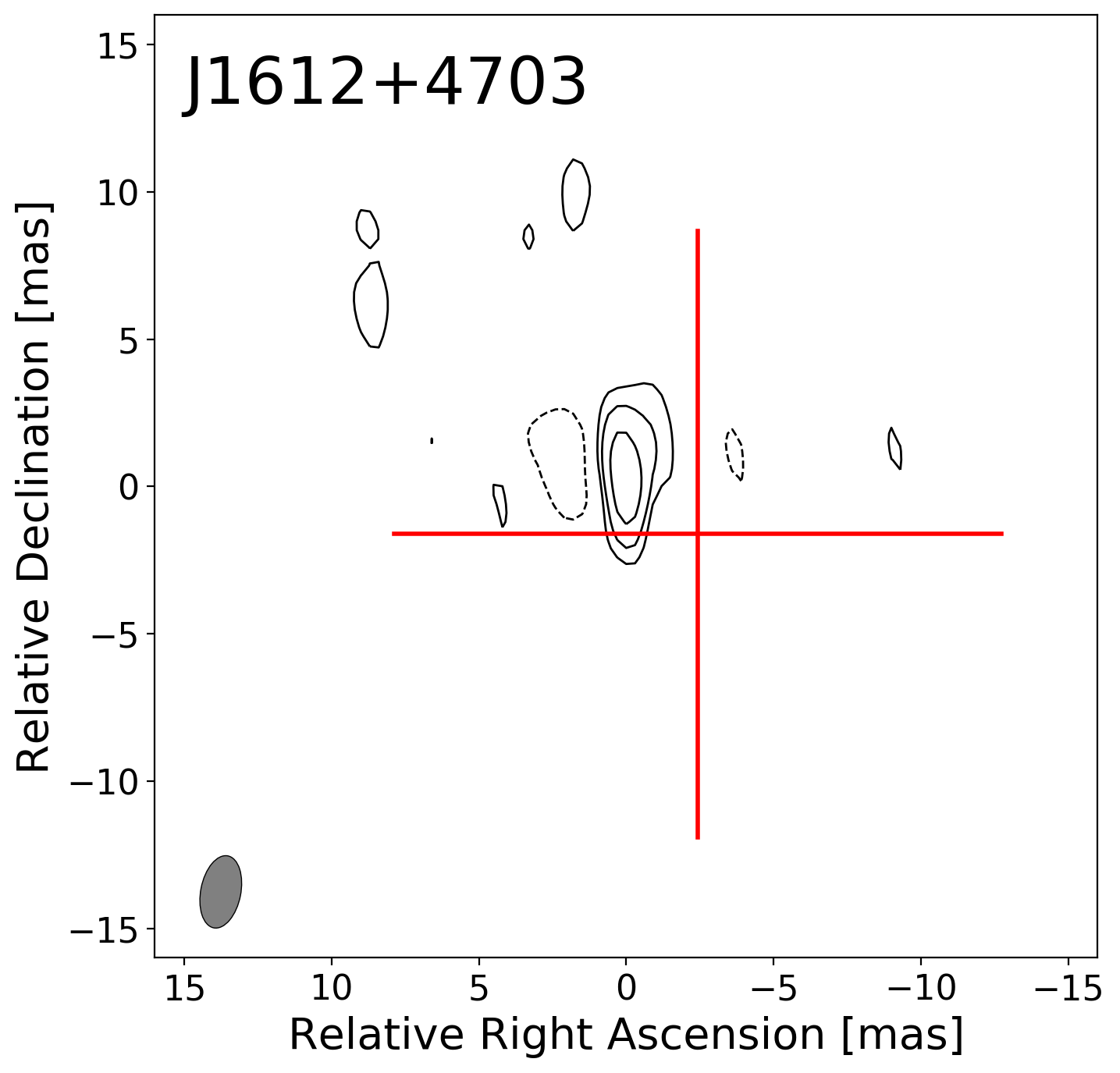}
            }
    \caption{$Continued$}
\end{figure*}
\begin{figure*}
    \setcounter{figure}{0}
\gridline{  \includegraphics[width=0.40\textwidth]{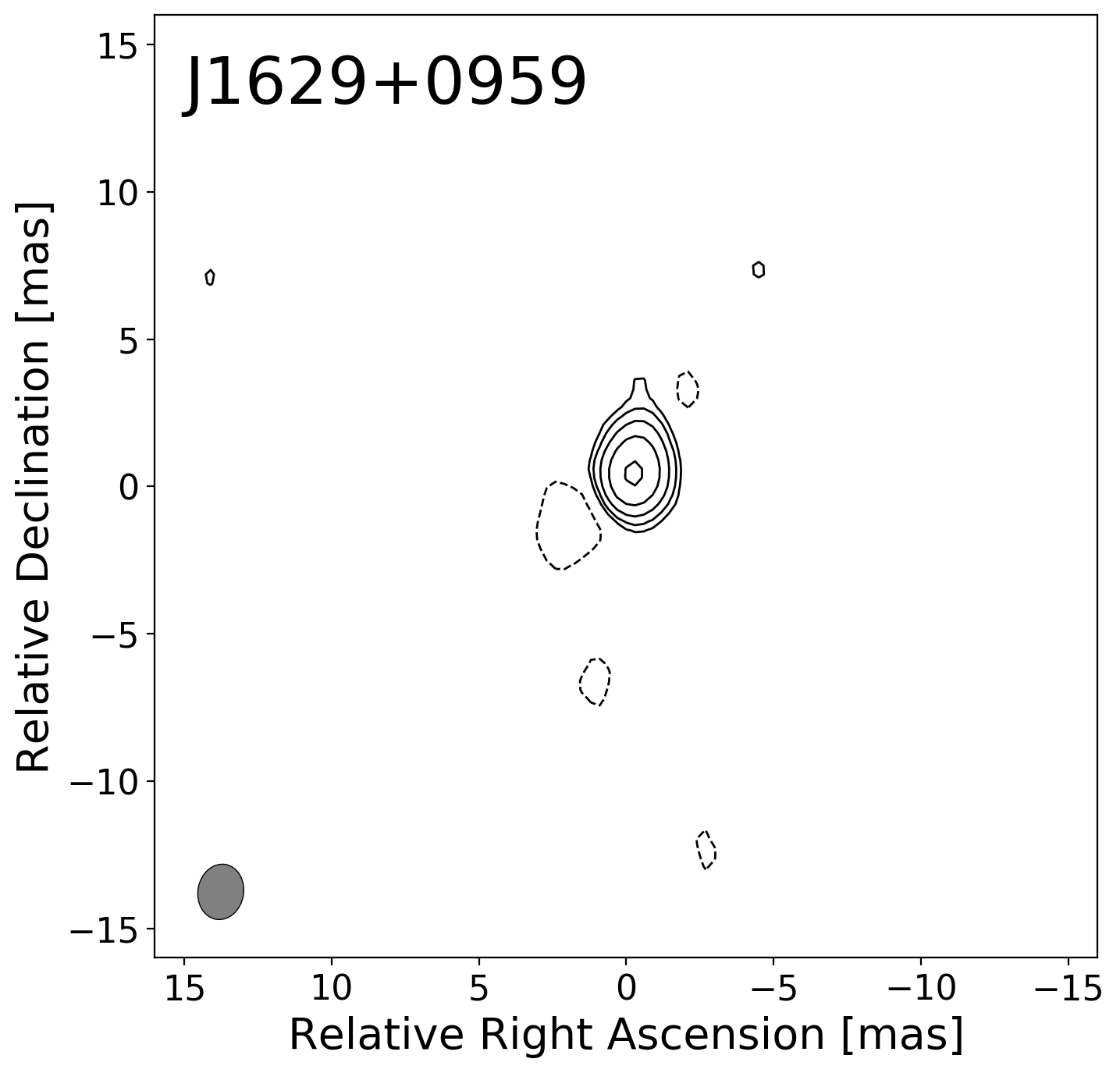}
            \includegraphics[width=0.40\textwidth]{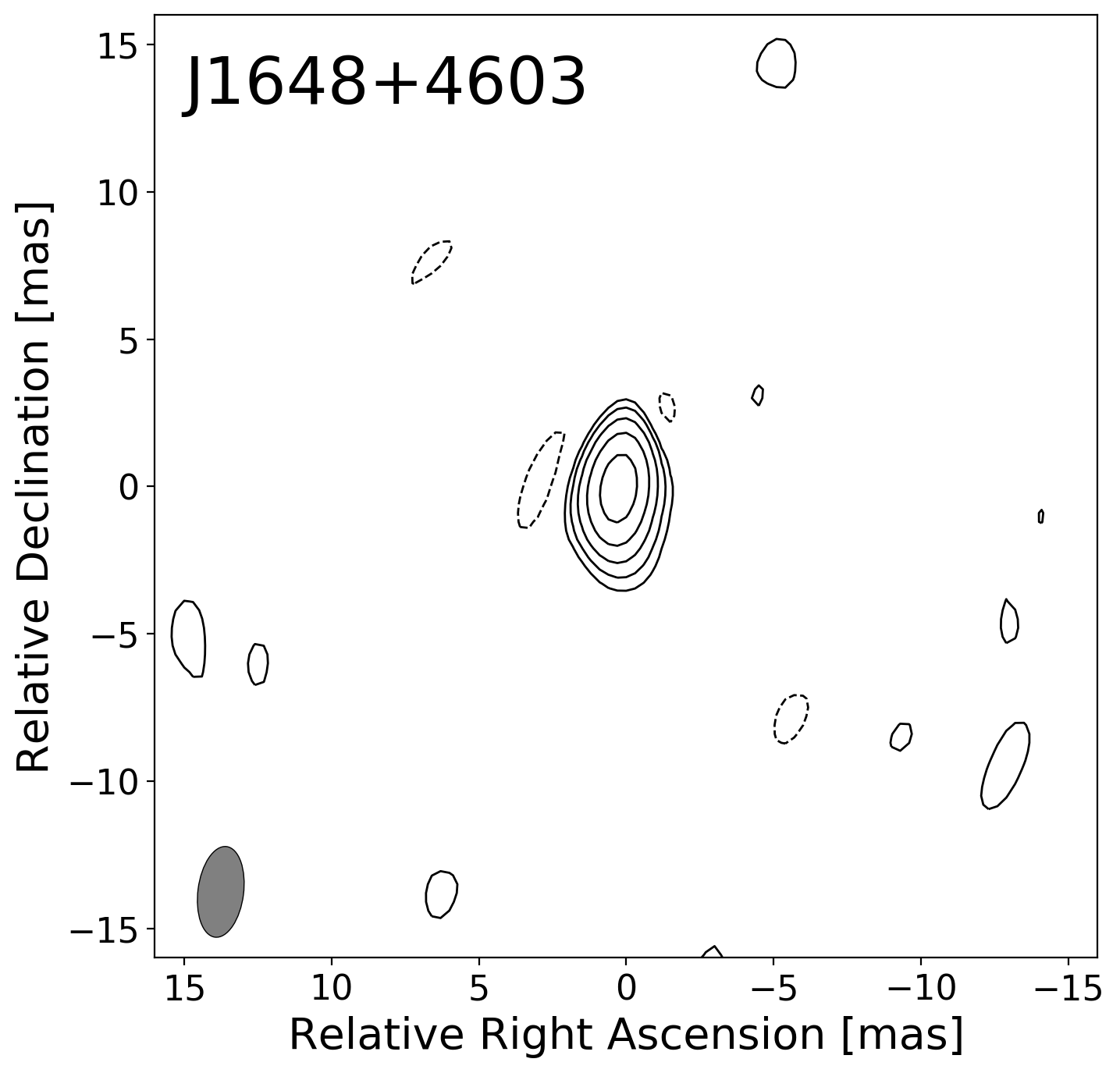}
            }
\gridline{  \includegraphics[width=0.40\textwidth]{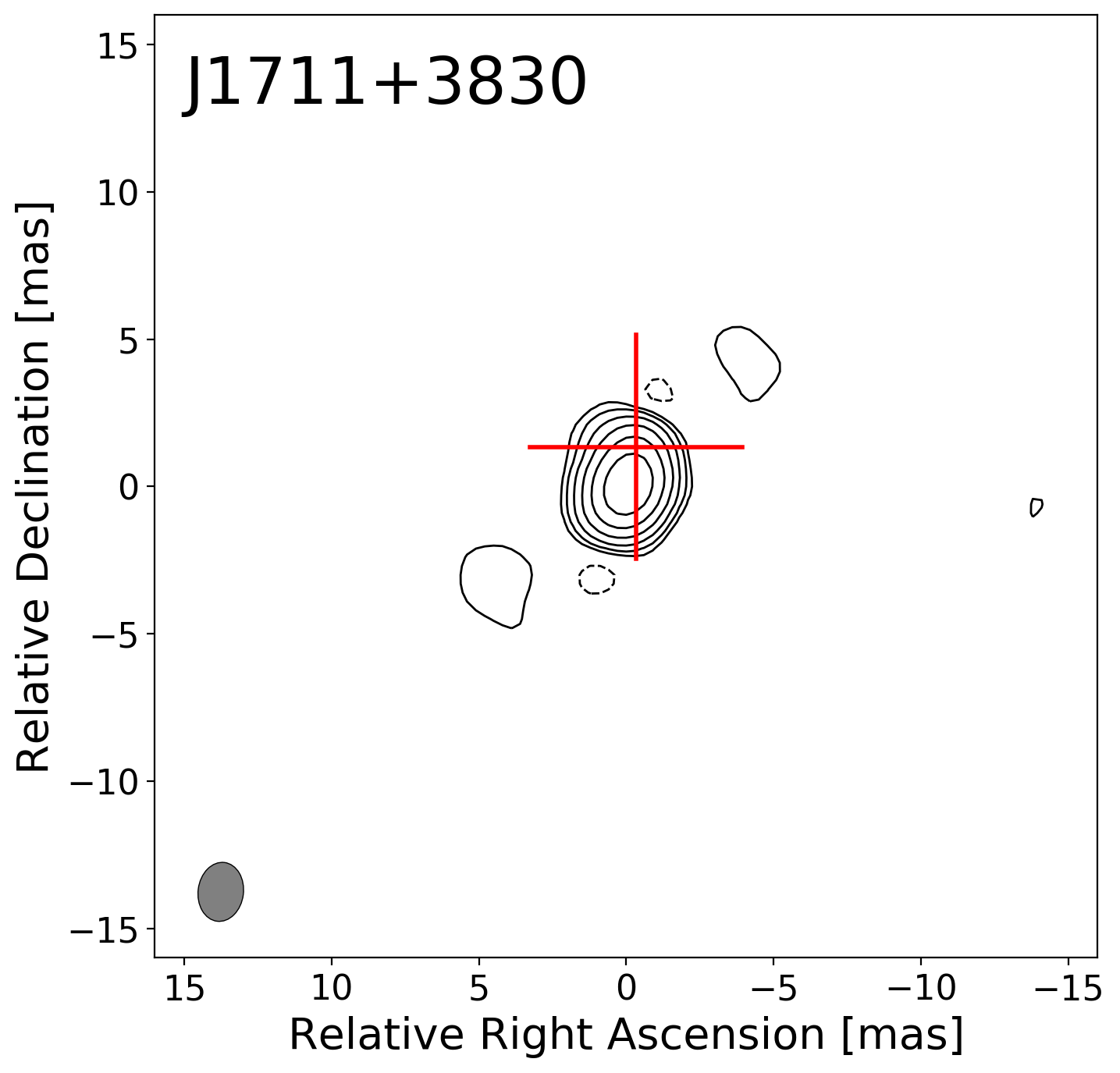}
            \includegraphics[width=0.40\textwidth]{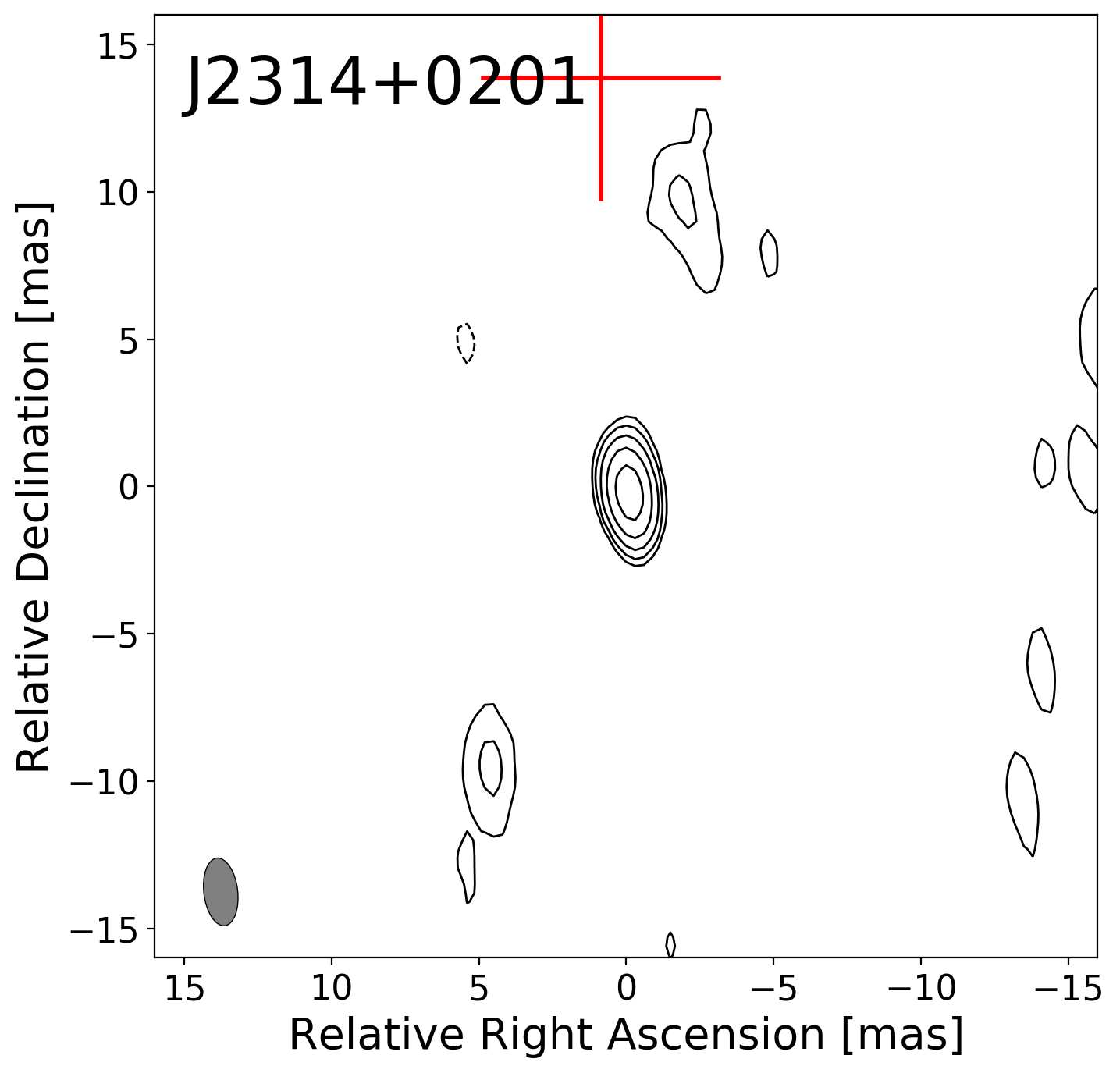}
            }
\gridline{  \includegraphics[width=0.40\textwidth]{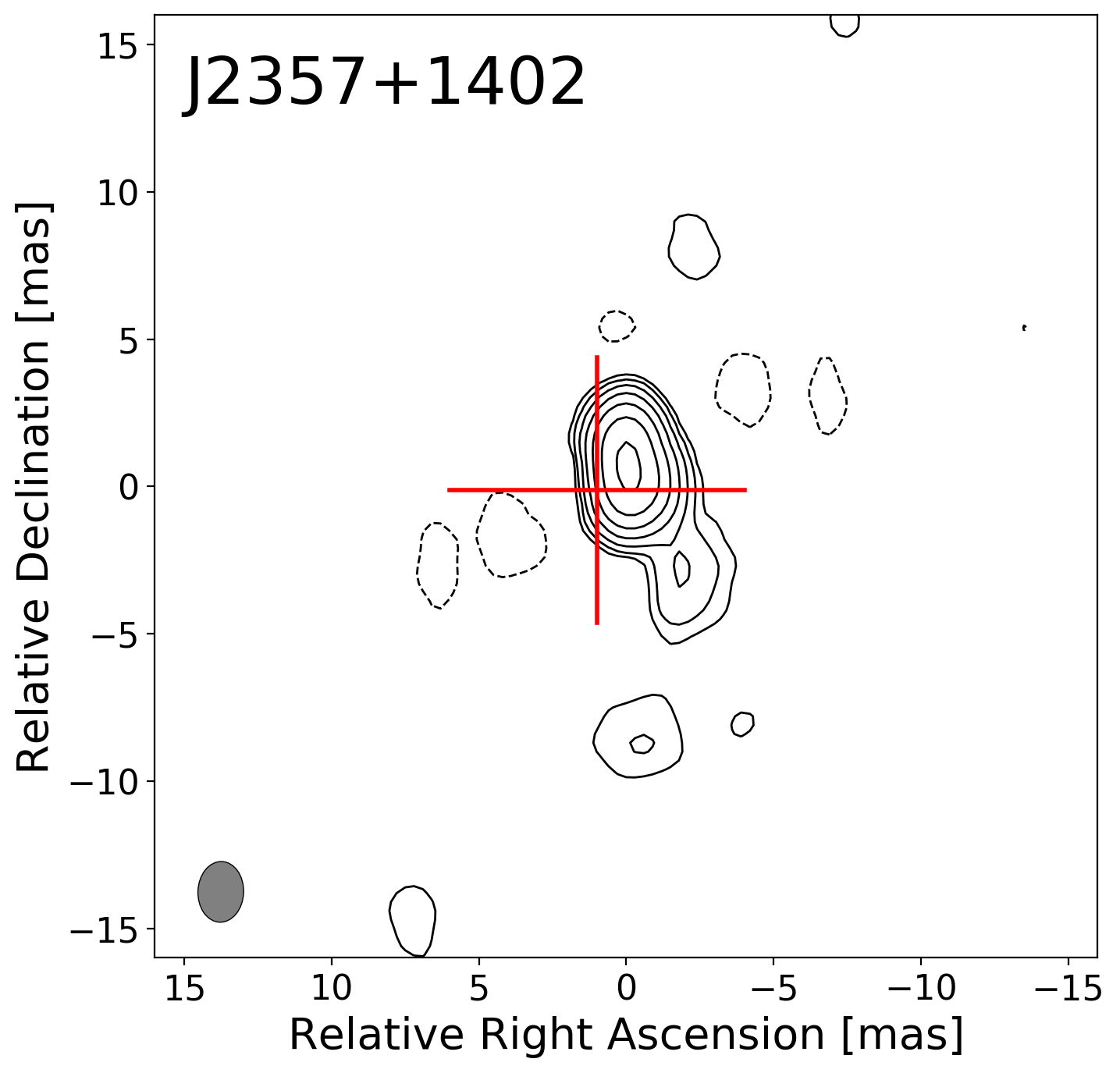}
            }
    \caption{$Continued$}
\end{figure*}

\section{Discussion} \label{discussion}
\subsection{The Origin of the Radio Emission and Doppler Boosting} \label{origin}

For all detected sources, the derived brightness temperatures exceed $T_\mathrm{b} = 10^6$~K by at least three orders of magnitude (Table~\ref{tab:phyparams}), indicating non-thermal radio emission related to AGN activity \citep{1992ARA&A..30..575C,2000ApJ...530..704K,2011A&A...526A..74M}. In general, AGN-associated activity in radio incudes relativistic jets, outflows, and coronal emission \citep{2019NatAs...3..387P}. The $T_\mathrm{b} > 10^9$~K values determined in our sample indicate the presence of relativistic plasma jets in these sources. There is also evidence for AGN activity based on the X-ray analysis of these objects \citep{2019MNRAS.489.2732I}. The Doppler factor can be calculated as \begin{equation} \label{eq:doppler}
    \delta = \frac{T_\mathrm{b}}{T_\mathrm{b,int}}.
\end{equation}
The emission is considered Doppler-boosted if $\delta > 1$. Because of the frequency dependence of $T_\mathrm{b,int}$ \citep{2014JKAS...47..303L}, and our observed 5-GHz frequency corresponds to $\sim27$-GHz rest-frame frequency at the average $z$ of the sample, $T_\mathrm{b,int}$ should be around $2 \times 10^{10}$~K. We adopted $T_\mathrm{b,int} = 2 \times 10^{10}$~K to calculate the Doppler factors using Equation~\ref{eq:doppler}. The $\delta$ value is lower than unity in $5$ of our sources. For what concerns the rest of the sample, $11$ objects appear Doppler-boosted (Table~\ref{tab:phyparams}). The unresolved source, J1612$+$4703, with $\delta > 0.2$ could still be Doppler-boosted. When $\delta \leq 1$, it is because either the jet inclination angle is large, or the flux density is measured at a frequency far away from that of the spectral peak. 

\subsection{Radio Spectra} \label{sed}
Total flux density data from single-dish and low-resolution interferometric measurements are available for all of the selected quasars. The frequencies range from $\sim 100$~MHz to $\sim 10$~GHz, allowing us to build their total flux density spectra. Most of the data are from the Green Bank surveys such as 87GB \citep{1991ApJS...75.1011G} and GB6 \citep{1996ApJS..103..427G}, the VLA surveys, such as FIRST \citep{1997ApJ...475..479W}, NVSS \citep{1998AJ....115.1693C}, VLASS \citep{2020PASP..132c5001L,2020RNAAS...4..175G}, and CLASS \citep{2003MNRAS.341....1M}, as well as RACS \citep{2020PASA...37...48M,2024PASA...41....3D,2025PASA...42...38D} and the TGSS \citep[Tata Institute of Fundamental Research Giant Metrewave Radio Telescope Sky Survey,][]{2017AA...598A..78I}. Table~\ref{tab:sed} lists all the spectral information collected from the literature with their respective references. 

We fitted the observed spectral points to characterize the shape of the continuum spectra. The slope of a power-law spectrum is charaterized by the spectal index $\alpha$. The spectrum is considered steep if $\alpha < -0.5$, flat if $-0.5 \leq \alpha \leq 0$, and inverted if $\alpha > 0$. Out of the $18$ spectra, $14$ were best characterized with a power-law function. The spectra of the remaining $4$ sources (J0031$+$1507, J0257$+$4338, J0918$+$0637, and J1325$+$1123) show a curved shape, where best fitted with a log-parabolic function in the form of 
\begin{equation} \label{eq:logp}
\log S = a (\log \nu - \log~\nu_0)^2 + b,
\end{equation} 
resulting in lower reduced $\chi^2$ values compared to a power-law fit. In the formula above, $\nu_0$ is the turnover frequency corresponding to the peak flux density $S_0$, while $a$ and $b$ are numerical constants \citep[e.g.][]{2000A&A...363..887D,2017MNRAS.467.2039C}. In bright $z>3$ radio quasars, it is common to find peaked continuum radio spectra \citep[e.g.][]{2021MNRAS.508.2798S,2022A&A...659A.159S,2025A&A...698A.158I}. A few peaked-spectrum sources were found among high-$z$ radio quasars in our previous studies \citep{2022ApJS..260...49K,2024AA...690A.321K}. We can estimate the physical scale of the peaked-spectrum sources using the correlation between the rest-frame peak frequency and the linear size \citep{1997AJ....113..148O}. The four peaked-spectrum sources are J0031$+$1507, J0257$+$4338, J0918$+$0637, and J1325$+$1123, and their calculated sizes are in between $2$ and $25$ pc, which suggest very young radio jets.

\begin{figure*}
\centering
\gridline{  \includegraphics[width=0.33\textwidth]{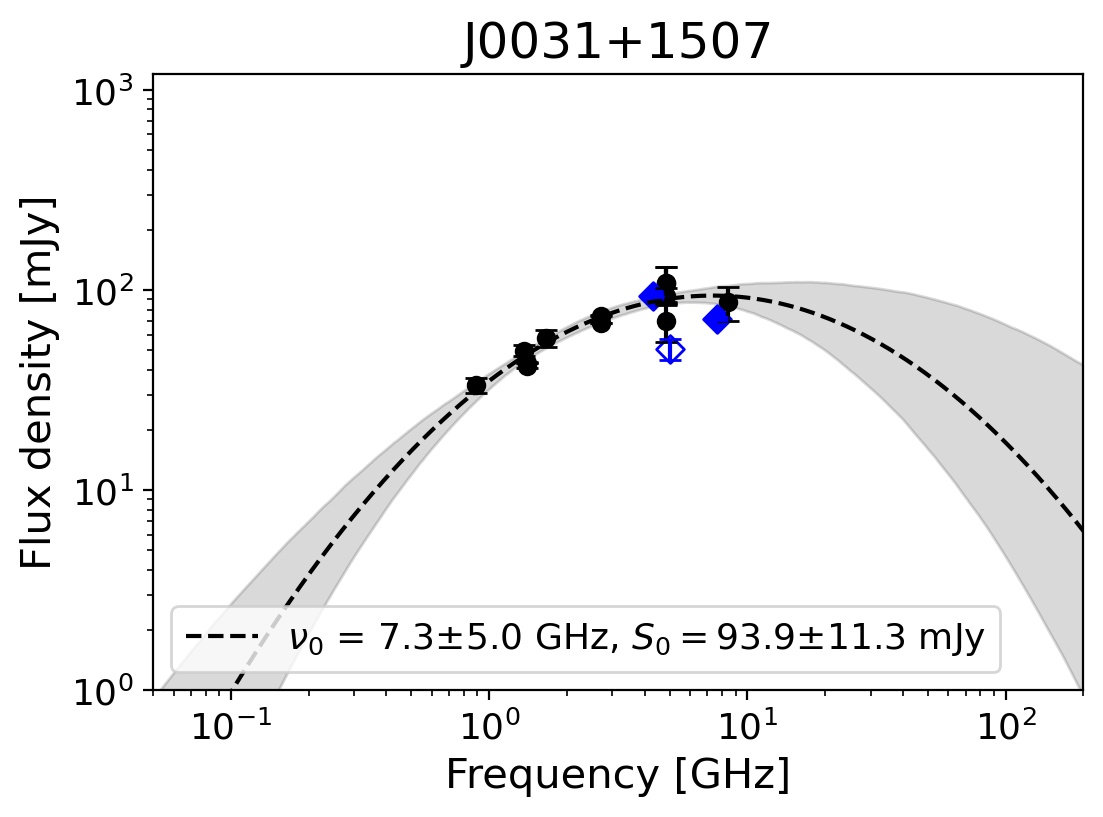}
            \includegraphics[width=0.33\textwidth]{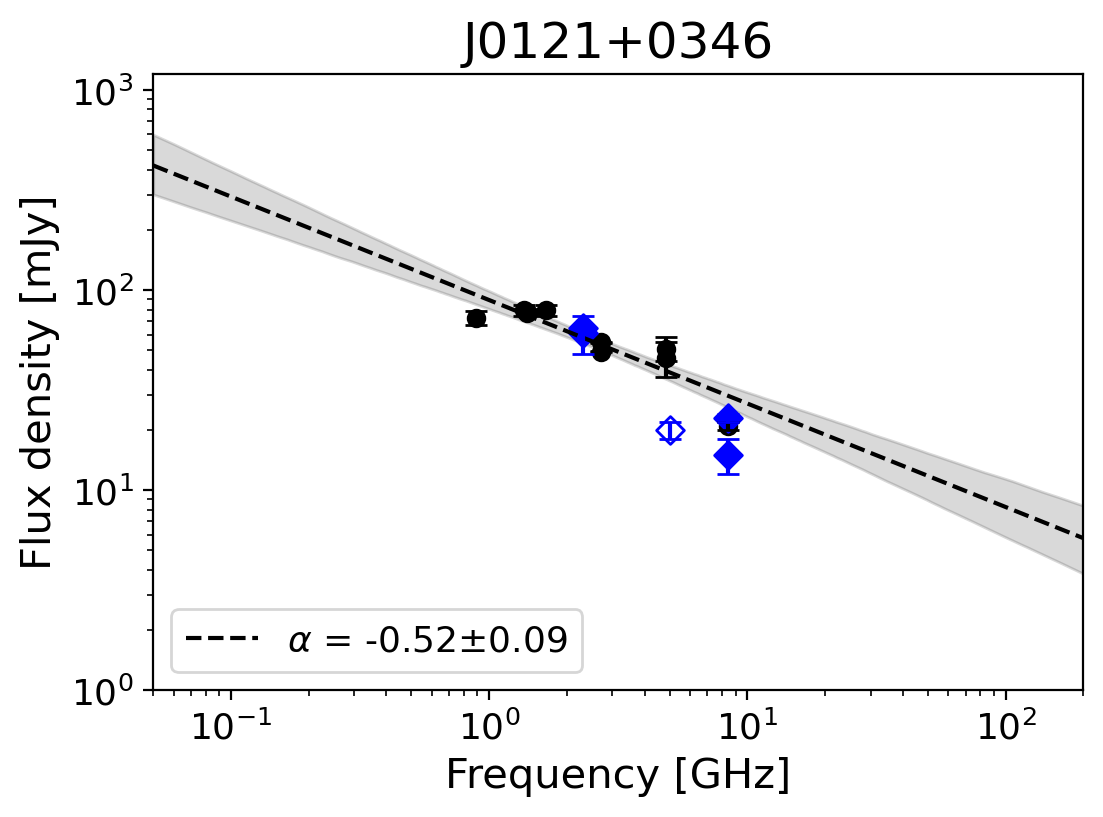}
            \includegraphics[width=0.33\textwidth]{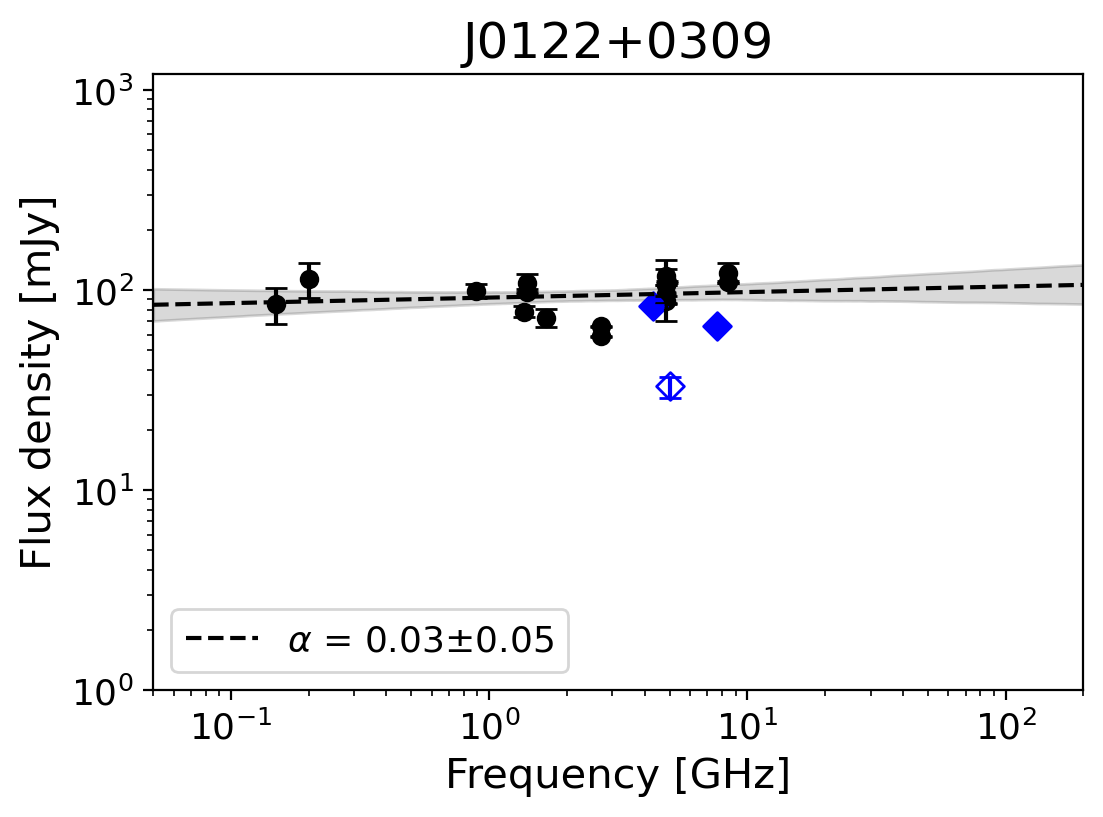}
            }
\gridline{  \includegraphics[width=0.33\textwidth]{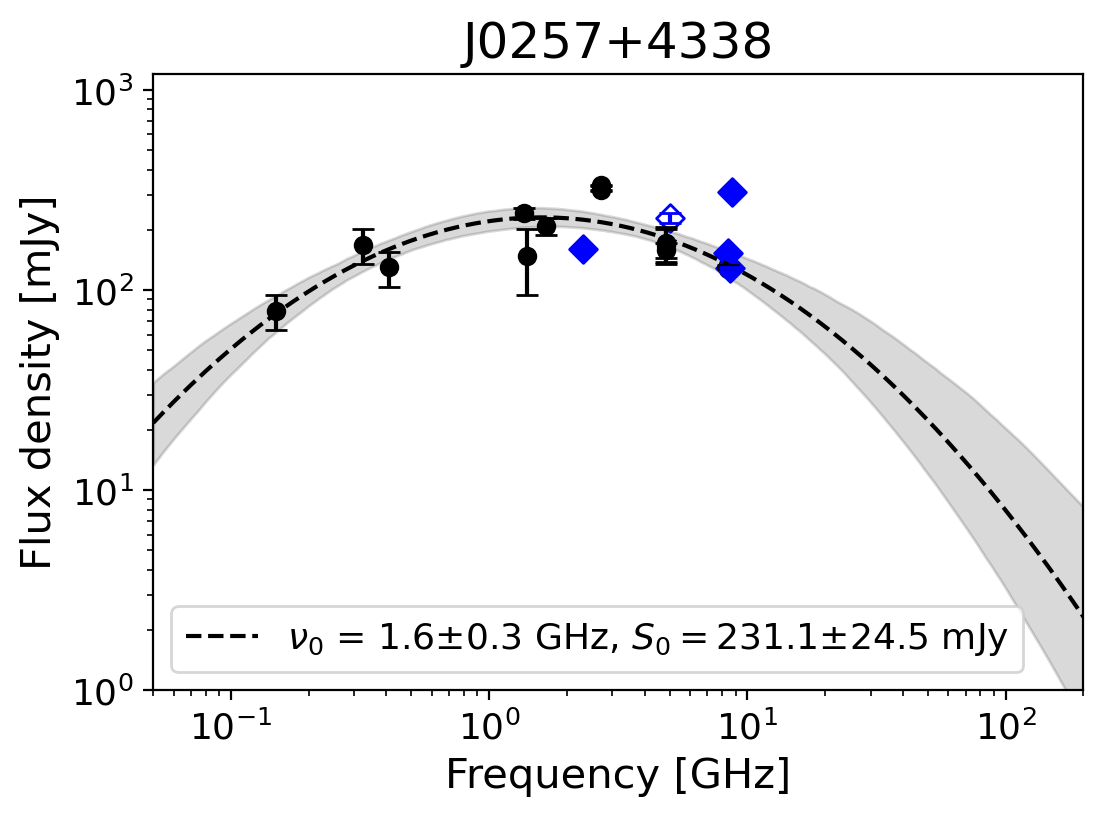}
            \includegraphics[width=0.33\textwidth]{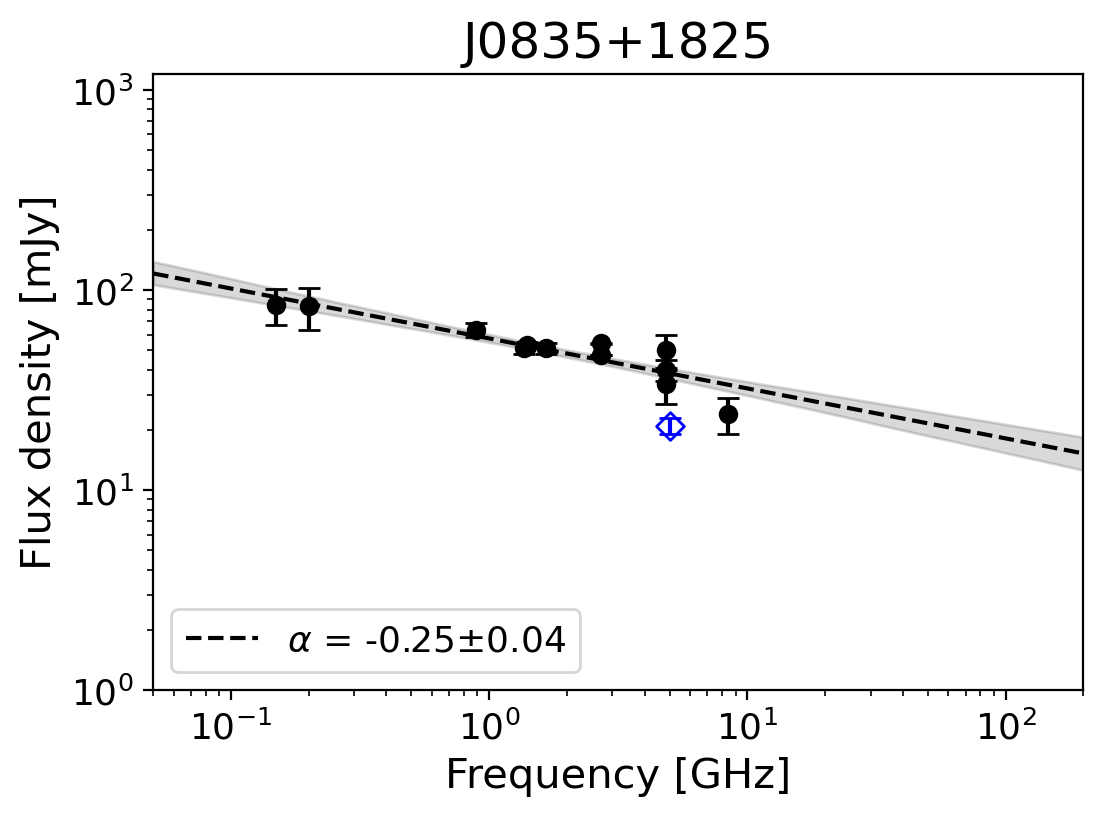}
            \includegraphics[width=0.33\textwidth]{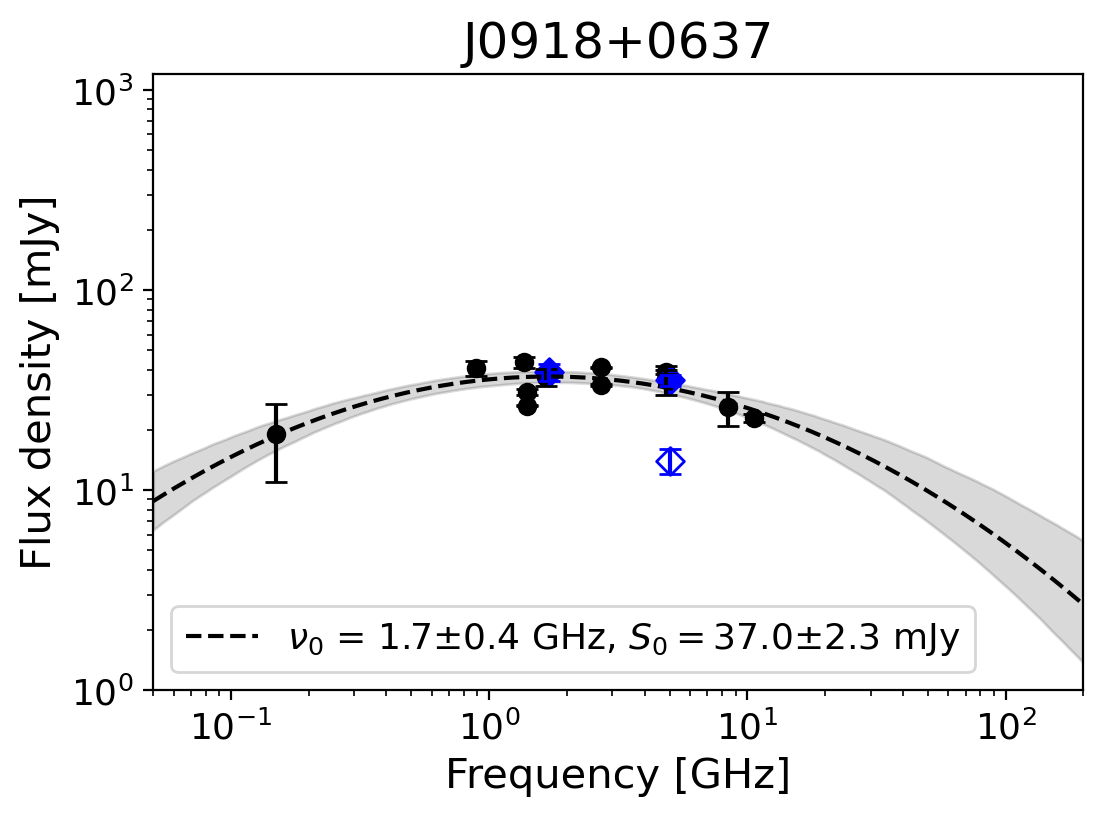}
            }
\gridline{  \includegraphics[width=0.33\textwidth]{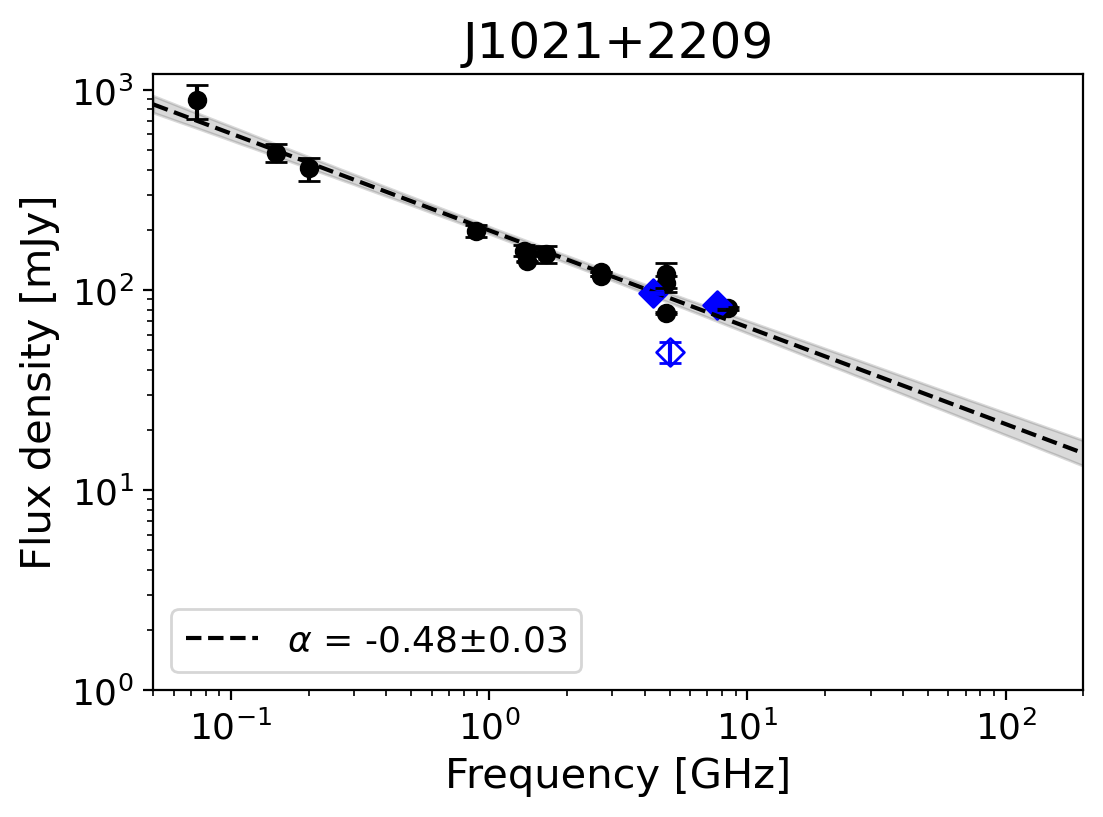}
            \includegraphics[width=0.33\textwidth]{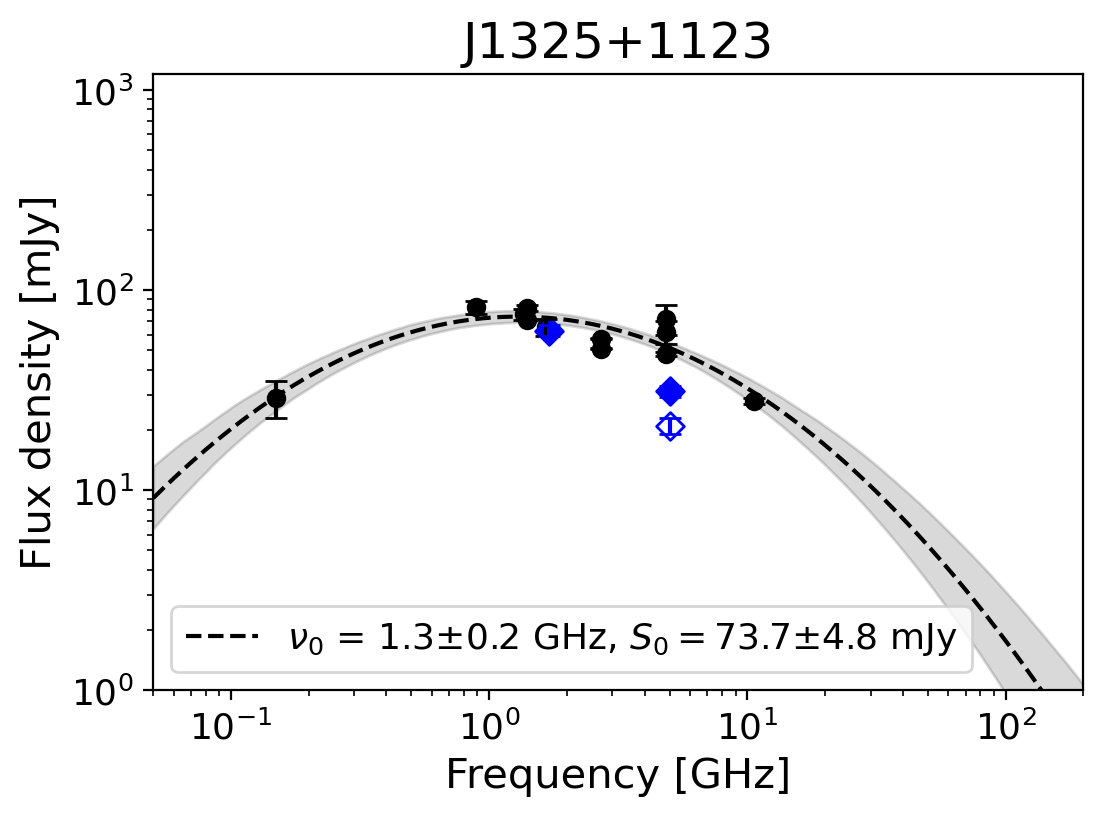}
            \includegraphics[width=0.33\textwidth]{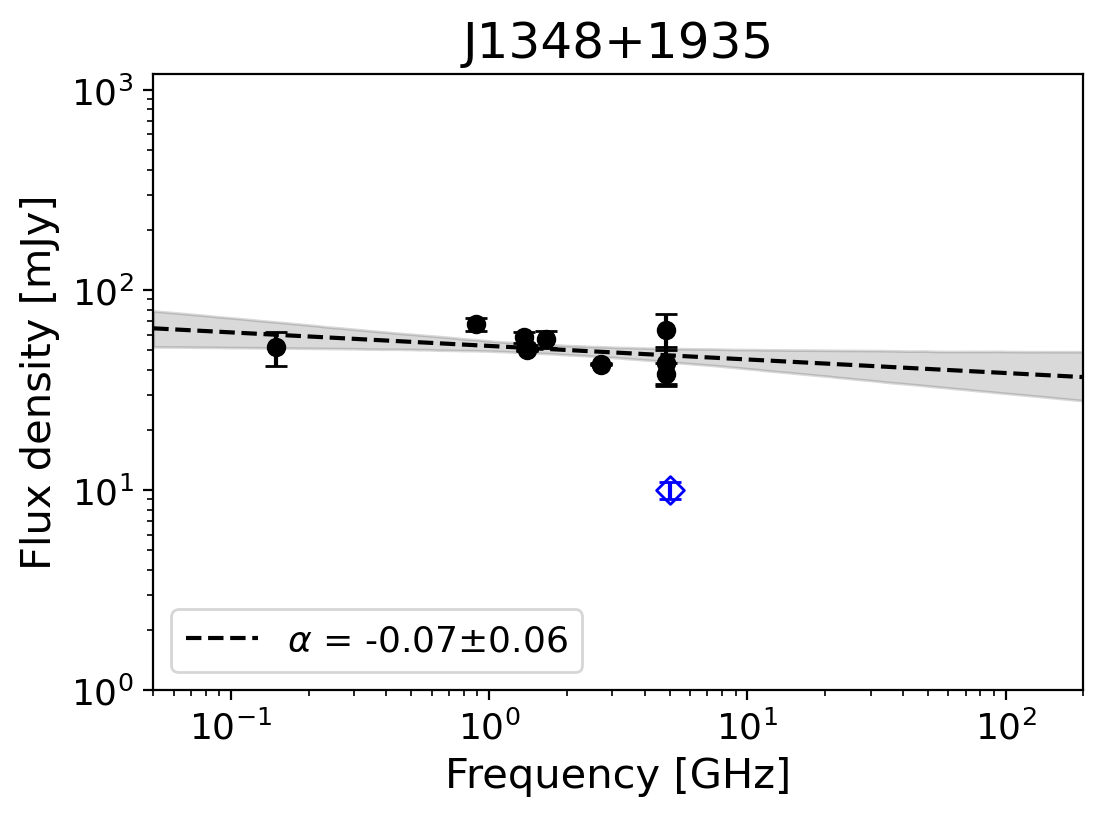}
            }
    \caption{Radio contiuum spectra of the target sources. Black points indicate total flux density data from single-dish and low-resolution interferometric measurements. The black dashed lines show the best-fit spectrum (power-law or log-parabolic). The shaded area indicates the $1\sigma$ error of the fits. The blue filled diamonds are the VLBI flux densities, while the empty diamonds are the $5$-GHz EVN measurements reported in this paper. The parameters of the spectral fits are given in the insets as well as in Table~\ref{tab:phyparams}, and the individual spectral points with references are listed in Table~\ref{tab:sed}.} 
    \label{fig:spec}
\end{figure*}
\begin{figure*}
    \setcounter{figure}{1}
\gridline{  \includegraphics[width=0.33\textwidth]{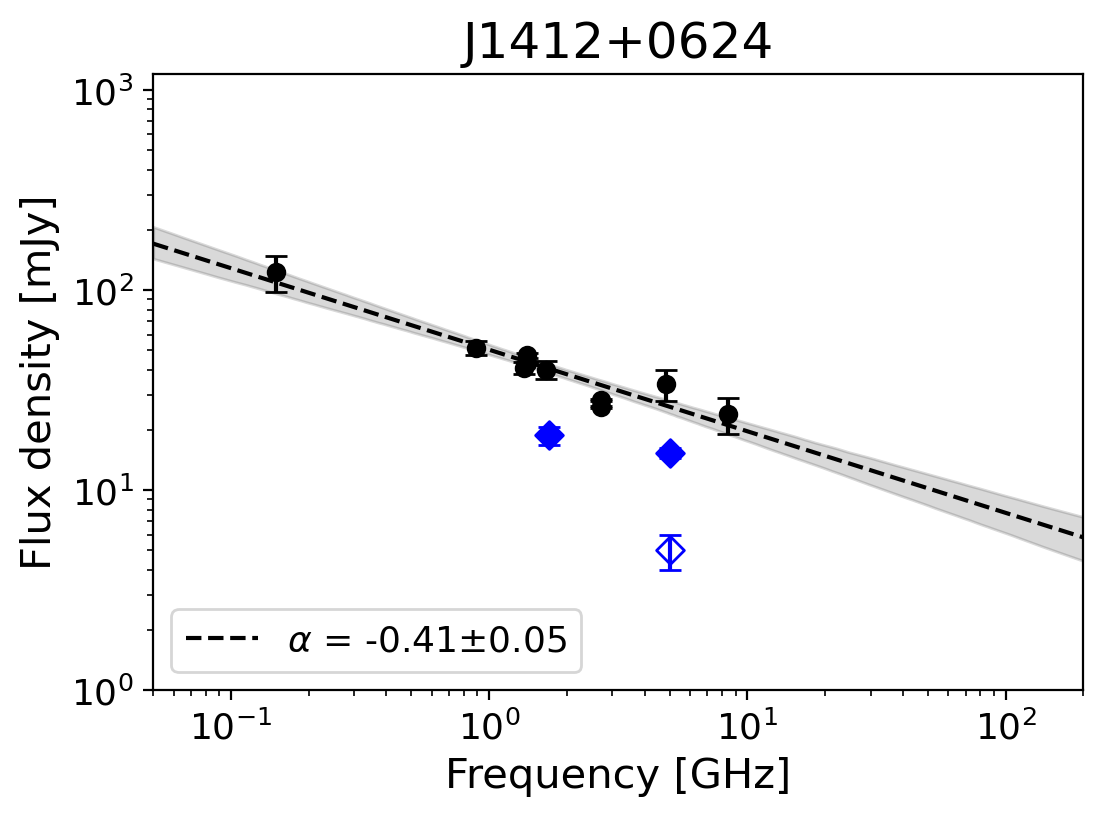}
            \includegraphics[width=0.33\textwidth]{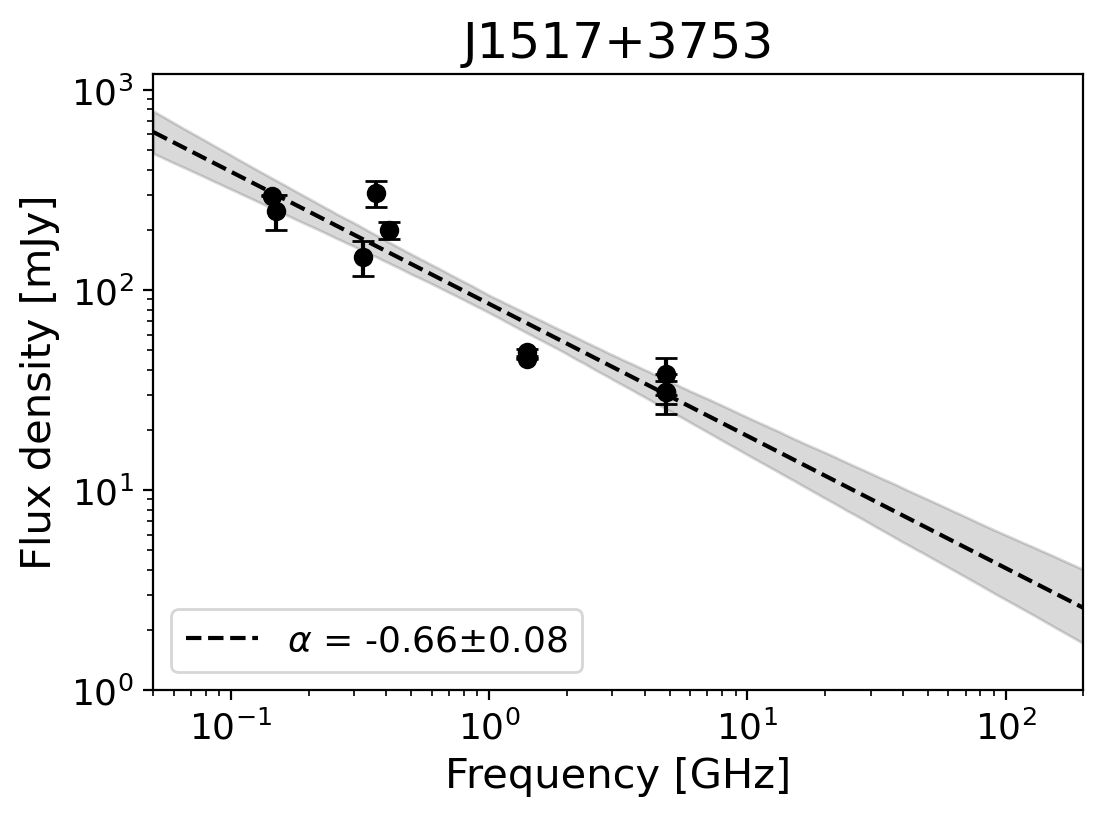}
            \includegraphics[width=0.33\textwidth]{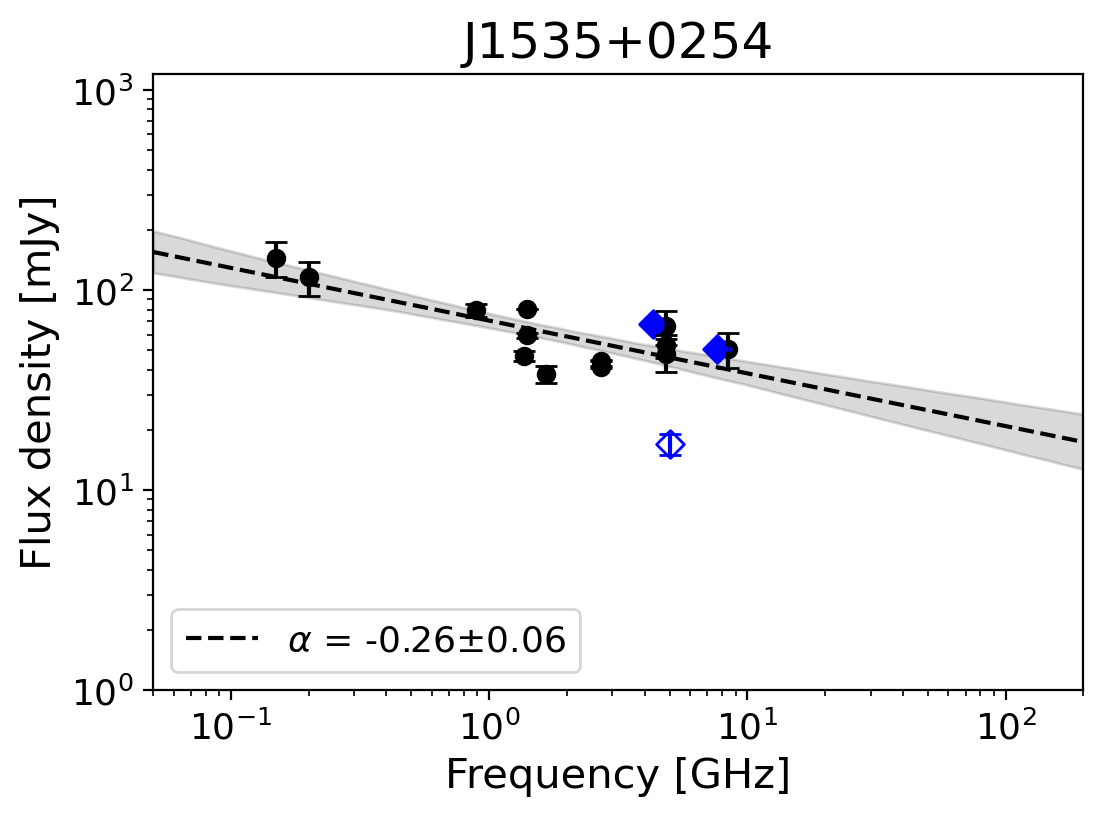}
            }
\gridline{  \includegraphics[width=0.33\textwidth]{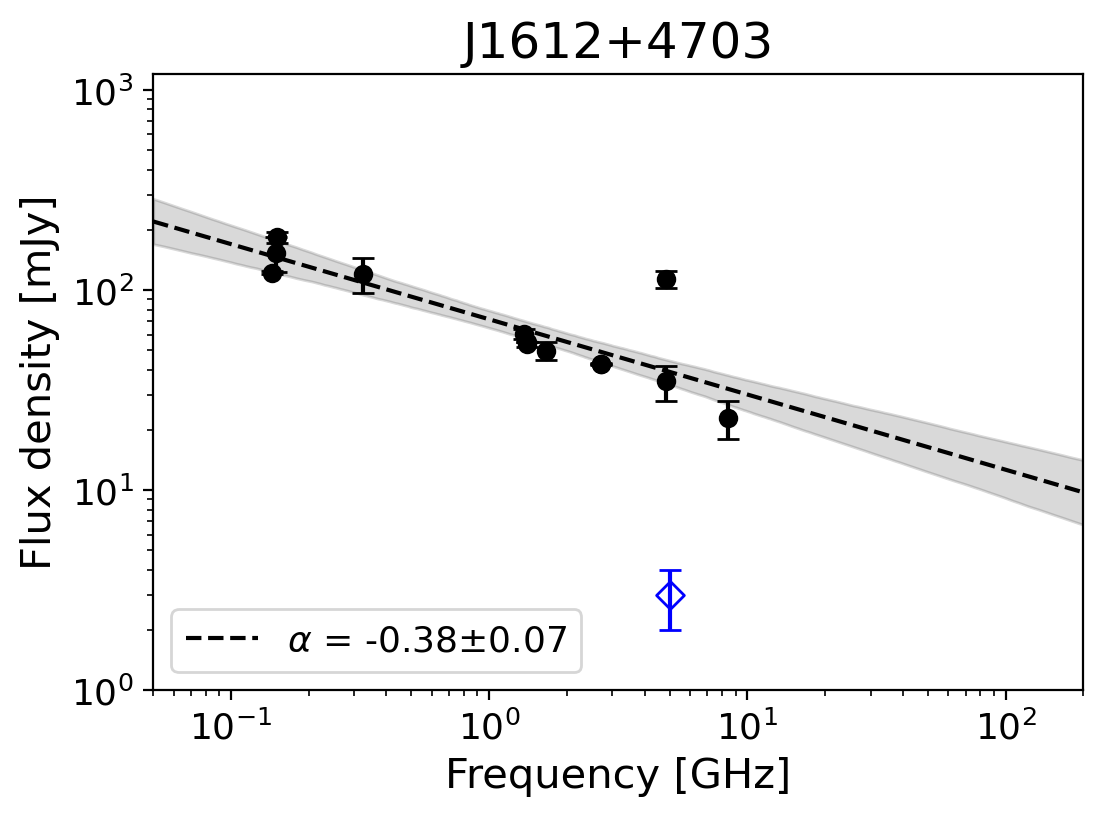}
            \includegraphics[width=0.33\textwidth]{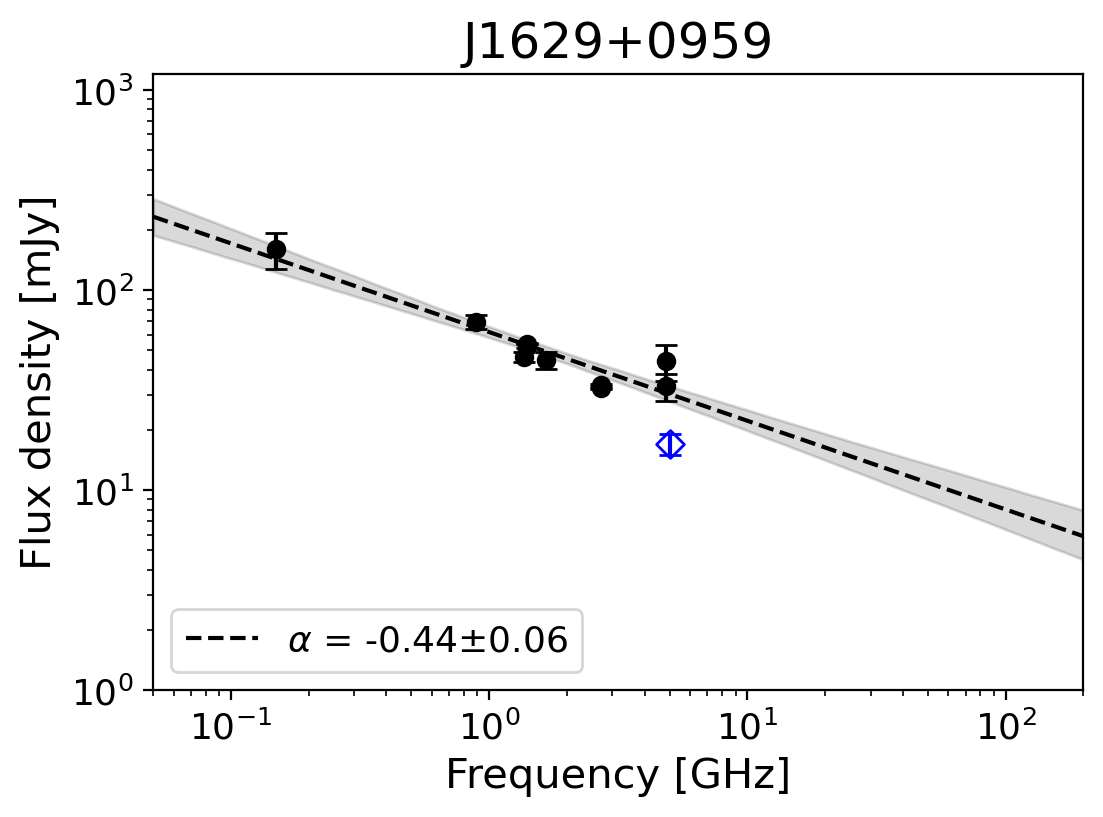}
            \includegraphics[width=0.33\textwidth]{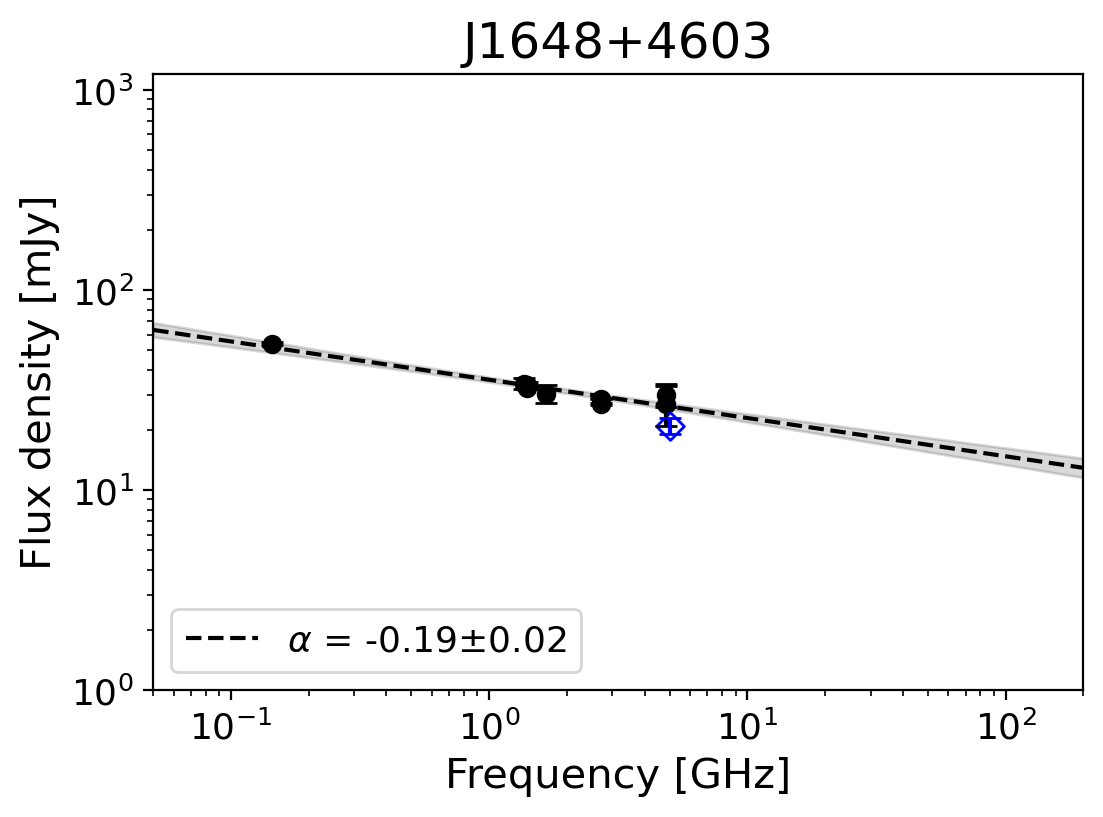}
            }
\gridline{  \includegraphics[width=0.33\textwidth]{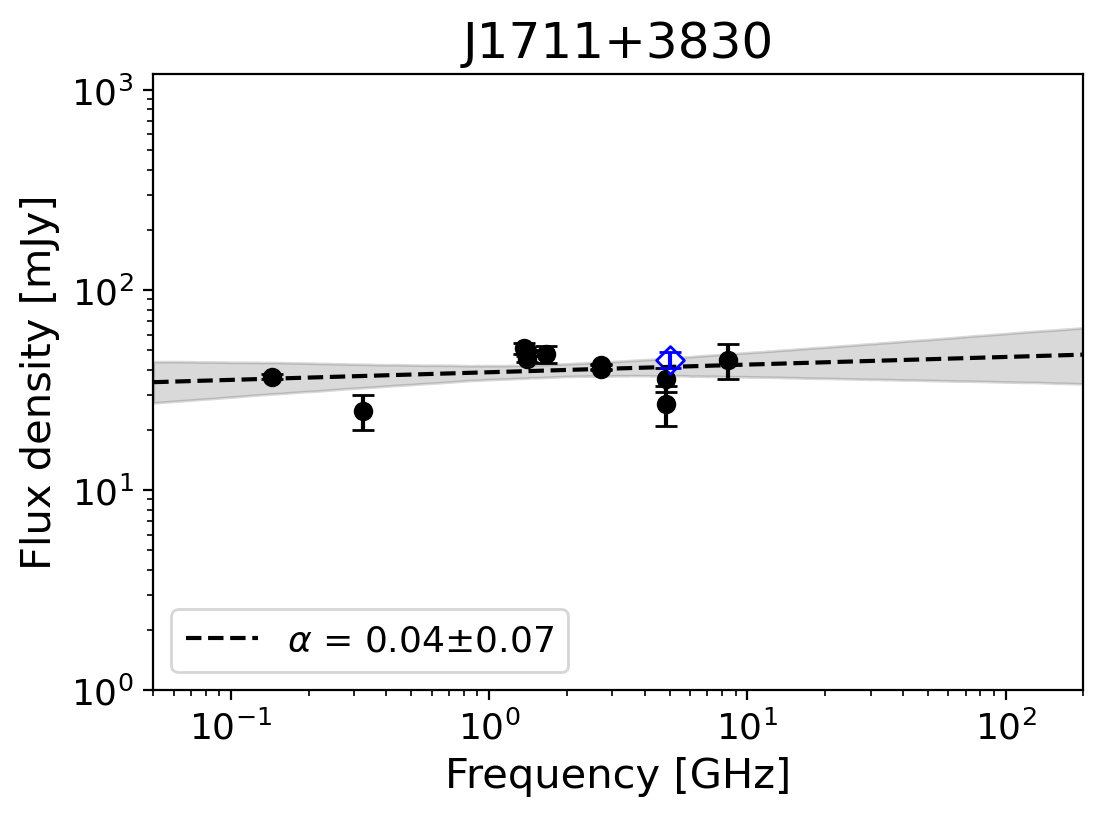}
            \includegraphics[width=0.33\textwidth]{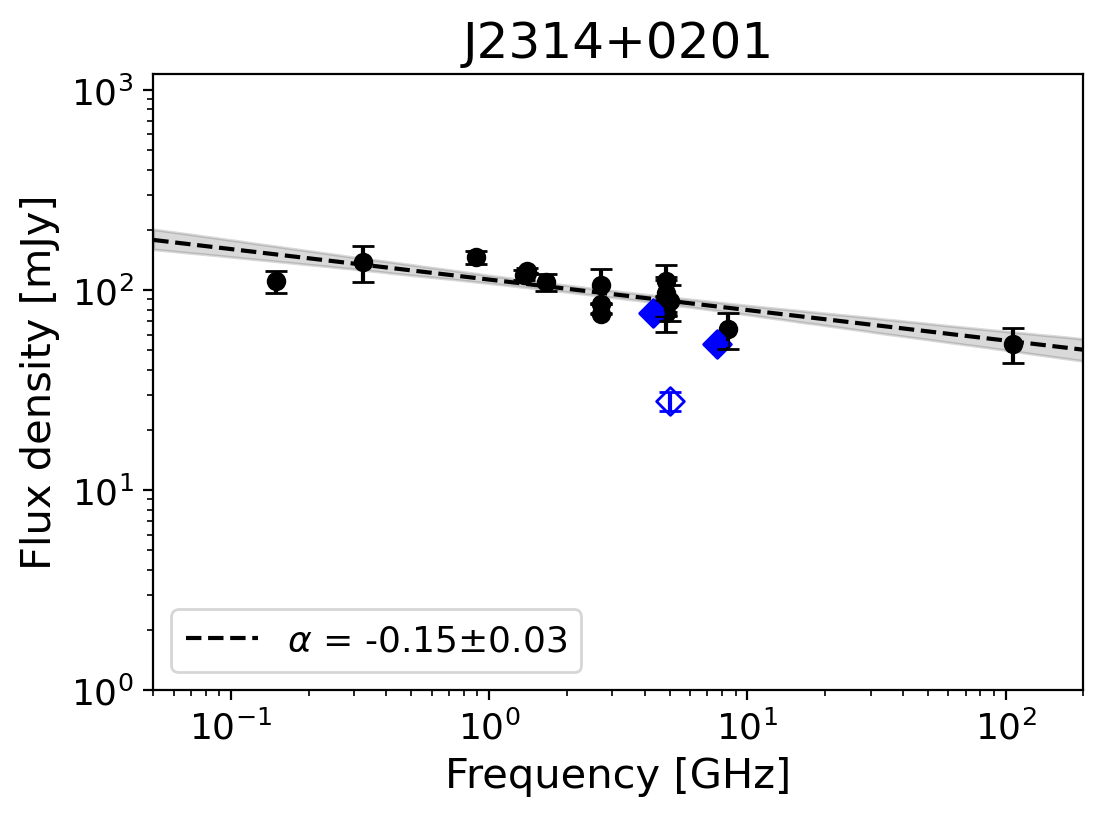}
            \includegraphics[width=0.33\textwidth]{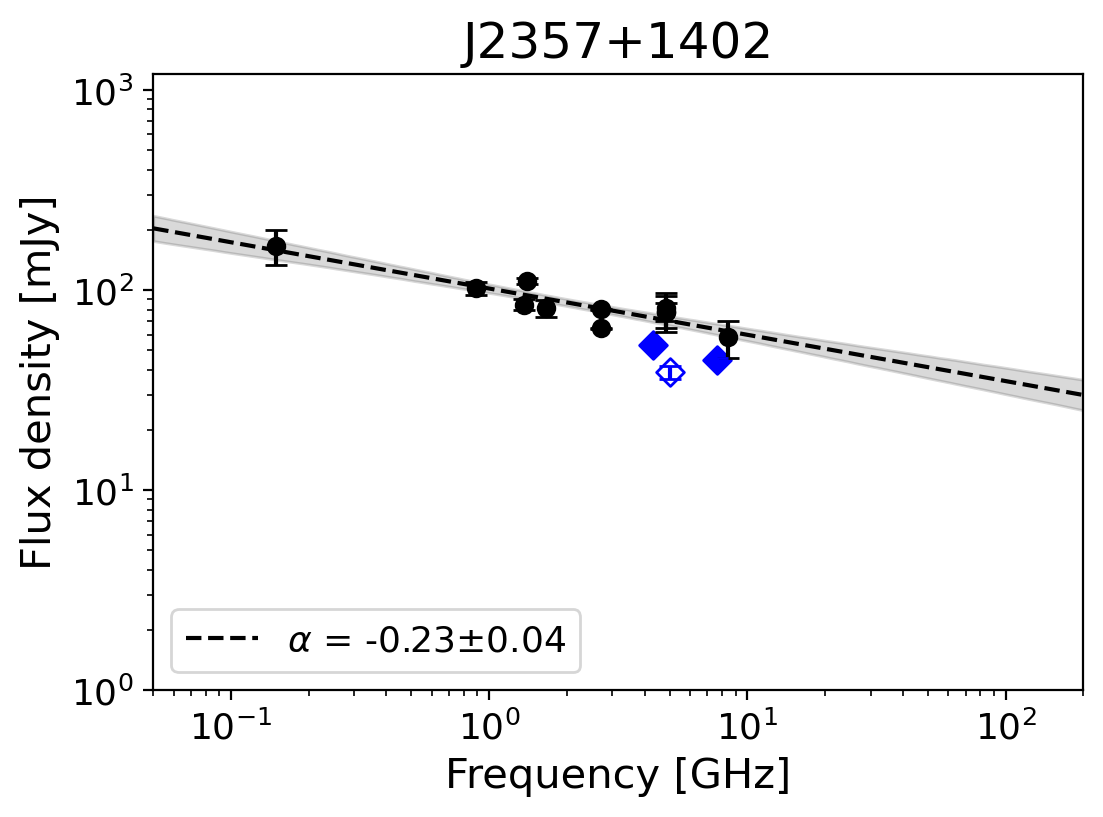}
            }
    \caption{$Continued$}
\end{figure*}

There are also other independent VLBI measurements available for some sources, as mentioned in Section~\ref{sample}. Eleven of our targets have previous VLBI observations by either the VLBA or the EVN (see Table~\ref{tab:sed}). Two simultaneous spectral points would be enough to build the compact continuum spectrum if a power-law shape is assumed. However, these spectral data were measured at different epochs and by different network compositions, therefore it is not feasible to fit them together (including our 5-GHz data) to determine the compact spectral indices at mas scales. In addition, some of the sources appear highly variable. In the future, revisiting these sources with simultaneous dual-frequency VLBI observations could reveal the compact spectral shape. In our previous study \citep{2022ApJS..260...49K}, we determined two-point spectral indices for J0918$+$0637 ($\alpha^5_{1.7} = -0.25 \pm 0.11$), J1325$+$1123 ($\alpha^5_{1.7} = -0.61 \pm 0.09$), and J1412$+$0624 ($\alpha^5_{1.7} = -0.18 \pm 0.10$). Additionally \citet{2010AA...520A.113B} reported $\alpha^{8.4}_{2.3} = -1.08$ for J0121$+$0346. 

\subsection{Flux Density Variability} \label{variability}
Flux density variability is a strong indicator of AGN with powerful relativistic jets, also at high redshifts \citep[e.g.][]{2024Galax..12...25S}. Variability can be detected by observing the target at the same frequencies and angular resolution at different epochs. As there is no long-term monitoring on going for the high-$z$ quasars in our sample, we have to rely on sparse flux density measurements made in various radio sky surveys. Another way to infer variability is to compare high-resolution interferometric data with low-resolution ones, including total flux density measurements. For example, \citet{2016MNRAS.463.3260C} showed that the $S_\mathrm{FIRST}/S_\mathrm{NVSS}$ ratio can be used as an indicator of variability. FIRST and NVSS surveys were performed at $1.4$~GHz but at different epochs, and in different VLA configurations. The B configuration used for FIRST provided $\sim 5\arcsec$ angular resolution, while the most compact D configuration used for NVSS led to $\sim 45\arcsec$ resolution. As mentioned in Section~\ref{results}, if $S_\mathrm{FIRST}/S_\mathrm{NVSS} \geq 1.1$, the source can be considered as variable. There are three such sources in our list: J0122$+$0309, J1535$+$0254, and J1711$+$3830 (Table~\ref{tab:phyparams}). 
The FIRST flux densities can also be compared to the $1.37$-GHz RACS-mid \citep{2024PASA...41....3D} flux densities. The resolution in the RACS-mid survey is $\sim 25^{\prime\prime}$, and it is between the FIRST and the NVSS resolutions. The ratio $S_\mathrm{FIRST}/S_\mathrm{RACS}$ behaves just like $S_\mathrm{FIRST}/S_\mathrm{NVSS}$, as discussed before. Three sources found to be variable by this criterion: J0122$+$0310, J1535$+$0254, and J1629$+$0959. 
The cases of $S_\mathrm{FIRST}/S_\mathrm{NVSS} \ll 0.9$ and/or $S_\mathrm{FIRST}/S_\mathrm{RACS}  \ll 0.9$ can be caused either by variability or the presence of an extended emission region which remains unresolved on NVSS or RACS, but becomes resolved out in FIRST. It is certainly possible that a source is variable, even with $S_\mathrm{FIRST}/S_\mathrm{NVSS} \approx 1$ and/or $S_\mathrm{FIRST}/S_\mathrm{RACS} \approx 1$, just being unresolved with nearly equal flux densities at the two epochs when the measurements were made. Without flux density monitoring, no additional conclusion can be drawn on the variability on these sources.

The flux density ratio $S_\mathrm{VLBI}/S_\mathrm{total}$ can also be used as an indicator of variability. Due to the lack of proper VLBI spectral information, only the $5$-GHz compactness can be estimated. Such estimations were carried out before at $1.4$-GHz \citep[e.g.][]{2016MNRAS.463.3260C,2022ApJS..260...49K,2024AA...690A.321K}. We consider the $S_\mathrm{VLBI}/S_\mathrm{total} \geq 1.1$ ratio that indicates significant variability. Two sources satisfy this definition in our sample (J0257$+$4338 and J1711$+$3830, see Table~\ref{tab:phyparams}). 

We consider flux density variability significant for a given source if either of the ratios described above exceeds unity by more than $10\%$. By these criteria, we find J0122$+$0309, J0257$+$4338, J1535$+$0254, J1629$+$0959, and J1711$+$3830 to be variable. Among the overlapping sample sources in \citet{2022ApJS..260...49K}, J0918$+$0637 turned out to be variable, while it could not be determined with certainty for J1325$+$1123 and J1412$+$0624.

\subsection{Astrometric Positions} \label{gaia}
The high-accuracy optical position can pinpoint the location of the central region of the quasar. It was shown \citep{2017A&A...598L...1K,2019ApJ...871..143P} that the optical position is generally linked to the thermal emission originating from the accretion disk. In some cases, this position can also be related to the synchrotron emission of bright optical jets \citep[e.g.][]{2024A&A...684A.202L}. In turn, the radio position measured by VLBI corresponds to the brightest compact region of the jet. This region is usually the synchrotron self-absorbed base of the jet called the radio core \citep[e.g.][]{1997AJ....114.2284F}, where the optical depth $\tau_{\nu}$ = 1 at the given $\nu$ frequency. The brightest feature might also be associated with a compact hotspot in a shock front between the jet and the surrounding medium, typically further away from the central engine. 

The accuracy of the astrometric VLBI measurements is in the order of $(0.1-10)$~mas \citep{2025ApJS..276...38P}. The only optical instrument that is capable of achiving this level of accuracy is the \textit{Gaia} astrometric space telescope \citep{2016A&A...595A...1G, 2023A&A...674A...1G}. The most recent \textit{Gaia} DR3 catalogue contains the most accurate optical positions available nowadays for AGN, including many distant quasars, too. Comparing \textit{Gaia} and VLBI positions could reveal further information about the nature of the radio source. Finding \textit{Gaia}--VLBI positional mismatch exceeding the typical values in high-redshift quasars could help constrain the source classification and the nature of compact radio emission. Typically, up to a few mas offset between radio and \textit{Gaia} positions can occur and its position angle is found to statistically coincide with the VLBI jet direction \citep{2019MNRAS.482.3023P}. A significant offset would suggest an extended structure with a quasar jet misaligned with respect to the line of sight. 

The \textit{Gaia} DR3 optical coordinates, whenever available, are shown with crosshairs in the EVN images in Figure~\ref{fig:imgs}. Twelve of our targets are found in the \textit{Gaia} catalogue \citep{2023A&A...674A...1G}. The uncertainties of the optical positions ($\sigma_\mathrm{Gaia}$) are all within 3~mas. The astrometric excess noise is significant in the cases of J1535$+$0254 and J1612$+$4703 only, where we added it to the formal position error in quadrature. We consider the \textit{Gaia}--VLBI offset significant if it exceeds $3\sigma_{\rm pos}$, where $\sigma_{\rm pos} = \sqrt{\sigma_\mathrm{VLBI}^2+\sigma_\mathrm{Gaia}^2}$. 
The VLBI coordinate errors ($\sigma_\mathrm{VLBI}$) are given in Table~\ref{tab:imgparams} and calculated by adding in quadrature three main error contributions. The first one is the formal International Celestial Reference Frame positional uncertainty of the phase-refrence calibrator source \citep[ICRF,][]{2020A&A...644A.159C}. The second contribution is determined by the angular resolution of the interferometer network and the image signal-to-noise ratio \citep[see][]{1990AJ.....99.1663L}. The third effect to consider is the target--calibrator angular separation. Since many of our calibrators were located $> 2^\circ$ away from the targets (Table~\ref{tab:targets}), this is the dominant factor for VLBI relative astrometric errors in our case. To estimate this in the most conservative way, we considered the results of \citet{2004ApJ...604..339C} who determined the astrometric errors as a function of the calibrator angular separation by comparing measured and modeled pulsar positions. In addition, VLBI astrometric errors can be caused by other observational effects. By doing phase referencing, our target positions are linked to the ICRF. However, the ICRF positions of the calibrators are determined at X-band ($8$~GHz), while our observations were carried out at C-band ($5$~GHz). Thus we have to consider the effect of frequency-dependent core shift, which can reach $\sim 0.5$~mas \citep[e.g.][]{2011A&A...532A..38S} between these frequencies. Additionally, there could be up to $\sim 0.2$~mas offset between the phase-referenced and group-delay positions along the jet direction, caused by opacity effects on the emission at the jet base \citep{2009A&A...505L...1P}. We also added these contributions in quadrature to $\sigma_\mathrm{VLBI}$.

We found that the \textit{Gaia}--VLBI offset is significant ($> 3\sigma_\mathrm{pos}$) in two sources, J0918$+$0637, and especially J2314$+$0201 (Figure~\ref{fig:imgs}). The remaining sources have marginal, $(1-3) \times \sigma_\mathrm{pos}$, or no offset ($<1 \sigma_\mathrm{pos}$). Their detected radio features can be related to the inner jet (core), located close to the central engine. It is also possible that an offset indicates a dual AGN system where only one of the components is a radio emitter (detected by VLBI), while the other is much brighter in the optical (detected by \textit{Gaia}). A similar case was found by e.g. \citet{2011ApJ...736..125K} but with arcsec-scale separation. The absence of detected double-peaked narrow or blended broad optical emission lines that would suggest the presence of a second SMBH \citep[e.g.][]{2011ApJ...735...48S,2011ApJ...740L..44F} in our sources is against this scenario. Note that jet precession can be a marker of smaller-separation (not resolvable) binary\footnote{Following the convention used by \citet{2019NewAR..8601525D}, we call a gravitationally-bound sub-pc separation systems as binary, and larger-separation systems as dual AGN.} SMBHs \citep[e.g.][]{2014MNRAS.445.1370K,2023MNRAS.526.4698K,2018MNRAS.478.3199B}, but without prominent jets in our high-redshift sources (Figure~\ref{fig:imgs}), it is not possible to study.

\subsection{Distinguishing Between Blazars and Misaligned Sources: VLBI versus X-rays} \label{classy}

The most common approaches to classify blazars are through multi-band SED-fitting \citep[e.g.][]{2015MNRAS.446.2483S,2019AA...627A..72G,2021AA...655A..95S,2022AA...663A.147S}, and low-resolution radio and/or high-resolution VLBI observations \citep[e.g.][]{2016MNRAS.463.3260C,2017MNRAS.467..950C,2022ApJS..260...49K}. 
Alternatively, the X-ray-to-optical flux ratio has also been used to effictively classify oriented sources \citep{2019MNRAS.489.2732I}. Like in the SED modelling, this method is based to the fact that, in sources oriented close to the line of sight, the X-ray emission is particularly strong due to Doppler boosting. For this reason, these two methods give very similar and consistent results \citep[see discussion in][]{2024AA...684A..98C}.
In many cases, relying on a single classification method might not be sufficient, as the as the different method could disagree in some cases \citep[e.g.][]{2016MNRAS.463.3260C,2017MNRAS.467..950C,2018evn..confE..31G,2021AN....342.1092G,2022ApJS..260...49K}. High-resolution imaging with VLBI revealed that some sources have radio cores with low brightness temperatures, and extended structures with steep radio spectrum on scales of $\sim 10-100$~mas. These properties are at odds with the blazar classification. Source samples with limited size and with incorrect classifications may lead to wrong conclusions about the space density of high-$z$ radio-loud quasars. 
For this reason, it is extremely important to compare the classifications obtained with different methods. As mentioned before, SED modelling and the method based on the X-ray-to-optical flux ratios are not completely independent and, indeed, they give similar results. Conversely, the VLBI classification based on the brightness temperature is a completely different technique so the consistency with the other two methods is not necessarily granted. For this reason, in the following we want to compare the VLBI classifications obtained for the 18 CLASS quasars with those based on the X-ray-to-optical flux ratio that are available for the totality of the sample and reported in \citet{2019MNRAS.489.2732I}. Following \citet{2019MNRAS.489.2732I}, we introduce the two-point spectral index $\tilde{\alpha}_{\rm ox} = -0.3026\log(L_{\mathrm{10keV}}/L_{\mathrm{2500\AA}})$. The errors of $\tilde{\alpha}_{\rm ox}$ introduced by \citet{2019MNRAS.489.2732I} were recalculated because they did not take into account the uncertainty caused by the possibly different optical spectral slopes \citep[see more details in Appendix C of][]{2024AA...684A..98C}. For our source sample, the $\tilde{\alpha}_{\rm ox}$ values and the new error bars are listed in Table~\ref{tab:phyparams}. Based on the criterion $\tilde{\alpha}_{\rm ox} < 1.355$, the sources are considered blazars \citep{2019MNRAS.489.2732I}. 

In the context of our sample, under blazars we mean FSRQs. In general, blazars are divided into two main sub-classes, FSRQs that show prominent optical and ultraviolet emission lines in their spectra, and BL Lac objects with weak or no emission lines \citep[e.g.][]{1999ASPC..159..351F,2015Ap&SS.357...75M}. In the absence of emission lines, the redshifts of BL Lacs are often hard to determine, thus most of the known BL Lacs are at $z<2$ \citep{2008AJ....135.2453P}. Misaligned sources are the unbeamed gigahertz-peaked spectrum (GPS), compact steep-spectrum (CSS) sources \citep{2021A&ARv..29....3O}, and high-frequency peakers \citep[HFP, e.g.][]{2000A&A...363..887D}.

\begin{figure}
    \centering
    \includegraphics[width=0.6\linewidth]{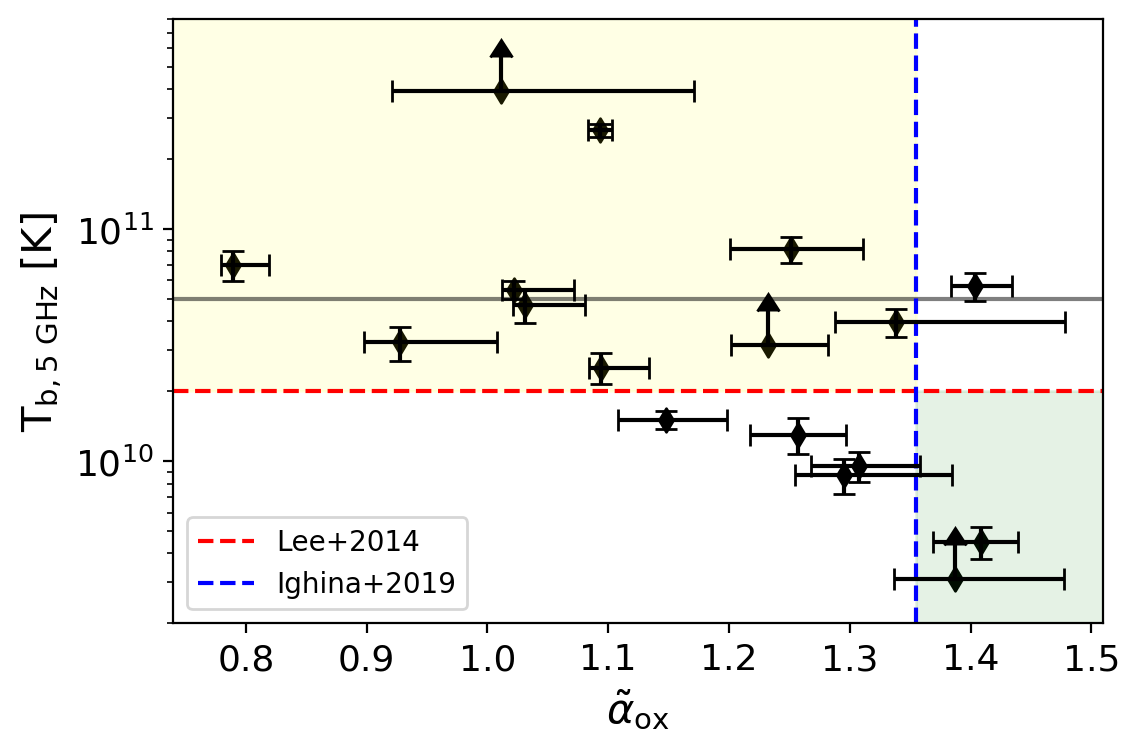}
    \caption{The $\tilde{\alpha}_{\rm ox}-T_{\rm b,5~GHz}$ diagnostic plot} for the $17$ high-redshift quasars studied here. The $\tilde{\alpha}_{\rm ox}$ values are taken from \citet{2019MNRAS.489.2732I} and $\tilde{\alpha}_{\rm ox} = 1.355$ serves as the threshold (vertical blue dashed line) to separate the blazar$-$non-blazar populations. The horizontal red dashed line represents the intrinsic brightness temperature at $5$~GHz ($T_{\rm b, int} = 2 \times 10^{10}$~K) taken from \citet{2014JKAS...47..303L}. The grey line is the theoretical equipartition brightness temperature \cite[$T_{\rm b, eq} \approx 5 \times 10^{10}$~K,][]{1994ApJ...426...51R}. The sources in the yellow shaded area are considered as blazars, since they fulfil both criteria. The green shaded area reprensents the location of misaligned sources. The blazar classification of sources falling in any of the two white regions is uncertain. The proposed source classifications are summarized in Table~\ref{tab:summary}.
    \label{fig:aox-tb}
\end{figure}

Blazars show Doppler-boosted emission on pc (VLBI) scales and the Doppler boosting influences the observed brightness temperature. If both $T_{\rm b}$ and $\tilde{\alpha}_{\rm ox}$ are proxies of orientation, we expect that the two quantities are correlated. Indeed, as shown in Figure~\ref{fig:aox-tb}, the values of $\tilde{\alpha}_{\rm ox}$ and $T_{\rm b,5~GHz}$ measured for the $17~z>4$ targets studied here show hint of correlation: objects with the highest values of $T_{\rm b,5~GHz}$ have on average flatter $\tilde{\alpha}_{\rm ox}$ (mostly $<1.35$), i.e. they are more X-ray loud, compared to objects with lower brightness temperature. The Pearson correlation coefficient for the data points is $-0.36$, which suggests a not very significant correlation. Filling this parameter space would be essential to better constrain the relation between $\tilde{\alpha}_{\rm ox}$ and $T_{\rm b,5~GHz}$. Once again, we note that source variability might severely affect the location of individual points, but not the general trend. The thresholds applied for distinguishing blazars from misaligned sources (i.e. $\tilde{\alpha}_{\rm ox} = 1.355$ and $T_{\rm b, int} = 2 \times 10^{10}$~K) are also indicated. The parameter space where blazars reside based on both criteria is shown with the yellow shaded area. On the other hand, misaligned sources can be found in the green shaded area. For sources falling in other sections of the parameter space, the classification is inconclusive since the two criteria lead to contradictory results. 

If we consider $T_{\rm b} = 2 \times 10^{10}$~K as the intrinsic brightness temperature, among the 14 sources that were previously classified as blazars by \citet{2019MNRAS.489.2732I}, 10 show Doppler-boosted radio emission and 4 appear non-boosted based our $5$-GHz VLBI results. On the other hand, there is only one object that appears Doppler-boosted with VLBI but is not considered as blazar based on the $\tilde{\alpha}_{\rm ox}$ criterion (Table~\ref{tab:summary}). The blazar/non-blazar classification for the 5 sources outside the shaded box in Figure~\ref{fig:aox-tb} should be considered as uncertain. However, these differences do not necessarily imply a discrepancy between the two classification methods, rather shed light on their inherent uncertainties. Since blazars are known to be highly variable in time, and spectral indices are usually calculated from non-simultaneous measurements, the values should be considered indicative only. Also, measured brightness temperatures can depend on observing frequency and vary with time. Finally, $T_{\rm b}$ and $\tilde{\alpha}_{\rm ox}$ measurements are not contemporaneous either.

Table~\ref{tab:summary} summarizes the classification of the $17$ sources, taking optical, X-ray, low- and high-resolution radio data into account. The discussion of individual sources is presented in Section~\ref{notes}. We found two sources of misaligned type, while ten are highly possibly blazars. The remaining five of them have uncertain classification (Table~\ref{tab:summary}). Among them, J0918$+$0637 and J1325$+$1123 were previously classified as FSRQs \citep{2022ApJS..260...49K,2019MNRAS.489.2732I}, but based on our current $5$-GHz VLBI data, they do not show Doppler boosting. This could be the result of high variability, especially for J0918$+$0637, which makes it more likely to be a blazar. The case of J1711$+$3830, which is not Doppler-boosted at $5$-GHz, is similar, since this source is highly variable and has an inverted radio spectrum. 

\begin{deluxetable*}{lcccccc}
\tablecaption{Classification of the target sample.}
\tablewidth{0pt}
\tablehead{
\colhead{Source ID} & \colhead{Total Flux Density} &  \colhead{Variable} &  \colhead{Optical--Radio} &  \colhead{Doppler-boosted} & \colhead{X-ray Class} &  \colhead{Proposed Class}  \\
 & \colhead{Spectrum} & &  \colhead{Offset} & &
}
\decimalcolnumbers
\startdata
J0031$+$1507 & Peaked & Possibly & Marginal & Yes & Non-blazar & Uncertain  \\
J0121$+$0346 & Steep (Flat) & Possibly & Marginal & No & Non-blazar & Misaligned   \\ 
J0122$+$0309 & Inverted (Flat) & Yes & -- & Yes & Blazar & FSRQ  \\
J0257$+$4338 & Peaked & Yes & Marginal & Yes & Blazar & FSRQ  \\ 
J0835$+$1825 & Flat & Possibly & -- & Yes & Blazar & FSRQ \\ 
J0918$+$0637 & Peaked & Yes & Yes & No & Blazar & Uncertain  \\ 
J1021$+$2209 & Flat (Steep) & Possibly & -- & Yes & Blazar & FSRQ  \\
J1325$+$1123 & Peaked & Possibly & No & No & Blazar & Uncertain  \\ 
J1348$+$1935 & Flat & Possibly & Marginal & Yes & Blazar & FSRQ  \\
J1412$+$0624 & Flat & Possibly & Marginal & No & Blazar & Uncertain  \\
J1535$+$0254 & Flat & Yes & No & Yes & Blazar & FSRQ   \\
J1612$+$4703 & Flat & Possibly & No & Possibly & Non-blazar & Misaligned  \\
J1629$+$0959 & Flat & Yes & -- & Yes & Blazar & FSRQ  \\
J1648$+$4603 & Flat & Possibly & -- & Yes & Blazar & FSRQ  \\
J1711$+$3830 & Inverted (Flat) & Yes & Marginal & No & Blazar & Uncertain  \\
J2314$+$0201 & Flat & Possibly & Yes & Yes & Blazar & FSRQ  \\ 
J2357$+$1402 & Flat & Possibly & No & Yes & Blazar & FSRQ  \\
\enddata
\tablecomments{Col.~1 -- source designation; Col.~2 -- spectral classification based on the shape of the total flux density spectrum (in bracets we give the spectral class that is also possible within the uncertainties); Col.~3 -- source variability; Col.~4 -- radio–optical positional offset; Col.~5 -- Doppler-boosting (“Possibly” means T$_{\rm b,5~GHz}$ is lower limit); Col.~6 -- Classification based on the $\tilde{\alpha}_{\rm ox}$ parameter; Col.~7 -- the proposed classification of the source.}
\label{tab:summary}
\end{deluxetable*}


\section{Summary and Conclusions} \label{conclusion}
In this paper, we investigated 17 high-redshift blazar candidates located at $4 \le z \le 5.36$. The main goal of this study was to confirm the blazar nature of these quasars using high-resolution VLBI imaging observations, and to compare the findings to previous classifications by \citet{2019MNRAS.489.2732I} which were based on optical and X-ray data. The targets were observed with the EVN at $5$~GHz in 2017 and 2018. There are 7 of the 18 targets that were not yet observed with VLBI at any frequency, and 2 other targets were observed with VLBI at $5$~GHz for the first time. The list of targets included an additional source, J1517$+$3753, which later turned out to have low redshift ($z = 0.56$). It was also the only source not detected with the EVN. 

In addition to the new high-resolution VLBI imaging data, we collected complementary low-resolution radio flux density measurements from the literature, to draw a more complete picture about the source properties, such as the origin of the radio emission, flux-density variability, and spectral indices. The VLBI data were used to estimate the core brightness temperatures and assess the presence of Doppler boosting, a characteristic property of blazar-type sources with low-inclination relativistic jets. Precise \textit{Gaia} DR3 optical astrometric positions were available for as many as 12 of the targets, assisting the constrains on the source classification.

After comparing all the collected data, we propose a classification as blazar or misaligned radioquasars. The classifications based on the radio data are generally in good agreement with the X-ray classes proposed by \citet{2019MNRAS.489.2732I}. 
Two objects are classified as misaligned sources both from the radio and the X-ray point of view, while ten of them belong to a type of blazars, FSRQs, with core--jet or compact core morphologies. Finally, there are five objects for which the two classification methods do not agree. These objects could be either at the border of the blazar classification, as discussed in \citet{2024AA...684A..98C}, or variability could be affecting one of the two classifications. Although some of our uncertain sources show strong characteristics of FSRQs, additional data would be needed to classify them with absolute certainty. This study suggest that the observation of the uncertain sources should be repeated or complemented with further data since variability might affect the classification from blazar to non-blazar (and vice versa) according to the level of activity of the core.

Studies in the past have reached the conclusion that the ratio of blazars to misaligned sources is $\sim 0.5$ in high-redshift blazar candidate samples \citep[e.g.][]{2016MNRAS.463.3260C,2017MNRAS.467..950C,2022ApJS..260...49K}. We found that at least  60\% of our candidates are confirmed blazars but, as already discussed, variability may have affected the classification of a number of other possible blazars. Only the combination of different methods is able to properly recover all the oriented sources present in the sample. As this sample is observed at $5$~GHz only, follow-up dual-frequency VLBI observations would help to better constrain their classification. Our VLBI imaging revealed a few sources (J1021$+$2209, J1348$+$1935, J1612$+$4703, and J2357$+$1402) that can be subject to future jet proper motion studies. The blazar J2314$+$0201 has the largest positional offset between its \textit{Gaia} optical and VLBI radio positions, which might indicate a dual AGN system. Additional follow-up work, e.g. using optical spectroscopic studies to look for double-peaked emission lines, or deeper radio imaging to reveal faint emission from the putative radio-quiet component in the \textit{Gaia} position could help investigating this intriguing possibility.

The use of multi-band (radio, optical, and X-ray) data, and the combination of different classification methods (i.e. $T_{\rm b}$, astrometry, and $\tilde{\alpha}_{\rm ox}$) proved to be generally feasible and effective. However, there are still cases where individual sources cannot be securely classified, as these objects may be at the border of the classes. Acquiring multi-band data for large samples is a challenging task, especially because quasi-simultaneous measurements would be preferred, but the combined method presented here can play an important role in the classification process in the future. 

\begin{longrotatetable}
\begin{deluxetable*}{rp{1.6cm}|rrrrrrrrrr}
\tablecaption{Archival flux densities from low-resolution (top) and VLBI (bottom) radio measurements.}
\tablewidth{0pt}
\tabletypesize{\scriptsize}
\tablehead{\colhead{$\nu$ (GHz)} & \colhead{Instrument} & \colhead{J0031$+$1507} & \colhead{J0121$+$0347} & \colhead{J0122$+$0309} & \colhead{J0257$+$4338} &  \colhead{J0835$+$1825} & \colhead{J0918$+$0637} & \colhead{J1021$+$2209} & \colhead{J1325$+$1123} & \colhead{J1348$+$1935} & \colhead{Reference}}
\decimalcolnumbers
\startdata
0.074 & VLSS &  &  &  &  &  &  & 890 (170) &  &  & \citep{2007AJ....134.1245C} \\
0.15 & TGSS &  &  & 85 (17) & 79 (16) & 84 (17) & 19 (5) &  & 29 (6) & 52 (10) & \citep{2017AA...598A..78I} \\
0.2 & GLEAM &  &  & 114 (23) &  & 83 (20) &  &  &  &  & \citep{2017MNRAS.464.1146H} \\
0.325 & WN &  &  &  & 169 (34) &  &  &  &  &  & \citep{1997AAS..124..259R} \\
0.408 & B3 &  &  &  & 130 (26) &  &  &  &  &  & \citep{1985AAS...59..255F} \\
0.888 & RACS & 33.6 (2.9) & 72.9 (7.8) & 98.9 (7.8) & & 63.2 (4.9) & 40.6 (3.4) & 197.9 (14.5) & 81.9 (6.3) & 67.5 (5.3) & \citep{2020PASA...37...48M} \\
1.37 & RACS & 49.9 (3.0) & 79.2 (4.8) & 78.1 (4.8) & 242.0 (14.0) & 51.2 (31.1) & 43.5 (2.7) & 157.7 (9.6) & 75.8 (4.6) & 58.3 (3.6) & \citep{2024PASA...41....3D} \\
1.40 & FIRST & 43.4 (0.1) & 76.6 (0.2) & 109.0 (0.1) &  & 52.4 (0.1) & 26.5 (0.1) & 139.3 (0.45) & 71.1 (0.1) & 49.9 (0.2) & \citep{1997ApJ...475..479W} \\
1.40 & NVSS & 41.9 (1.3) & 78.1 (2.4) & 98.4 (3.0) & 148.1 (5.3) & 53.2 (1.6) & 30.9 (1.0) & 153.0 (4.6) & 81.4 (2.5) & 51.6 (1.6) & \citep{1998AJ....115.1693C} \\
1.66 & RACS & 57.4 (5.7) & 79.2 (4.8) & 78.2 (7.3) & 209.0 (21.0) & 51.2 (31.1) & 36.7 (3.7) & 152.0 (15.0) & 65.2 (6.5) & 56.9 (5.8) & \citep{2025PASA...42...38D} \\
2.70 &  VLASS 1 & 68.5 (0.2) & 49.2 (0.4) & 65.8 (0.3) & 315.7 (1.1) & 54.3 (0.2) & 41.2 (0.2) & 122.9 (0.5) & 51.0 (0.3) & 42.4 (0.3) & \citep{2020RNAAS...4..175G} \\
2.70 & VLASS 2 & 74.6 (0.3) & 54.9 (0.3) & 58.9 (0.3) & 333.8 (1.0) & 47.5 (0.2) & 33.5 (0.3) & 118.1 (0.3) & 57.3 (0.3) & 42.9 (0.4) & \\ 
4.85 & GB6 & 93 (9) & 51 (7) & 96 (10) & 159 (14) & 40 (5) & 36 (6) & 108 (10) & 62 (8) & 38 (5) & \citep{1996ApJS..103..427G} \\
4.85 & MITG & 70 (15) &  & 106 (21) &  &  &  & 77) &  &  & \citep{1991ApJS...75..801G} \\
4.85 & BWE & 108 (22) & 46 (9) & 88 (18) & 169 (34) & 34 (7) &  &  &  & 43 (9) & \citep{1991ApJS...75....1B} \\
4.85 & PMN &  &  & 118 (24) &  &  &  &  &  &  & \citep{1996ApJS..103..145W} \\
4.85 & 87GB & 109 (22) &  &  & 173 (35) & 50 (10) &  & 120 (17) & 72 (12) & 63 (13) & \citep{1991ApJS...75.1011G} \\
4.85 & Effelsberg &  &  &  &  &  & 39 &  & 48 &  & \citep{2004MNRAS.348..857H} \\
4.9 & MASIV &  &  & 110 &  &  &  &  &  &  & \citep{2011AJ....142..108K} \\
8.40 & CLASS & 87 (17) & 21 & 122 (24) & 133 & 24 (5) & 26 (5) & 81 &  &  & \citep{2003MNRAS.341....1M} \\ 
8.4 & MASIV &  &  & 110 &  &  &  &  &  &  & \citep{2011AJ....142..108K} \\
8.4 & VLA &  &  &  & 128 &  &  &  &  &  & \citep{1992MNRAS.254..655P} \\
10.6 & Effelsberg &  &  &  &  &  & 23 &  & 28 &  & \citep{2004MNRAS.348..857H} \\
2.3 & VLBA &  & 65$^{\dagger}$ &  & 160$^{\dagger}$ &  &  &  &  &  & \citep{2025ApJS..276...38P} \\
4.3 &  & 93$^{\dagger}$ &  & 83$^{\dagger}$ &  &  &  & 97$^{\dagger}$ &  &  &  \\
7.6 &  & 72$^{\dagger}$ &  & 66$^{\dagger}$ &  &  &  & 84$^{\dagger}$ &  &  &  \\
8.4 &  &  & 23$^{\dagger}$ &  & 154$^{\dagger}$ &  &  &  &  &  &  \\
8.6 &  &  &  &  & 129$^{\dagger}$ &  &  &  &  &  &  \\
8.7 &  &  &  &  & 311$^{\dagger}$ &  &  &  &  &  &  \\
1.7 & EVN &  &  &  &  &  & 38.5 (3.9) &  & 62.7 (3.9) &  & \citep{2022ApJS..260...49K} \\
2.3 &  &  & 61 (13) &  &  &  &  &  &  &  & \citep{2010AA...520A.113B} \\
5 &  &  &  &  &  &  & 35.5 (2.2) &  & 31.2 (1.9) &  & \citep{2022ApJS..260...49K} \\
8.4 &  &  & 15 (3) &  &  &  &  &  &  &  & \citep{2010AA...520A.113B} \\
\enddata
\tablecomments{$^{\dagger}$ median flux density. Col.~1 -- frequency of the flux density measurement; Col.~2 -- instrument or survey name; Cols.~3--11 -- flux density measurements of the given source in mJy (errors are given in parentheses where available); Col.~12 -- reference of the flux density measurement.}
\label{tab:sed}
\end{deluxetable*}
\end{longrotatetable}

\begin{longrotatetable}
\begin{deluxetable*}{rp{1.6cm}|rrrrrrrrrr}
\tablenum{6}
\tablecaption{\textit{Continued}}
\tablewidth{0pt}
\tabletypesize{\scriptsize}
\tablehead{\colhead{$\nu$ (GHz)} & \colhead{Instrument} & \colhead{J1412$+$0624} & \colhead{J1535$+$0254} & \colhead{J1612$+$4703} & \colhead{J1517$+$3753} & \colhead{J1629$+$0959} & \colhead{J1648$+$4603} & \colhead{J1711$+$3830} & \colhead{J2314$+$0201} & \colhead{J2357$+$1402} & \colhead{Reference}
}
\decimalcolnumbers
\startdata
0.144 & LOTSS &  & 296 (2) &  & 112 (2) &  & 54 (1) & 37 (1) &  &  & \citep{2022AA...659A...1S} \\
0.15 & TGSS & 123 (25) & 249 (50) & 145 (29) & 154 (31) & 160 (32) &  &  &  & 167 (33) & \citep{2017AA...598A..78I} \\
0.151 & 7C &  &  &  & 184 (11) &  &  &  &  &  & \citep{1995AAS..110..419V} \\
0.2 & GLEAM & 116 (23) &  &  &  &  &  &  &  &  & \citep{2017MNRAS.464.1146H} \\
0.325 & WN &  & 147 (29) &  & 121 (24) &  &  & 25 (5) & 138 (28) &  & \citep{1997AAS..124..259R} \\
0.365 & Texas &  & 307 (46) &  &  &  &  &  &  &  & \citep{1996AJ....111.1945D} \\
0.408 & B3 &  & 200 (20) &  &  &  &  &  &  &  & \citep{1985AAS...59..255F} \\
0.888 & RACS & 51.3 (4.2) &  & 79.4 (6.1) &  & 69.6 (5.4) &  &  & 146.5 (10.8) & 102.3 (7.8) & \citep{2020PASA...37...48M} \\
1.37 & RACS & 41.0 (2.7) &  & 46.9 (2.9) & 60.3 (3.6) & 46.4 (2.8) & 34.1 (2.1) & 51.1 (3.1) & 119.0 (7.2) & 84.7 (5.1) & \citep{2024PASA...41....3D} \\
1.40 & FIRST & 43.5 (0.1) & 45.4 (0.2) & 80.3 (0.2) & 54.9 (0.2) & 53.8 (0.1) & 32.3 (0.1) & 49.9 (0.1) & 121.5 (0.1) &  & \citep{1997ApJ...475..479W} \\
1.40 & NVSS & 47.2 (1.5) & 48.9 (1.9) & 59.5 (1.8) & 53.5 (1.7) & 52.8  (1.6) & 33.7 (1.1) & 45.3 (1.4) & 124.8 (3.8) & 111.1 (3.4) & \citep{1998AJ....115.1693C} \\
1.66 & RACS & 40.0 (4.0) &  & 38.0 (3.8) & 49.9 (5.0) & 44.8 (4.5) & 30.3 (3.1) & 47.8 (4.8) & 110.0 (11.0) & 81.3 (8.1) & \citep{2025PASA...42...38D} \\
2.70 &  VLASS 1.1 & 25.9 (0.3) & & 44.3 (0.3) & 42.7 (0.2) & 32.3 (0.3) & 28.7 (0.2) & 42.4 (0.2) & 79.4 (0.3) & 80.1 (0.3) &  \\
2.70 & VLASS 2.2 & 28.3 (0.3) &  & 41.3 (0.3) & 42.8 (0.3) & 33.5 (0.3) & 26.9 (0.3) & 40.2 (0.3) & 85.5 (0.3) & 64.6 (0.3) &  \\ 
2.7 & RASS &  &  &  &  &  &  &  & 106 (21) &  & \citep{2000AA...363..141R} \\
4.85 & GB6 & 34 (6) & 31 (4) & 53 (7) & 114 (11) & 33 (5) & 30 (4) & 36 (5) & 84 (9) & 78 (8) & \citep{1996ApJS..103..427G} \\
4.85 & MITG &  &  &  &  &  &  &  & 93 (19) &  & \citep{1991ApJS...75..801G} \\
4.85 & BWE &  & 31 (7) & 48 (9) & 35 (7) &  &  &  & 77 (15) & 78 (16) & \citep{1991ApJS...75....1B} \\
4.85 & PMN &  &  &  &  &  &  &  & 97 (19) &  & \citep{1996ApJS..103..145W} \\
4.85 & 87GB &  & 38 (8) & 66 (13) &  & 44 (9) & 27 (6) & 27 (6) & 111 (22) & 81 (16) & \citep{1991ApJS...75.1011G} \\
5 & RASS &  &  &  &  &  &  &  & 88 (18) &  & \citep{2000AA...363..141R} \\
8.40 & CLASS & 24 (5) &  & 51 (10) & 23 (5) &  &  & 45 (9) & 64 (13) & 58 (12) & \citep{2003MNRAS.341....1M} \\ 
107 & RASS &  &  &  &  &  &  &  & 54 (11) &  & \citep{2000AA...363..141R} \\ \hline\hline
4.3 &  &  &  & 68$^{\dagger}$ &  &  &  &  & 77$^{\dagger}$ & 53$^{\dagger}$ &  \\
7.6 &  &  &  & 51$^{\dagger}$ &  &  &  &  & 54$^{\dagger}$ & 45$^{\dagger}$ &  \\
1.7 & EVN & 18.8 (1.9) &  &  &  &  &  &  &  &  & \citep{2022ApJS..260...49K} \\
2.3 & &  &  &  &  &  &  &  &  &  & \citep{2010AA...520A.113B} \\
5 &   & 15.4 (0.9) &  &  &  &  &  &  &  &  & \citep{2022ApJS..260...49K} \\
\enddata
\tablecomments{$^{\dagger}$ median flux density. Col.~1 -- frequency of the flux density measurement; Col.~2 -- instrument or survey name; Cols.~3--11 -- flux density measurements of the given source in mJy (errors are given in parentheses where available)}; Col.~12 -- reference of the flux density measurement.
\end{deluxetable*}
\end{longrotatetable}

\begin{acknowledgments}
The EVN is a joint facility of independent European, African, Asian, and North American radio astronomy institutes. Scientific results from data presented in this publication are derived from the following EVN projects: EC062 and EC066.
This work presents results from the European Space Agency (ESA) space mission Gaia. Gaia data are being processed by the Gaia Data Processing and Analysis Consortium (DPAC). Funding for the DPAC is provided by national institutions, in particular the institutions participating in the Gaia MultiLateral Agreement (MLA). The Gaia mission website is \url{https://www.cosmos.esa.int/gaia}. The Gaia archive website is \url{https://archives.esac.esa.int/gaia}.
This research has made use of the NASA/IPAC Extragalactic Database (NED) which is operated by the Jet Propulsion Laboratory, California Institute of Technology, under contract with the National Aeronautics and Space Administration.
This research has made use of the VizieR catalog access tool, CDS, Strasbourg, France (DOI: 10.26093/cds/vizier). The original description  of the VizieR service was published in \citet{2000A&AS..143...23O}. 
We used in our work the Astrogeo VLBI FITS image database, DOI: 10.25966/kyy8-yp57, maintained by Leonid Petrov. 
This research was supported by the Hungarian National Research, Development and Innovation Office (NKFIH), grant number OTKA K134213, and by the NKFIH excellence grant TKP2021-NKTA-64.
This research was also supported by HUN-REN.
AC, LI and AM acknowledge financial support from INAF under the projects “Quasar jets in the early Universe” (Ricerca Fondamentale 2022) and “Testing the obscuration in the early Universe” (Ricerca Fondamentale 2023)
\end{acknowledgments}





%
\facilities{EVN}

\software{{\sc Aips} \citep{2003ASSL..285..109G},
        {{\sc ParselTongue}} \citep{2006ASPC..351..497K},
        {\sc Difmap} \citep{1997ASPC..125...77S},
        {\sc astropy} \citep{2013A&A...558A..33A,2018AJ....156..123A,2022ApJ...935..167A}, {\sc Matplotlib} \citep{2007CSE.....9...90H}
          }


\appendix

\section{Notes on Individual Sources} \label{notes}

\noindent\textit{J0031$+$1507:} The 5-GHz EVN image shows the core component extending towards the northwestern direction. The \textit{Gaia} position is slightly offset in the same direction, with $\sim 5$~mas separation ($< 3\sigma_\mathrm{pos}$). The contradicting properties make the classification of this source uncertain. However, its total radio spectrum is peaked, and  based on the spectral turnover frequency ($>5$~GHz), this source might belong to the subclass of HFP \citep[e.g.][]{2000A&A...363..887D} that display compact blazar-like pc-scale structure \citep{2012MNRAS.424..532O}. 
\\
\textit{J0121$+$0347:} The total radio spectrum is steep, but can be considered flat within the uncertainties. A marginal \textit{Gaia}--VLBI positional offset is found. The lack of Doppler-boosted radio emission and the X-ray class suggest that this is a misaligned jetted quasar, more specifically a CSS source.
\\
\textit{J0122$+$0309:} The EVN image shows a compact core component, which has an inverted (or flat within the uncertainties) spectrum and Doppler-boosted radio emission. The source found to be variable, supporting its blazar classification.
\\
\textit{J0257$+$4338:} This is the brightest object in our sample on mas scale, its $5$-GHz flux density is an order of magnitude higher than for the rest of our sample. The core component shows an extension in the northeastern direction, similarly to other, $8$-GHz VLBA images \citep{2025ApJS..276...38P}. As the lower-resolution $2$-GHz VLBA images show an outer jet component in the northwestern direction, this suggests an overall bendt jet structure. \citet{2025Univ...11...91G} measured the apparent speed of the inner jet component based on the $8$-GHz VLBA data and obtained $(0.99 \pm 0.22)$~c. They estimated the Doppler factor ($\delta = 1.95$) and Lorentz factor ($\Gamma = 1.48$), as well as the viewing angle ($\iota = 27.6^{\circ}$). The \textit{Gaia} position has a slight offset from the 5-GHz radio position. The total radio spectrum is peaked, while the scatter of VLBI flux densities measured at different epochs suggests a highly variable nature. As variability affects the total flux density values at different epochs, the spectral shape could also be flat rather than peaked. According to our results, the radio emission is significantly Doppler-boosted and this source is a blazar.
\\
\textit{J0835$+$1825:} The $5$-GHz EVN image shows a single compact component with Doppler-boosted emission. The total radio spectrum is flat. Considering the X-ray class as well, this source appears as a blazar (FSRQ).
\\
\textit{J0918$+$0637:} The source displays a single compact core at $5$~GHz. The \textit{Gaia} optical position has a significant offset (about $5.5$~mas) from the radio core. The total radio spectrum is peaked, however, a flat power-law fit gives only a slightly higher reduced $\chi^2$ value. This source is variable and possibly Doppler boosted, as also found by \citet{2022ApJS..260...49K}. These new data are somewhat contradictory to the previous blazar classifications, therefore we rather consider this source as unceratain.
\\
\textit{J1021$+$2209:} The source shows a bright core component and a jet extension in the north-eastern direction. The modeled flux density of the jet component is $\sim 25\%$ of the core flux density. The total radio spectrum is flat with $\alpha_{\rm total} = -0.48 \pm 0.03$. The core emission is highly Doppler-boosted ($\delta \geq 19.7$). This is a blazar-type source with core--jet structure also seen in archival $8$-GHz VLBA images \citep{2025ApJS..276...38P}.   
\\
\textit{J1325$+$1123:} We deteted a compact core component without any extended feature. The $5$-GHz radio and the \textit{Gaia} optical positions agree within errors. The total radio spectrum is peaked, and the VLBI spectrum is steep, as was also reported by \citet{2022ApJS..260...49K}, who proposed this source as a blazar with gigahertz-peaked spectrum. However, here we found no evidence for Doppler boosting, in contrast to \citet{2022ApJS..260...49K}. As it was classified as blazar in the X-ray class, contradicting the radio data, we consider its classification as uncertain.
\\
\textit{J1348$+$1935:} The source shows a core--jet structure, where a faint jet component is visible in the northeast. The modeled flux density of the jet component is only $20\%$ compared to the core. The core brightness temperature is a lower limit, but it exceeds the intrinsic value, indicating Doppler boosting. The total radio spectrum is flat. This source is classified as a core--jet blazar.
\\
\textit{J1412$+$0624:} A faint compact component is detected. We did not detect the spurious extended features at $5$~GHz seen by \citet{2022ApJS..260...49K}. The \textit{Gaia} position has a marginal offset with respect to the radio position. Both the total and the VLBI spectra are flat, as reported also by \citet{2022ApJS..260...49K}. This source was previously classified as a blazar based on X-ray and VLBI data, but it is on the border of the blazar definition, due to the lack of Doppler boosting and observed variability. The class is uncertain.
\\
\textit{J1535$+$0254:} We detected a single compact core. The \textit{Gaia} measurement has a large excess noise, therefore, the $3\sigma_\mathrm{pos}$ positional error approaches $\sim 10$~mas. The total radio spectrum is flat, as well as the VLBI spectrum based on the VLBA and our EVN flux densities. This source is an FSRQ type.
\\
\textit{J1612$+$4703:} This is the faintest detected source in the sample with $3.6$~mJy~beam$^{-1}$ peak brightness. There is an extension to the north of the core, which can be fitted with a component with $\sim 1$~mJy flux density. The \textit{Gaia} measurement also has a large excess noise, resulting in high positional uncertainties, close to $\sim 10$~mas, like in the case of J1535$+$0254. The total radio spectrum is flat. The core brightness temperature is a lower limit. We classify this source as a misaligned one, which agrees with what \citet{2019MNRAS.489.2732I} proposed based on the X-ray data.
\\
\textit{J1629$+$0959:} We detected a single component that shows slightly Doppler-boosted emission. The source has a flat total radio spectrum. Its blazar classification is supported by both the radio and the X-ray data.
\\
\textit{J1648$+$4603:} Similarly to most of the targets, only a single component is present in the $5$-GHz EVN image. The core is Doppler-boosted, however, slightly extended to the south. The total radio spectrum is flat. Based on the radio data and the previous X-ray classification, we propose this source as a blazar (FSRQ).
\\
\textit{J1711$+$3830:} This source has an inverted radio spectrum (or flat within the uncertainties). Evidence for Doppler boosting is not found in the single core component detected, but the source is variable, and its X-ray class is blazar. We consider it uncertain, because of the contradicting properties found in the radio.
\\
\textit{J2314$+$0201:} We detected a compact core component at $5$~GHz. There are two symmetrical noise peaks to the north and south, at $\sim 10$~mas from the core. The northern peak is in the direction of the extended part of the central component seen in the archival $4$-GHz VLBA image \citep{2025ApJS..276...38P}, albeit with poor resolution. The total radio spectrum of the source is flat, and Doppler boosting is detected. In agreement with the X-ray class, this source is a blazar. Notably, the \textit{Gaia} optical position is also in this direction, offset by $\sim 15$~mas from the radio core, which is a highly significant value. The large optical--radio positional offset is challenging the blazar definition. One might speculate about a dual AGN system whose components are a radio-loud but optically faint (represented by the VLBI source), and a radio-quiet but optically bright one (pinpointed by \textit{Gaia}), respectively. There is no other supporting evidence for this scenario, but if it is true, the projected separation of the AGN companions would be just $\lesssim 100$~pc. Recent observational results indicate that the fraction dual AGN with $\lesssim 10$~kpc separations can reach up to $20-30\%$ in the early Universe, even more than predicted by cosmological simulations \citep{2025A&A...696A..59P}.
\\
\textit{J2357$+$1402:} This source has the most complex mas-scale radio structure, as it is resolved into multiple components. The jet starts in the southwestern direction and then appears to bend toward the south. The southwestern jet component is located $3.2$~mas to the core and modeled with $\sim 3.9$~mJy flux density. An archival $4$-GHz VLBA image is also consistent with this jet extension \citep{2025ApJS..276...38P}. In the future, jet proper motion could potentially be detected using multiple-epoch VLBI imaging of this source, complementing a rather limited $z>4$ sample studied to date \citep[e.g.][]{2025Univ...11...91G}. The spectra constructed from total flux density and VLBI measurements are both flat. The core coincides with the \textit{Gaia} position and has Doppler-boosted emission at $5$~GHz, indicating that this source is a core--jet blazar. 
\\
\textit{J1517$+$3753:} This source turned out to be a low-redshift quasar ($z = 0.56$). Also, this is the only source in our sample not detected with the EVN at $5$~GHz. Applying $(u,v)$ taper in \textsc{Difmap} to reduce the weights of the long baselines did not reveal any extended emission in the vicinity of its FIRST position. The $5\sigma$ upper limit to its brightness is $0.18$~mJy\,beam$^{-1}$. Its radio spectrum is steep, with $\alpha_\mathrm{total} = -0.66 \pm 0.08$. The FIRST image shows faint extended features, which is consistent with a source resolved out on mas scales. 


\bibliography{main}
\bibliographystyle{aasjournalv7}



\end{document}